\documentclass[journal,comsoc]{IEEEtran}

\usepackage[T1]{fontenc}%
\usepackage[utf8]{inputenc}
\usepackage{textcomp}
\usepackage{hyperref}
\usepackage{cite}
\usepackage[pdftex]{graphicx}
\usepackage{amsmath}
\usepackage[cmintegrals]{newtxmath}
\usepackage{trfsigns}
\usepackage{upgreek}

\usepackage{tikz}
\usetikzlibrary{calc,shapes,arrows,positioning,arrows.meta,bending,automata,quotes,angles}

\tikzstyle{sum} = [thick,draw, circle,inner sep=-1pt,outer sep=0pt,font=\normalsize]

\pgfmathsetmacro{\blockwidth}{30}
\pgfmathsetmacro{\blockheight}{17}

\tikzstyle{comblock} = [thick,black,fill=white,draw=black, rectangle,minimum height=\blockheight pt,minimum width=\blockwidth pt]
\tikzstyle{comtriag} = [regular polygon, regular polygon sides=3,
              draw=black, black,  text width=0.90em,
              inner sep=0.9mm, outer sep=0mm,
              shape border rotate=-90]

\tikzstyle{input} =  [coordinate]
\tikzstyle{output} = [coordinate]

\tikzset{%
Double/.style={%
    to path={%
      ($(\tikztostart)!2pt!90:($(\tikztotarget)!5pt!(\tikztostart)$)$) -- ($($(\tikztotarget)!2.3pt!(\tikztostart)$)!2pt!270:(\tikztostart)$)
      ($(\tikztostart)!2pt!270:($(\tikztotarget)!4pt!(\tikztostart)$)$) -- ($($(\tikztotarget)!2.3pt!(\tikztostart)$)!2pt!90:(\tikztostart)$)
      ($($(\tikztotarget)!5pt!(\tikztostart)$)!5pt!90:(\tikztostart)$)
       .. controls
          ($($(\tikztotarget)!3pt!(\tikztostart)$)!3pt!90:(\tikztostart)$) and
          ($(\tikztotarget)!0.1pt!(\tikztostart)$)
       .. (\tikztotarget)
       .. controls
          ($(\tikztotarget)!0.1pt!(\tikztostart)$) and
          ($($(\tikztotarget)!3pt!(\tikztostart)$)!3pt!270:(\tikztostart)$)
       ..
     ($($(\tikztotarget)!5pt!(\tikztostart)$)!5pt!270:(\tikztostart)$)
    }
  }
}

\usepackage[exponent-product = \times]{siunitx}
\usepackage{graphicx}

\usepackage{cleveref}
\usepackage{mathtools}
\usetikzlibrary{patterns}

\usepackage{pgfplots}
\pgfplotsset{compat=1.14}
\usepackage{pgfplotstable}

\usepackage{booktabs}
\usepackage{trfsigns}
\usepackage[b]{esvect}
\usepackage{blkarray}
\usepackage{multirow,array}

\makeatletter
\newcommand*\bigcdot{\mathpalette\bigcdot@{.5}}
\newcommand*\bigcdot@[2]{\mathbin{\vcenter{\hbox{\scalebox{#2}{$\m@th#1\bullet$}}}}}
\makeatother

\newcommand{\cvv}[1]{\reflectbox{\ensuremath{\vv{\reflectbox{\ensuremath{#1}}}}}}

\DeclareMathOperator*{\T}{T}

\DeclareMathOperator*{\argmax}{arg\,max}
\DeclareMathOperator*{\argmin}{arg\,min}
\DeclareMathOperator*{\E}{\mathbb{E}}
\DeclareMathOperator{\sinc}{sinc}
\DeclareMathOperator{\trace}{tr}

\newcommand{\moment}[1]{\ensuremath{\mathrm{\mu}_{4}}}

\newcommand{\uarg}{\ensuremath{{\bm{\cdot}}}}

\newcommand{\dimC}[1]{\ensuremath{\in \mathbb{C}^{#1}}}
\newcommand{\dimR}[1]{\ensuremath{\in \mathbb{R}^{#1}}}
\newcommand{\covm}[2]{\ensuremath{\boldsymbol{\mathbf{#1}}_{{#2}}}}
\newcommand{\mean}[1]{\ensuremath{\boldsymbol{\mathbf{\upmu}}_{{#1}}}}
\newcommand{\mident}[1]{\ensuremath{\mathrm{\textbf{I}}_{#1}}}
\newcommand{\mnull}[1]{\ensuremath{\mathrm{\textbf{0}}_{#1}}}
\newcommand{\mones}[1]{\ensuremath{\mathrm{\textbf{1}}_{#1}}}

\newcommand{\He}[1]{\ensuremath{h\!\left({#1}\right) }}

\newcommand{\MI}[2]{\ensuremath{I\!\left({ \mathrm{#1} ; \mathrm{#2}}\right) }}
\newcommand{\MIsub}[3]{\ensuremath{I{#3}\!\left({ \mathrm{#1} \,; \mathrm{#2}}\right) }}

\newcommand{\Div}[2]{\ensuremath{D\!\left({ #1 \,\|\, #2}\right) }}

\newcommand{\Prob}[2]{\ensuremath{p_{\mathrm{#1}}({#2}) }}

\newcommand{\GaussDist}[3]{\ensuremath{\mathcal{N}\!\left(#1;\,#2\,,#3\right) }}

\newcommand{\bm}[1]{\ensuremath{\boldsymbol{\mathbf{#1}}}}

\definecolor{mycolor2}{RGB}{196,7,27}
\definecolor{mycolor1}{RGB}{0,101,189}
\definecolor{bblack}{RGB}{088,088,090}
\definecolor{oorange}{RGB}{255,180,000}
\definecolor{mycolor3}{RGB}{255,180,000}
\definecolor{mycolor4}{HTML}{3A7E55}
\definecolor{mycolor5}{HTML}{3A7E55}

\tikzstyle{recty} = [draw, rectangle, scale=1.2,font=\normalsize,minimum height=2.5em,minimum width=3em]

\usepackage{graphicx}
\usepackage{caption}
\usepackage{subcaption}

\usepackage{etoolbox}

\begin{document}

\title{Achievable Rates for Short-Reach Fiber-Optic Channels with Direct Detection}
\author{Daniel Plabst, 
Tobias Prinz,~\IEEEmembership{Student Member,~IEEE}, Thomas Wiegart,~\IEEEmembership{Student Member,~IEEE}, \\
Talha Rahman,~\IEEEmembership{Member,~IEEE},
Neboj\v{s}a Stojanovi\'{c},~\IEEEmembership{Member,~IEEE},
Stefano Calabr\`o, \\ Norbert Hanik,~\IEEEmembership{Senior Member,~IEEE},
and Gerhard Kramer,~\IEEEmembership{Fellow,~IEEE}%
\thanks{Date  of  current  version  \today.}
\thanks{Accepted to \textit{J. Lightw. Technol.} on January 22, 2022.}
\thanks{D.~Plabst, T.~Prinz, T.~Wiegart, N.~Hanik, and G.~Kramer are with the Institute for Communications Engineering, Technical University of Munich, 80333 Munich, Germany (e-mail: $\lbrace$daniel.plabst,\, tobias.prinz,\, thomas.wiegart,\, norbert.hanik,\, gerhard.kramer$\rbrace$@tum.de). T.~Rahman, N.~Stojanovi\'{c}, and S.~Calabr\`o, are with the Huawei Munich Research Center, 80992 Munich, Germany (e-mail:
$\lbrace$talha.rahman,\, nebojsa.stojanovic,\, stefano.calabro$\rbrace$@huawei.com).}}
 
\markboth{}%
{Shell \MakeLowercase{\textit{et al.}}: Bare Demo of IEEEtran.cls for IEEE Communications Society Journals}

\makeatletter
\patchcmd{\@maketitle}
  {\addvspace{0.5\baselineskip}\egroup}
  {\addvspace{-1.0\baselineskip}\egroup}
  {}
  {}
\makeatother

\maketitle

\begin{abstract}
Spectrally efficient communication is studied for short-reach fiber-optic links with chromatic dispersion (CD) and receivers that employ direction-detection and oversampling. 
Achievable rates and symbol error probabilities are computed by using auxiliary channels that account for memory in the sampled symbol strings. 
Real-alphabet bipolar and complex-alphabet symmetric modulations are shown to achieve significant energy gains over classic intensity modulation.
Moreover, frequency-domain raised-cosine (FD-RC) pulses outperform time-domain RC (TD-RC) pulses in terms of spectral efficiency for two scenarios. First, if one shares the spectrum with other users then inter-channel interference significantly reduces the TD-RC rates. Second, if there is a transmit filter to avoid interference then the detection complexity of FD-RC and TD-RC pulses is similar but FD-RC achieves higher rates.
\end{abstract}


\IEEEpeerreviewmaketitle

\section{Introduction}
Fiber-optic communication with direct detection (DD) is a cost-effective approach for intra-datacenter communication up to several dozen kilometers~\cite{chagnon_optical_comms_short_reach_2019}. DD receivers measure the amplitude or \emph{intensity} of signals, e.g., by using a photodiode (PD). This reduces the hardware complexity, cost and form factors as compared to coherent receivers~\cite{chagnon_optical_comms_short_reach_2019,zhong_dsp_for_short_reach_2018,qian_imdd_beyond_bw_lim_dcoi_2019}. However, the spectral efficiency (SE) is reduced and compensating transmission impairments is challenging~\cite{kramer_SE_2003,randel_dmt_80km_2015,wiener_filter_plabst2020}.

DD receivers are usually paired with intensity modulation (IM) transmitters to allow for easy reconstruction of the transmitted data~\cite{dissanayake_comparison_ofdm_imdd_2013,chen_performance_analysis_ofdm_imdd_2012,mecozzi_imdd_capacity_optical_amp2001}. More generally, one may use several amplitudes, e.g., $Q$-PAM with $Q\ge2$, together with DD and receiver signal processing such as linear equalization and machine learning~\cite{wiener_filter_plabst2020,wettlin_dsp_short_reach_2020,karanov_deep_learning_2018}. IM constellations have a positive mean that aids DD but carries no information and consumes energy. 

\subsection{Oversampling}
Oversampling at the receiver increases the capacity of DD systems~\cite{mecozzi_capacity_amdd_2018}. In fact, if optical amplification (OA) noise is the dominant impairment then DD can operate within \SI{1}{bit/s/Hz} of the coherent detection capacity, and this motivates using complex-valued constellations~\cite{mecozzi_capacity_amdd_2018,SecondiniDirectDetectionBPAM2020,tasbihi_capacity_waveform_channels_time-limited_2020,tasbihi2021direct}. For example, if the transmit signal is minimum phase, then its phase can be reconstructed from its intensity by using a Kramers-Kronig (KK) receiver~\cite{KK_receiver_mecozzi_2016}. After recovering the phase, one can mitigate linear impairments such as chromatic dispersion (CD) with standard methods such as linear equalizers or orthogonal frequency division multiplexing (OFDM). However, minimum phase signals require inserting an optical carrier that decreases the SE~\cite{SecondiniDirectDetectionBPAM2020,toward_practical_kk_receiver_bo_2019,li_joint_optimization_resampling_cspr_dd_kk_2017}. 

\begin{figure}[t!]
    \centering 
    \begin{tikzpicture}[xscale=1.2,yscale=0.9]
    \draw[-stealth, thick] (-1.2, 0) -- (5.3, 0) node[right,font=\footnotesize] {$t/T_\text{s}$};
    \draw[-stealth, thick] (0, -1.0) -- (0, 1.25) node[left,font=\footnotesize] {$X(t)$};
    
    \draw[-] (1, 0.1) -- (1, -0.1) node[below] {$1$};
    \draw[-] (2, 0.1) -- (2, -0.1) node[below] {$2$};
    \draw[-] (3, 0.1) -- (3, -0.1) node[below] {$3$};
    \draw[-] (4, 0.1) -- (4, -0.1) node[below] {$4$};
    \draw[-] (5, 0.1) -- (5, -0.1) node[below] {$5$};
    \draw[-] (-1, 0.1) -- (-1, -0.1) node[below] {$-1$};
    \draw[-] (0.1, 1) -- (-0.1, 1) node[right,xshift=0.15cm] {$1$};
    
    \draw [gray, densely dotted, smooth,samples=10,domain=-1:0] plot(\x,{-(\x+1)});
    \draw [gray, densely dotted, smooth,samples=10,domain=0:1] plot(\x,{-(-\x+1)});
    
    \draw [gray, dashed, smooth,samples=10,domain=0:1] plot(\x,{(\x)});
    \draw [gray, dashed, smooth,samples=10,domain=1:2] plot(\x,{(-\x+2)});
    
    \draw [gray, dashed, smooth,samples=10,domain=1:2] plot(\x,{(\x-1)});
    \draw [gray, dashed, smooth,samples=10,domain=2:3] plot(\x,{(-\x+3)});
    
    \draw [gray, densely dotted, smooth,samples=10,domain=2:3] plot(\x,{-(\x-2)});
    \draw [gray, densely dotted, smooth,samples=10,domain=3:4] plot(\x,{-(-\x+4)});
    
    \draw [gray, dashed, smooth,samples=10,domain=3:4] plot(\x,{(\x-3)});
    \draw [gray, dashed, smooth,samples=10,domain=4:5] plot(\x,{(-\x+5)});
    
   \draw [red, thick]  (0,-1) -- (1,1) -- (2,1) -- (3,-1) -- (4,1); 
   
   \node[font=\footnotesize,right] at(0,-1){$s_0\!=\!-1$}; 
   \node[font=\footnotesize,above] at(1,1){$s_1\!=\!+1$}; 
   \node[font=\footnotesize,above] at(2,1){$s_2\!=\!+1$}; 
   \node[font=\footnotesize,right] at(3,-1){$s_3\!=\!-1$}; 
\node[font=\footnotesize,above] at(4,1){$s_4\!=\!+1$}; 
   
\end{tikzpicture}
    \begin{tikzpicture}[xscale=1.2,yscale=1,
    dot/.style    = {anchor=base,fill,circle,inner sep=2pt}]
    \draw[-stealth, thick] (-1.2, 0) -- (5.3, 0) node[right,font=\footnotesize] {$t/T_\text{s}$};
    \draw[-stealth, thick] (0, -1.0) -- (0, 1.25) node[left,font=\footnotesize] {$|X(t)|^2$};
    
    \draw[-] (1, 0.1) -- (1, -0.1) node[below] {$1$};
    \draw[-] (2, 0.1) -- (2, -0.1) node[below] {$2$};
    \draw[-] (3, 0.1) -- (3, -0.1) node[below] {$3$};
    \draw[-] (4, 0.1) -- (4, -0.1) node[below] {$4$};
    \draw[-] (5, 0.1) -- (5, -0.1) node[below] {$5$};
    \draw[-] (-1, 0.1) -- (-1, -0.1) node[below] {$-1$};
    \draw[-] (0.1, 1) -- (-0.1, 1) node[right,xshift=0.15cm] {$1$};
    
    \draw [gray, densely dotted, smooth,samples=10,domain=-1:0] plot(\x,{-(\x+1)});
    \draw [gray, densely dotted, smooth,samples=10,domain=0:1] plot(\x,{-(-\x+1)});
    
    \draw [gray, dashed, smooth,samples=10,domain=0:1] plot(\x,{(\x)});
    \draw [gray, dashed, smooth,samples=10,domain=1:2] plot(\x,{(-\x+2)});
    
    \draw [gray, dashed, smooth,samples=10,domain=1:2] plot(\x,{(\x-1)});
    \draw [gray, dashed, smooth,samples=10,domain=2:3] plot(\x,{(-\x+3)});
    
    \draw [gray, densely dotted, smooth,samples=10,domain=2:3] plot(\x,{-(\x-2)});
    \draw [gray, densely dotted, smooth,samples=10,domain=3:4] plot(\x,{-(-\x+4)});
    
    \draw [gray, dashed, smooth,samples=10,domain=3:4] plot(\x,{(\x-3)});
    \draw [gray, dashed, smooth,samples=10,domain=4:5] plot(\x,{(-\x+5)});
    
    \draw [red, thick, samples=50, domain=0:1] plot(\x, {(2*\x-1)^2});
    \draw [red, thick, samples=50, domain=1:2] plot(\x, 1);
    \draw [red, thick, samples=50, domain=2:3] plot(\x, {(-2*\x+5)^2});
    \draw [red, thick, samples=50, domain=3:4] plot(\x, {(2*\x-7)^2});
    
    \draw (0.5, 0) coordinate[blue, dot];
    \draw (1.5, 1) coordinate[blue, dot];
    \draw (2.5, 0) coordinate[blue, dot];
    \draw (3.5, 0) coordinate[blue, dot];
    
    \draw (0, 1) coordinate[red, dot, inner sep=1.5pt];
    \draw (1, 1) coordinate[red, dot, inner sep=1.5pt];
    \draw (2, 1) coordinate[red, dot, inner sep=1.5pt];
    \draw (3, 1) coordinate[red, dot, inner sep=1.5pt];
    \draw (4, 1) coordinate[red, dot, inner sep=1.5pt];
    
    \draw[dotted,-stealth] (0,0) -- (0,-1.3) node[below,xshift=0.08cm]{\textcolor{red}{$\diamondsuit_1$}}; 
    \draw[dotted,-stealth] (0.5,0) -- (0.5,-1.3) node[below,xshift=0.08cm]{\textcolor{blue}{$\spadesuit_1$}}; 
    \draw[dotted,-stealth] (1,0) -- (1,-1.3) node[below,xshift=0.08cm]{\textcolor{red}{$\diamondsuit_2$}}; 
    \draw[dotted,-stealth] (1.5,0) -- (1.5,-1.3) node[below,xshift=0.08cm]{\textcolor{blue}{$\spadesuit_2$}}; 
    \draw[dotted,-stealth] (2,0) -- (2,-1.3) node[below,xshift=0.08cm]{\textcolor{red}{$\diamondsuit_3$}}; 
    \draw[dotted,-stealth] (2.5,0) -- (2.5,-1.3) node[below,xshift=0.08cm]{\textcolor{blue}{$\spadesuit_3$}}; 
    \node[below] at (3,-1.3){$\ldots$};
    \node[below,font=\footnotesize] at (-0.8,-1.3){Samples:};
\end{tikzpicture}
    \caption{Phase retrieval in DD.}
    \label{fig:dd_intro_oversampling}
\end{figure}
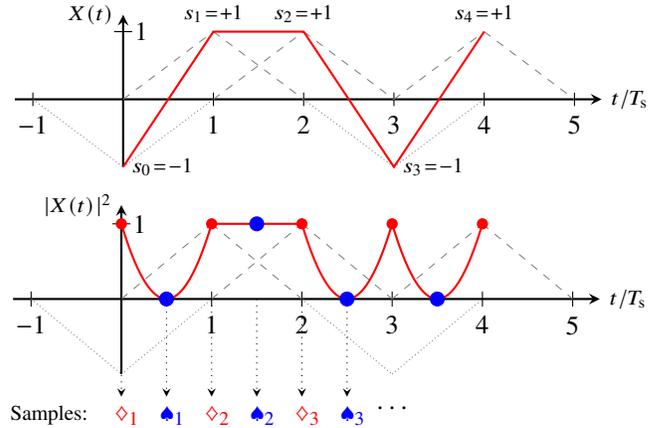

Real-valued \emph{bipolar} constellations ($Q$-ASK) were investigated in~\cite{SecondiniDirectDetectionBPAM2020}. For example, Fig.~\ref{fig:dd_intro_oversampling} depicts symbol reconstruction when the baseband pulse is triangular over two symbol periods $2 T_\text{s}$. The solid curve in the first subplot shows the received baseband signal $X(t)$ when using binary phase-shift keying (BPSK) and the transmit symbol string $(-1,+1,+1,-1,+1)$. The second subplot shows the DD output. Clearly, the samples $\diamondsuit_\kappa$ at symbol times $t = \kappa T_\text{s}$, $\kappa \in \mathbb{Z}$, carry no information. However, the samples $\spadesuit_\kappa$ at half-symbol times $t = (\kappa+\frac{1}{2})T_\text{s}$ experience intersymbol interference (ISI) and give information on phase changes. Thus, one can reconstruct the transmit string by using differential phase decoding~\cite{SecondiniDirectDetectionBPAM2020}.

\subsection{Pulse Shaping and Line Coding}
The triangular pulse in Fig.~\ref{fig:dd_intro_oversampling} was chosen to limit the oversampled ISI but without considering the spectral occupancy. Two classic methods to control the ISI and spectrum, and other signal properties, are \emph{pulse shaping} and \emph{line coding}.

A fundamental approach to pulse shaping with simultaneous timewidth and bandwidth constraints is based on prolate spheroidal wave functions (PSWFs)~\cite{slepian_pswf}. The application of PSWFs to data transmission was already pointed out in~\cite[Sec.~5.2]{slepian_pswf_2}. In practice, however, one usually uses pulses that satisfy the Nyquist ISI criterion. Perhaps the most popular choice is the class of frequency-domain raised-cosine (FD-RC) functions\footnote{For memoryless channels one often uses frequency-domain root-raised-cosine pulses with a matched filter so that the end-to-end pulse is FD-RC. However, the CD and DD make this choice less interesting.}~\cite[Eq.~(6.17)-(6.18)]{gallager2008principles}. For example, the paper~\cite{SecondiniDirectDetectionBPAM2020} uses an FD-RC pulse with a unit roll-off factor to limit the oversampled ISI to the half-symbol times $t = \pm T_\text{s}/2$. However, such pulses are spectrally inefficient. 
The paper~\cite{tasbihi2021direct} instead uses a time-domain RC (TD-RC) pulse\footnote{TD-RC pulses are called ``Tukey waveforms'' in~\cite{tasbihi2021direct} and they have been studied, e.g., for deep-space communication in~\cite{McEliece-Palanki-02}.} that extends up to two symbol periods $2T_s$ so that the ISI is limited to the half-symbol times $t=\pm T_\text{s}/2$, as in~\cite{SecondiniDirectDetectionBPAM2020} and Fig.~\ref{fig:dd_intro_oversampling}. We show that the SEs of FD-RC pulses with small roll-off factors are better than those of TD-RC pulses, see Sec.~\ref{subsec:fdrc-tdrc} below.

We remark that using TD-RC pulses is a special case of partial-response signaling~\cite{Lender-Spectrum-66,Kabal-Pasupathy-COMM75} or faster-than-Nyquist signaling~\cite{MacColl-36,Lender-TCOM63,mazo75bstj}. Another closely related approach is continuous phase modulation~\cite{Aulin-TCOM81-1,Aulin-TCOM81-2}. As pointed out in~\cite{Massey-IZS84}, each of these systems can be understood as a coded modulation where the code simultaneously shapes the signal spectrum and provides coding gain. In the fiber-optic literature, several authors have designed such codes for CD channels and DD receivers. For example, the duobinary signaling of~\cite{Lender-TCOM63} was generalized in~\cite{Penninckx-PTL97} and~\cite{Stark-OFC99} under the names Phase-Shaped Binary Transmission (PSBT) and Phased Amplitude-Shift Signaling (PASS), respectively. These methods were further generalized in~\cite{Forestieri-JLT01} and the extensions are called Combined Amplitude-Phase Shift (CAPS) codes in~\cite{Forestieri-JLT17,Morsy-Osman-OE18}.

\subsection{Contributions and Organization}
We focus on short-reach fiber-optic channels without OA where thermal noise limits performance~\cite[p.~154]{AgrawalFourthEdFiberOptics}. Our goal is to study reliable communication for channels with CD, DD and oversampling, and with spectrally efficient pulses. In particular, we focus on FD-RC pulses with small roll-off factor. Such pulses introduce significant ISI when oversampling so that an optimal receiver is complex. However, we find that simplified receivers based on auxiliary (mismatched) channel models can perform well. Moreover, the resulting SEs are benchmarks for designing near-optimal receiver algorithms and for improving signaling.

We remark that many short-reach fiber-optic systems are based on low-cost and low-SE designs with a few wavelengths, short error-control codes, small modulation alphabets, and DD with coarse-quantization and hard-decision decoders. However, the continuing growth in demand for high data rates should result in the usual evolution to systems with high SE via many wavelengths, sophisticated error-control codes, higher-order modulation, and receivers with fine quantization and soft-decision decoders. One purpose of this paper is to study the performance of such systems with one simplifying hardware constraint, namely the use of DD receivers, and with sophisticated signaling and signal processing.

This paper is organized as follows. Sec.~\ref{sec:system-model} develops continuous- and discrete-time models for short-reach fiber-optic channels. Sec.~\ref{sec:achievable_rates} reviews lower and upper bounds on achievable rates based on optimized auxiliary channel densities~\cite{topsoe1967information,arnoldsimulationmi}. Sec.~\ref{sec:symbolwise_app_map} reviews symbol-wise maximum \emph{a posteriori} (MAP) detection based on the forward-backward algorithm~\cite{bcjr_1974}. Sec.~\ref{sec:simulation_results} gives numerical results on the SEs of different constellations, transmit pulses and link lengths. Sec.~\ref{sec:conclusions} provides concluding remarks and suggests research problems. The Appendix optimizes an auxiliary channel model.

\subsection{Notation} 
Vectors and matrices are written using bold letters. 
The transposes of the vector $\bm{a}$ is written as $\bm{a}^{\T}$.
The $n \times 1$ all-ones and all-zeros vectors are $\mones{n}$ and $\mnull{n}$, respectively, and
the $n \times n$ identity matrix is $\mident{n}$. The diagonal matrix with entries taken in order from $\bm{a}$ is written as $\text{diag}(\bm{a})$.
The notation $\bm{a}\otimes\bm{b}$ refers to the Kronecker product of $\bm{a}$ and $\bm{b}$.
The determinant and trace of the square matrix $\bm{A}$ are written as $\det \bm{A} $ and $\trace \bm{A}$, respectively. We use the string notation $x_\kappa^n=(x_\kappa,\ldots,x_n)$. Accordingly, we use $\mathbf{x}_\kappa^n= (\mathbf{x}_\kappa,\ldots,\mathbf{x}_n)$ for a string of vectors. 

The sinc function is defined as $\sinc(t) = \sin(\pi t)/(\pi t)$. The signal $a(t)$ and its Fourier transform $A(f)$ are related by the notation $a(t)$ \laplace\, $A(f)$. The expression $g(t)*h(t)$ refers to the convolution of $g(t)$ and $h(t)$ and $\|a(t)\|^2 = \int_{-\infty}^{\infty} \big|\, a(t) \big|^2 \mathrm{d}t $ is the energy of $a(t)$.

Random variables (RVs) are written with upper case letters and their realizations with lower-case letters. $\E[\cdot]$ denotes expectation. The mean of the random vector $\bm{A}$ is $\mean{\bm{A}}=\E[\bm{A}]$ and the covariance matrix of two real random vectors $\bm{A}$, $\bm{B}$ is $\covm{C}{\bm{A}\bm{B}}=\E[\bm{A}\bm{B}^{\T}]$.  Univariate and multivariate real Gaussian distributions are written as $\GaussDist{x}{\mu}{\sigma^2}$ and $\GaussDist{\mathbf{x}}{\boldsymbol{\upmu}}{\mathbf{C}}$, respectively, where $\mu$ and $\sigma^2$ denote mean and variance and $\boldsymbol{\upmu}$ and $\mathbf{C}$ denote mean vector and covariance matrix.
The distribution or density of the random vector $\bm{X}$ is written as $p_{\bm{X}}$. The joint distribution or density of $\bm{X}$ and $\bm{Y}$ is written as $p_{\bm{XY}}$. Entropy, divergence and mutual information are defined as in~\cite{cover1991elementsofIT}:
\begin{align}
    & h(\bm{X}) = \E \left[-\log_2 p(\bm{X})\right] \label{eq:entropy} \\
    & \Div{p_{\bm{X}}}{p_{\bm{Y}}} = \E \left[\log_2 \frac{p_{\bm{X}}(\bm{X})}{p_{\bm{Y}}(\bm{X})}\right] \\
    & I(\bm{X};\bm{Y}) = \Div{p_{\bm{XY}}}{p_{\bm{X}}\,p_{\bm{Y}}}
\end{align}
where we measure the quantities in bits. As in \eqref{eq:entropy}, we often discard subscripts on distributions or densities if the arguments are uppercase or lowercase versions of their RVs. A basic property of divergence is $\Div{p_{\bm{X}}}{p_{\bm{Y}}}\ge0$ with equality if and only if $p_{\bm{X}}=p_{\bm{Y}}$ almost everywhere.

\section{System Model}
\label{sec:system-model}

Propagation in fiber is described by the Nonlinear Schr\"odinger Equation (NLSE)~\cite[p.~65]{AgrawalFourthEdFiberOptics}, \cite[Part~2]{essiambre2010jlt} that exhibits attenuation and CD, both linear effects, and a Kerr nonlinearity. We are interested in short-reach links where the launch power is sufficiently small so that one may neglect the Kerr non-linearity, and where the system is built without OA so that one may neglect optical noise. We next describe the system model in more detail.

\subsection{Continuous-Time Model}
\label{sec:time-continuous-model}
\begin{figure*}[t!]
    \centering
    \usetikzlibrary{decorations.markings}
\tikzset{node distance=2.3cm}

\pgfdeclarelayer{background}
\pgfdeclarelayer{foreground}
\pgfsetlayers{background,main,foreground}
\tikzset{boxlines/.style = {draw=black!20!white,}}
\tikzset{boxlinesred/.style = {densely dashed,draw=red!50!white,thick}}

\pgfmathsetmacro{\samplerwidth}{30}

\tikzset{midnodes/.style = {midway,above,text width=1.5cm,align=center,yshift=-0.2em}}
\tikzset{midnodesRP/.style = {midway,above,text width=1.5cm,align=center,yshift=-1.4em}}

\begin{tikzpicture}[]
    \node [input, name=input] {Input};
    \node [comblock,right of=input,node distance=2.1cm] (txfilter) {$g_\text{tx}(t)$};
    \node [comblock,right of=txfilter] (cir) {$h_L(t)$};
    \node [comblock,right of=cir] (sld) {$\left\lvert \,\cdot\, \right\rvert^2$};
    \node [sum,right of=sld,node distance=1.7cm] (sumnode) {$+$};
    \node [input, name=noise,above of=sumnode,node distance=0.7cm] {Input};
    \node [comblock,right of=sumnode,node distance=2cm] (rxfilter) {$g_\text{rx}(t)$};
    \node [comblock,right of=rxfilter,minimum width=\samplerwidth pt,node distance=2.1cm] (sampler) {};
    \node [comblock,right of=sampler,node distance=2.9cm] (dsp) {Detector};
    \node [input, name=output, right of=dsp,node distance=1.3cm] {Output};
    \draw[thick] (sampler.west) -- ++(\samplerwidth/4 pt,0) --++(\samplerwidth/2.7 pt,\samplerwidth/4.5 pt );
    \draw[thick] (sampler.east) -- ++(-\samplerwidth/3pt,0);

    \draw ($(sampler.west) + (\samplerwidth/4.5 pt,0.25)$)edge[out=0,in=100,-latex,thick] ($(sampler.east) + (-\samplerwidth/2.5 pt,-0.2)$);

    \draw[-latex,thick] (input) -- node[midnodes](s_ti){$X_\kappa$  } (txfilter);

    \draw[-latex,thick] (txfilter) -- node[midnodes](){$X(t)$\\[0.2em] } (cir);
    \draw[-latex,thick] (cir) -- node[midnodes](){$X_L(t)$\\[0.2em] } (sld);
    \draw[-latex,thick] (sld) -- node[midnodes](){$Z^\prime(t)$\\[0.2em] } (sumnode);
    \draw[-latex,thick] (sumnode) -- node[midnodes](){$Y'(t)$\\[0.2em] } (rxfilter);
    \draw[-latex,thick] (rxfilter) -- node[midnodes](){$Y(t)$\\[0.2em] } (sampler);
    \draw[-latex,thick] (sampler) --  node[midnodes](){$[Y_\kappa^\diamondsuit, Y_\kappa^\spadesuit]^\mathrm{T}$\\ [0.2em] } (dsp)  ;
    \draw[-latex,thick] (dsp) --  node[midnodes,xshift=0cm,align=center](){$\hat{X}_\kappa$\\[0.29em] }(output);

    \draw[-latex,thick] (noise) -- (sumnode);
    \node[above] () at (noise) {$N'(t) \in\, \mathbb{R}$};
    \node[above,yshift=1em] () at (sampler) {Sampler};
    \node[below,yshift=-1em] () at (sampler) {Rate $1/T_\text{s}^\prime$};

    \node[below,yshift=-1em] () at (cir) {Channel};
    \node[below,yshift=-1em] () at (sld) {SLD};

    \begin{pgfonlayer}{background}
        \draw[boxlinesred] ($( txfilter |- noise) + (0,+20pt)$) -- node[left,yshift=1.05cm,text width=1.2cm,rotate=90,font=\scriptsize  ,align=right,]{Electrical\\[0.2em] Optical}  ($( txfilter |- noise) + (0,-33pt)$);
    \end{pgfonlayer}
    \begin{pgfonlayer}{background}
        \draw[boxlinesred] ($( sld |- noise) + (0,+20pt)$) -- node[left,yshift=1.05cm,text width=1.2cm,rotate=90,font=\scriptsize  ,align=right,]{Optical\\[0.2em]Electrical}  ($(sld |- noise) + (0,-33pt)$);
    \end{pgfonlayer}

\end{tikzpicture}
    \caption{Communication system with DD.}
    \label{fig:continuous_detailed_system_model}
\end{figure*}
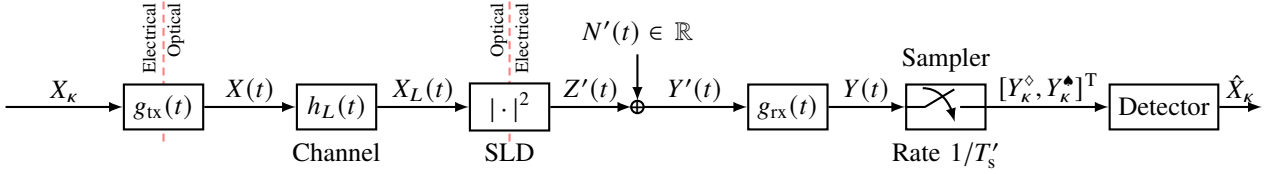

The continuous-time model is depicted in Fig.~\ref{fig:continuous_detailed_system_model}, cf.~\cite{wiener_filter_plabst2020}.

\subsubsection{Transmitter}
Consider a sequence of independent and identically distributed (i.i.d.) symbols $(X_\kappa)_{\kappa \in \mathbb{Z}} = \left( \ldots, X_1, X_2, X_3,\ldots \right)$ with finite alphabet $\mathcal{A}=\{a_1,\dots,a_Q\}$ of cardinality $Q$ and symbol probability mass function (PMF) $\Prob{}{x} = 1/Q$ for all $x\in\mathcal{A}$. We perform pulse shaping with $g_\text{tx}(t)$ to obtain the baseband waveform
\begin{align}
    X(t)  = \sum\limits_{\kappa} X_\kappa \cdot g_\text{tx}(t-\kappa T_\text{s})
\end{align}
with symbol-rate $B = 1/T_\text{s}$. Common choices for pulse shaping are FD-RC pulses which have the spectrum~\cite[Eq.~(6.17)]{gallager2008principles}
\begin{align}
  G_\text{tx}(f) = \begin{cases}
  1,
       & |f| \leq \frac{1 - \alpha}{2T_\text{s}} \\
  \frac{1}{2}\left[1 + \cos\left(\frac{\pi T_\text{s}}{\alpha}\left[|f| - \frac{1 - \alpha}{2T_\text{s}}\right]\right)\right],
       & \frac{1 - \alpha}{2T_\text{s}} < |f| \leq \frac{1 + \alpha}{2T_\text{s}} \\
  0,
       & \text{otherwise.}
 \end{cases}
 \label{eq:tx_fd-rc}
\end{align}
We focus mainly on the case $\alpha=0$ for which we obtain the sinc pulse 
\begin{align}
    g_\text{tx}(t) = \frac{1}{T_s} \cdot \sinc\left(\frac{t}{T_s}\right) \;\laplace\; G_\text{tx}(f)
    = 
    \begin{cases} 
        1, & \lvert f \rvert \leq \frac{1}{2T_s} \\
        0, & \text{otherwise.}
    \end{cases}
    \label{eq:tx_ps_filter}
\end{align}
We also consider TD-RC pulses that are time-domain versions of \eqref{eq:tx_fd-rc}, namely
\begin{align}
  g_\text{tx}(t) = \begin{cases}
  1,
       & |t| \le \frac{1 - \alpha}{2B}\\
  \frac{1}{2}\left[1 + \cos\left(\frac{\pi B}{\alpha}\left[|t| - \frac{1 - \alpha}{2B}\right]\right)\right],
       & \frac{1 - \alpha}{2B} < |t| \le \frac{1 + \alpha}{2B} \\
  0,
       & \text{otherwise.}
 \end{cases}
 \label{eq:tx_td-rc}
\end{align}

\subsubsection{Fiber-Optic Link}
The channel exhibits CD with response~\cite[Sec. II.B]{wiener_filter_plabst2020}
\begin{align}
    h_L(t) \;\laplace\; H_L(f) = \exp\left(\,\mathrm{j} \frac{\beta_2}{2}\omega^2 L\, \right) 
\end{align}
where $\beta_2$ is the CD parameter, $\omega=2\pi f$ is the angular frequency and $L$ is the fiber length.

\subsubsection{Receiver}
The receiver has optical and electrical components. First, a PD outputs the intensity $Z^\prime(t) = \lvert X_L(t) \rvert^2$ of the impinging field~\cite{wiener_filter_plabst2020}. We thus model the PD as having no active bandwidth constraint and we model the PD noise $N'(t)$ in Fig.~\ref{fig:continuous_detailed_system_model} by a real-valued white Gaussian random process with two-sided power spectral density $N_0/2$. 

Next, the receiver has a band-limited sampling device~\cite[Sec. III.B]{wiener_filter_plabst2020} with impulse response
\begin{align}
    g_\text{rx}(t) = 2B \cdot \sinc(2Bt) \;\laplace\; G_\text{rx}(f) 
    = 
    \begin{cases} 
        1, & \lvert f \rvert \leq B \\
        0, & \text{otherwise}
    \end{cases}
    \label{eq:rx_ps_filter}
\end{align}
that rejects out-of-band noise, avoids aliasing and accommodates the doubled signal bandwidth due to the square-law detector (SLD) operation $\lvert \,\uarg\, \rvert^2$. The filtered noise $N(t)=N'(t)*g_\text{rx}(t)$ is a zero-mean stationary Gaussian process with autocorrelation function
\begin{align}
\frac{N_0}{2} \cdot \left( g_\text{rx}(-\tau) * g_\text{rx}(\tau) \right) = \frac{N_0}{2} g_\text{rx}(\tau) 
.
\end{align}
We similarly define $Z(t)=Z'(t)*g_\mathrm{rx}(t)$.

\subsection{Discrete-Time Model}
\label{sec:time-discrete_model}
We formulate the discrete-time model for a receiver that samples at rate $1/T_\text{s}' = 2B$, i.e., the oversampling factor is $N_\text{os} = T_\text{s}/T_\text{s}'=2$ samples per transmit symbol\footnote{For a sinc pulse one should oversample at a rate slightly larger than $N_\text{os}=2$ to obtain sufficient statistics. However, we simply use $N_\text{os}=2$. For an FD-RC pulse with positive roll-off factor $\alpha$ the oversampling rate should be at least $N_\text{os}=2(1+\alpha)$ to obtain sufficient statistics.}.
Let $(X_k')_{k\in \mathbb{Z}} = ((0, X_\kappa) )_{\kappa \in \mathbb{Z}} = (\ldots, 0, X_1, 0, X_2, \ldots)$ and $Y_k=Y(k T_\text{s}')$, $k\in\mathbb{Z}$, be the upsampled transmitter and receiver sample strings, respectively.
The sampler output is
\begin{equation}
    \begin{aligned}
        Y_k &= Z_k + N_k
        \label{eq:scalar_system_equation}
    \end{aligned}
\end{equation}
for all $k$ where
\begin{align}
    Z_k
     = \left( \left| X_L(t) \right|^2 * g_\text{rx}(t) \right)_{t = k T_\mathrm{s}^\prime}
     \label{eq:z_time-discrete_sld_output_sampled_convolution}
\end{align}
and $N_1, N_2,\ldots$ is a white Gaussian process with density $\Prob{N}{n}=\GaussDist{n}{0}{\sigma^2_\text{N}}$ and $\sigma^2_\text{N}=N_0 B$.
Define the combined response of the transmit pulse and fiber as $\psi(t) = g_\mathrm{tx}(t) * h_L(t)$ and define the samples $\psi_k=\psi(k T_\text{s}')$, $k\in\mathbb{Z}$. We have $X_L(t) = \sum_\kappa X_\kappa \cdot \psi(t-\kappa T_\mathrm{s})$ and \begin{align}
    X_L(k T_\mathrm{s}^\prime)
    =\sum_m X_m \cdot \psi((k-2m)T_\text{s}')
    =\sum_m \psi_{m} X_{k-m}^\prime.
\end{align}

In the following, we focus on the transmitter pulse shape \eqref{eq:tx_ps_filter} for which $\left| X_L(t) \right|^2 * g_\mathrm{rx}(t) = \left| X_L(t) \right|^2$ and
\begin{align}
    Z_k
     =  \left| \, \sum\nolimits_{m = -\infty}^{\infty} \psi_m X^\prime_{k-m} \, \right|^2.
     \label{eq:z_time-discrete_sld_output}
\end{align}
We convert to vector-matrix notation and collect the upsampled strings for an even number $n$ of time steps: 
\begin{align}
    \mathbf{X}' &= \left[0,\,\;\;X_1,\; \; 0,\,\;\; X_2,\,\ldots  ,\; 0 ,\; X_{n/2} \right]^\mathrm{T} &&\dimC{n \times 1}\\
    \mathbf{Z} &= \left[Z_1,\; Z_2,\; Z_3,\;  \hspace{19pt}\ldots \hspace{18pt} Z_{n} \right]^\mathrm{T}   && \dimR{n \times 1}\\
    \mathbf{N} &= \left[N_1,\; N_2,\; N_3,\;   \hspace{17pt} \ldots \hspace{18pt} N_{n} \right]^\mathrm{T}  && \dimR{n \times 1}\\
    \mathbf{Y} &= \left[Y_1,\;\; Y_2,\;\; Y_3,\;  \hspace{20pt}\ldots \hspace{19pt} Y_{n} \right]^\mathrm{T}   && \dimR{n \times 1}.
    \label{eq:length_n_vector_stacking}
\end{align}
For illustration, suppose 
the channel values $\psi_m$ are zero outside the interval $[0,M-1]$ where $M$ is an odd positive integer that represents the oversampled channel memory. Collect these values in the time-reversed vector $\bm{\uppsi} = \left[\psi_{M-1}, \ldots,  \psi_0  \right]^\mathrm{T}$ and define the initial channel state as
\begin{align}
    \bm{s}_0 = \left[0,\, x_{1-\widetilde{M}},\,0,\, x_{2-\widetilde{M}},\, \quad \ldots,\, \quad 0,\, x_0\right]^{\T} \quad \dimC{(M-1) \times 1}
    \label{eq:initial_state_vector_known}
\end{align}
where $\smash{\widetilde{M}} = (M-1)/2$ is the channel memory in terms of the transmit symbols. The state $\bm{s}_0$ is assumed to be known by the receiver. Define the matrix $\bm{\Psi} \dimC{n \times (n + M-1)}$ as the Toeplitz matrix with successively right-shifted versions of $\bm{\uppsi}^\mathrm{T}$ as its rows. The output of the SLD is
\begin{align}
    \bm{Z} = \left\lvert \bm{\Psi}
    \begin{bmatrix}
    \bm{s}_0\\
    \bm{X}'
    \end{bmatrix}
    \right\rvert^{\circ 2} = \left\lvert \bm{\Psi} \,
    \tilde{\mathbf{X}}'
    \right\rvert^{\circ 2}
    \label{eq:Zvector}
\end{align}
where the notation $|\bm{X}'|^{\circ 2}$ refers to the vector $[|X_1'|^2,\dots,|X_n'|^2]$, and where $\tilde{\mathbf{X}}'=[\bm{s}_0^{\T}, (\bm{X}')^{\T}]^{\T}$.

\begin{figure}[t]
    \centering
    \hspace*{-1.9cm}
        \begin{tikzpicture}[scale=0.85,]
      \begin{axis}[width=8.6cm,height=14cm,yshift=0cm,
                  grid=none,xticklabel=\empty,yticklabel=\empty,
                  xtick={-180,180},
                  ytick={1},
                  yticklabels={},
                  ymin=-12.2,ymax=7.5,
                  xmin=-880,xmax=600,
                  rotate around={90:(current axis.origin)}, 
                  ticks=none,
                  axis line style={draw=none}
                  ]
      \addplot+[ycomb,black,mark options={black},samples at={  -450,  -360,  -270,  -180,   -90,     0.01,    90,   180,   270,   360,   450},line width=0.7pt] {2*sin(x)/ ((x)*(pi)/180) };

      \node [text width=5cm,font=\footnotesize] (diamond) at(axis cs:0,-2.2)  {
      \begin{equation*}
            \bullet \quad 
            \begingroup
            \renewcommand*{\arraystretch}{1.35}
            \begin{bmatrix}
              0 \\
              X_{\kappa-5} \\
              0 \\
              X_{\kappa-4} \\
              0 \\
              X_{\kappa-3} \\
              0 \\
              X_{\kappa-2} \\
              0 \\
              X_{\kappa-1} \\
              0 \\
            \end{bmatrix}
            \endgroup
      \end{equation*}
      }; 

      \node [text width=5cm,font=\footnotesize] (spade) at(axis cs:0,-5)  {
      \begin{equation*}
      \;\text{and}\;\;
            \begingroup
            \renewcommand*{\arraystretch}{1.35}
            \begin{bmatrix}
              X_{\kappa-5} \\
              0 \\
              X_{\kappa-4} \\
              0 \\
              X_{\kappa-3} \\
              0 \\
              X_{\kappa-2} \\
              0 \\
              X_{\kappa-1} \\
              0 \\
              X_\kappa
            \end{bmatrix}
            \endgroup
      \end{equation*}
      }; 
    
      \node [black] at(axis cs:-660,+1.25)   {Square-Law Detector};  

      \draw[latex-] (axis cs:-450,-6.4) -- (-450,-8) node[right]{$\kappa^\text{th}$ input };  
      
      \node [black] at(axis cs:-830,-0.65)   { Noiseless: $\mathbf{Z}_\kappa = \big[ Z^\diamondsuit_\kappa,$ };  
      \node [black] at(axis cs:-830,-5.6)  { $Z^\spadesuit_\kappa \big]^\mathrm{T}$ }; 
      
      \draw [decorate,thick,decoration={brace,amplitude=10pt},xshift=-4pt,yshift=0pt]
      (axis cs:530,-6.4) -- (-370,-6.4) node [midway, right,xshift=0.3cm, align=center, text width=3cm] 
      {Channel\\Memory:\\ $S_{\kappa-1} \triangleq \left( X_{\kappa-\widetilde{M}}, \ldots, X_{\kappa-1} \right) $};
      
      \draw[-,line width=0.7pt]  (axis cs:-450,0) -- (axis cs:450,0); %

      \draw[-,line width=0.7pt]  (axis cs:-560,4.3) -- (axis cs:-560,-11); %
      \draw[-,line width=0.7pt]  (axis cs:-760,4.3) -- (axis cs:-760,-11); %
  
      \node [recty,xshift=0.15cm, below=0.3cm of diamond,minimum height=0cm,minimum width=0cm,font=\footnotesize,thick] (sq_1) {$\lvert \,\uarg\, \rvert^2$};
      \node [recty,xshift=0.3cm, below=0.3cm of spade,minimum height=0cm,minimum width=0cm,font=\footnotesize,thick] (sq_2) {$\lvert \,\uarg\, \rvert^2$};
      
      \node [rotate=90] at(axis cs:0,3.1)   { FIR Filter}; 
      \node [rotate=90] at(axis cs:60,2.2)   {   }; %
      \node [rotate=90] at(axis cs:-450,0.9)   { $\uppsi_0$  }; 
      \node [rotate=90] at(axis cs:+450,0.9)   { $\uppsi_{M-1}$  };

       \end{axis}
    \end{tikzpicture}
    
    \caption{Signals for oversampling factor $N_\text{os} = 2$.}   
    \label{fig:Channel_Memory_Convolution}
\end{figure}
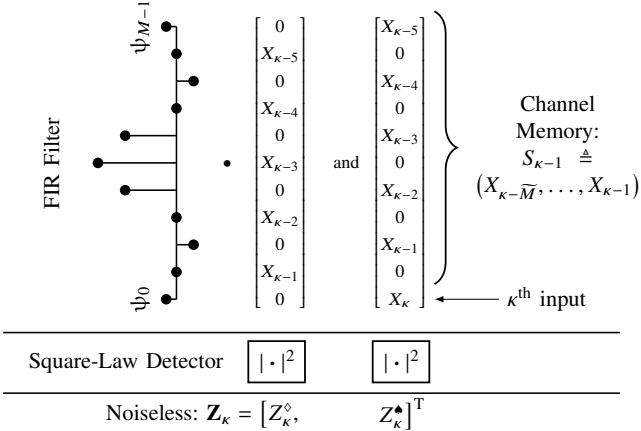

The channel's conditional probability density is Gaussian:
\begin{align}
p(\bm{y}|\bm{x}) = \GaussDist{\bm{y} - \left\lvert \bm{\Psi}
    \tilde{\bm{x}}'
\right\rvert^{\circ 2}} {\mathbf{0}_n}{\sigma_\mathrm{N}^2 \, \mident{n}}.
\label{eq:channel}
\end{align}
For clarity, we concatenate two subsequent outputs in a vector
\begin{align}
    \bm{Z}_\kappa &= 
        \left[ Z^\diamondsuit_\kappa,\; Z^\spadesuit_\kappa \right]^\mathrm{T}.
\end{align}
The channel output can similarly be written as
\begin{equation}
    \begin{aligned}
    \bm{Y}_\kappa =  \left[ Y^\diamondsuit_\kappa,\; Y^\spadesuit_\kappa \right]^\mathrm{T} 
    = \bm{Z}_\kappa + \left[ N^\diamondsuit_\kappa,\; N^\spadesuit_\kappa \right]^\mathrm{T}  
    .
    \end{aligned}
\end{equation}
The convolution and SLD operations are depicted in Fig.~\ref{fig:Channel_Memory_Convolution}. Observe that for link length $L=0$ the $Z^\diamondsuit_\kappa$ are ISI-free by the choice of $g_\text{tx}(t)$, while the $Z^\spadesuit_\kappa$ experience ISI. 

\section{Achievable Rates}
\label{sec:achievable_rates}
This section describes how to compute information-theoretic quantities for the underlying ISI channel~\cite[Sec. II b)]{PfisterAIRFiniteStateChan2001}. Define the entropy rate and mutual information rate as the respective
\begin{align}
    & h_n(\bm{Y}) = \frac{2}{n}  h(\bm{Y}) \label{eq:single_letter_h} \\
    & I_n(\bm{X};\bm{Y}) = \frac{2}{n} \, \MI{\bf{X}}{\bf{Y}}
    = h_n(\bm{Y}) -  h_n(\bm{N}).
    \label{eq:single_letter_mi}
\end{align}
The factor $2/n$ normalizes for the $n/2$ transmit symbols in $\mathbf{X}$. The rates are thus measured in bits per transmit symbol, also called bits per channel use (bpcu). The limiting rates are
\begin{align}
    \He{\mathcal{Y}} 
    = \lim_{n \rightarrow \infty} h_n(\bm{Y}), \quad
    \MI{\mathcal{X}}{\mathcal{Y}}
    = \lim_{n \rightarrow \infty} I_n(\bm{X};\bm{Y})
    \label{eq:limiting}
\end{align}
and these quantities exist if one uses stationary signaling~\cite[Thm.~4.2.1]{cover1991elementsofIT}. The expression $\MI{\mathcal{X}}{\mathcal{Y}}$ in \eqref{eq:limiting} is known to be an achievable rate for reliable communication~\cite{Shannon48} and one may attempt to maximize $\MI{\mathcal{X}}{\mathcal{Y}}$ over all input processes $\mathcal{X}$ to compute the channel capacity. However, the optimization is usually difficult for nonlinear channels with memory, and for computation and practical implementation one often resorts to i.i.d. signaling. The input process $\mathcal{X} = (X_\kappa)_{\kappa \in \mathbb{Z}}$ is hence stationary and the corresponding output process $\mathcal{Y} = (Y_k)_{k \in \mathbb{Z}}$ is cyclostationary with period $N_\text{os}$ (for every sample in $\mathcal{X}$ there are two samples in $\mathcal{Y}$).

\subsection{Computing Entropy and Probabilities}\label{sec:lower_bound}
To compute \eqref{eq:single_letter_h}, we may use Monte-Carlo simulation to approximate~\cite{arnoldsimulationmi}:
\begin{equation}
    h_n(\bm{Y})
    \approx - \frac{2}{n} \log_2 \Prob{}{\bm{y}}
    \label{eq:approx_mi} 
\end{equation}
where $\bm{y}$ is a realization of $\bm{Y}$. The approximation \eqref{eq:approx_mi} becomes exact in the limit $n\rightarrow\infty$ since $\mathcal{Y}$ is cyclostationary with period $N_\text{os}$ and ergodic. Define the channel state at time $\kappa$ as
\begin{align}
    s_{\kappa-1} = \left( x_{\kappa-\widetilde{M}},\ldots,\, x_{\kappa-1}  \right).
    \label{eq:state_space_sequence_rv}
\end{align}
The density value $\Prob{}{\bm{y}}$ can be computed by marginalizing:
\begin{align}
   \Prob{}{y_1^n} &= \sum\limits_{s_0} \sum\limits_{x_1^{n/2}}  \Prob{}{s_0,x_1^{n/2},y_1^n}
    \label{eq:py_vector_by_marginalization}
\end{align}
where we use string notation for clarity. We further compute
\begin{align}
    \Prob{}{x_1^{n/2},y_1^n\,|\,s_0}
    &=  \prod_{\kappa=1}^{n/2} \Prob{}{x_\kappa, \bm{y}_{\kappa} \,\lvert\, s_0, x_1^{\kappa-1} } \nonumber \\
    &=  \prod_{\kappa=1}^{n/2} \Prob{}{x_\kappa, \bm{y}_{\kappa} \,\lvert\, s_{\kappa-1}  }
    \label{eq:joint_output_density_before_marginalization}
\end{align}
where we use $\bm{Y}_\kappa = ( Y^\diamondsuit_\kappa,\,  Y^\spadesuit_\kappa )$ and that $( X_\kappa, \bm{Y}_{\kappa}) $ is conditionally independent of $\bm{Y}_1^{\kappa-1}$ given $(S_0,X_1^{\kappa-1})$. We thus have
\begin{equation}
    \Prob{}{y_1^n}
    = \sum\limits_{s_0} \sum\limits_{x_1^{n/2}}  p(s_0) \prod_{\kappa=1}^{n/2} \Prob{}{x_\kappa, \bm{y}_{\kappa}, s_\kappa \,\lvert\, s_{\kappa-1}  }
    \label{eq:marg1}
\end{equation}
where $s_\kappa$ is defined by $(s_{\kappa-1},x_\kappa)$. Note that $\Prob{}{x_\kappa, \bm{y}_{\kappa}, s_\kappa \lvert s_{\kappa-1}}$ does not depend on the time $\kappa$ and $s_0$ is known. 

\begin{figure*}[t]
    \vspace*{-5pt}
    \centering
    \scalebox{1.45}{\begin{tikzpicture}[thick,yscale=0.3, xscale=0.7, every node/.style={scale=0.62},auto, node distance=1.2cm,>=latex',double]
    \node [input, name=input] {};
    \node [input,right of=input] (box_1) {};
    \node [recty,right=1.2cm of box_1] (box_2) {};
    \node [recty,right=1.5cm of box_2] (box_3) {};
    \node [recty,right=1.5cm of box_3] (box_4) {};
    \node [recty,right=1.5cm of box_4] (box_5) {};
    \node [recty,right=1.5cm of box_5] (box_6) {};
    \node [input,right=1.2cm of box_6] (box_7) {};

    \node [input,node distance=1cm,above of=box_1] (i1) {};
    \node [input,node distance=1cm,above of=box_2] (i2) {};
    \node [input,node distance=1cm,above of=box_3] (i3) {};
    \node [input,node distance=1cm,above of=box_4] (i4) {};
    \node [input,node distance=1cm,above of=box_5] (i5) {};
    \node [input,node distance=1cm,above of=box_6] (i6) {};
    \node [input,node distance=1cm,above of=box_7] (i7) {};

    \node [input,node distance=1cm, below of=box_1] (o1) {};
    \node [input,node distance=1cm, below of=box_2] (o2) {};
    \node [input,node distance=1cm, below of=box_3] (o3) {};
    \node [input,node distance=1cm, below of=box_4] (o4) {};
    \node [input,node distance=1cm, below of=box_5] (o5) {};
    \node [input,node distance=1cm, below of=box_6] (o6) {};
    \node [input,node distance=1cm, below of=box_7] (o7) {};

    \draw (box_1) -- node[above,midway]{$S_0$} (box_2);
    \draw (box_2) -- node[above,midway]{$S_1$}(box_3);
    \draw (box_5) -- node[above,midway]{$S_{n/2-1}$}(box_6);

    \draw[white] (box_3) -- node[midway,black]{$\ldots$}(box_4);
    \draw[white] (box_4) -- node[midway,black]{$\ldots$}(box_5);

    \draw (box_6) -- node[above,midway]{$S_{n/2}$} (box_7);

    \draw (i2) node[above ]{$[0,X_1]$} --  (box_2) ;
    \draw (i3) node[above ]{$[0,X_2]$} -- (box_3) ;
    \draw (i4) node[above ]{$[0,X_\kappa]$} -- (box_4) ;
    \draw (i5) node[above ]{$[0,X_{n/2-1}]$} -- (box_5) ;
    \draw (i6) node[above ]{$[0,X_{n/2}]$} -- (box_6) ;

    \draw (o2) node[below]{$\mathbf{Y}_1 = [Y^\diamondsuit_1,Y^\spadesuit_1]$} -- (box_2);
    \draw (o3) node[below ]{$\mathbf{Y}_2 = [Y^\diamondsuit_2,Y^\spadesuit_2]$} -- (box_3);
    \draw (o4) node[below ]{$\mathbf{Y}_\kappa = [Y^\diamondsuit_\kappa,Y^\spadesuit_\kappa]$} -- (box_4);
    \draw (o5) node[below ]{$\mathbf{Y}_{n/2-1} = [Y^\diamondsuit_{n/2-1},Y^\spadesuit_{n/2-1}]$} -- (box_5);
    \draw (o6)  node[below ]{$\mathbf{Y}_{n/2} = [Y^\diamondsuit_{n/2},Y^\spadesuit_{n/2}]$} --(box_6);

    \draw[blue,draw=none] ($(box_2) - (-1,0.7)$)-- node[below,font=\large]{$\vv{\mu}(s_1)$} ($(box_3) - (1,0.7)$);
    \draw[blue,draw=none] ($(box_3) - (-1,0.7)$)-- node[below,font=\large]{$\vv{\mu}(s_{\kappa-1})$} ($(box_4) - (1,0.7)$);
    \draw[red,draw=none] ($(box_4) - (-1,-1.5)$)-- node[above,font=\large]{$\cvv{\mu}(s_{\kappa})$} ($(box_5) - (1,-1.5)$);
    \draw[red,draw=none] ($(box_5) - (-1,-1.5)$)-- node[above,font=\large]{$\cvv{\mu}(s_{n/2-1})$} ($(box_6) - (1,-1.5)$);

\end{tikzpicture}}
\begin{align}
p(x_\kappa,  \mathbf{y}_1^{n/2})  &= 
\sum\limits_{\substack{\vphantom{a}\\s_{\kappa-1}}} 
\overbrace{
\sum\limits_{\substack{x_{\kappa-1} \\s_{\kappa-2}}}   
\cdots
\sum\limits_{\substack{x_2 \\s_1}}   
\Prob{}{x_2, \bm{y}_{2}, s_2 \lvert s_{1}}\, 
\sum\limits_{\substack{x_1 \\s_0}}   
p(s_0)   \cdot \Prob{}{x_1, \bm{y}_{1}, s_1 \lvert s_{0}}}^{\text{\normalsize FW Path: \textcolor{blue}{$\vv{\mu}(s_{\kappa-1})$}}}
\cdot  \nonumber \\[-1.5em]
&\hspace{30pt}
\;\; \sum\limits_{\substack{\vphantom{a}\\s_{\kappa}}} 
\Prob{}{x_\kappa, \bm{y}_{\kappa}, s_\kappa \lvert s_{\kappa-1}} \;\; 
\overbrace{
\sum\limits_{\substack{x_{\kappa+1} \\s_{\kappa+1} } } \, 
\cdots
\sum\limits_{\substack{x_{n/2-1} \\s_{n/2-1} } } \, 
\Prob{}{x_{n/2-1}, \bm{y}_{n/2-1}, s_{n/2-1} \lvert s_{{n/2}-2}}
\sum\limits_{\substack{x_{n/2} \\s_{n/2} } } 
\Prob{}{x_{n/2}, \bm{y}_{n/2}, s_{n/2} \lvert s_{{n/2}-1}}}^{\text{\normalsize BW Path: \textcolor{red}{$\cvv{\mu}{(s_{\kappa})}$}}}
\label{eq:fw_bw_message_passing}
\end{align}
    \caption{Factor graph and associated computation for the FW and BW paths with two-fold oversampling.}
    \label{fig:factor_graph_forward_mp}
\end{figure*}
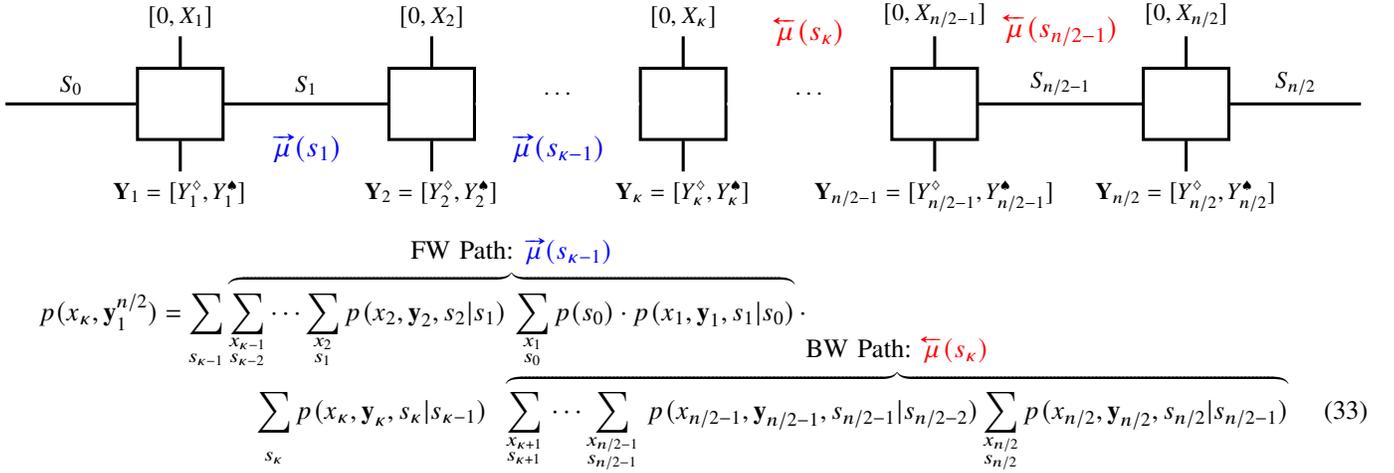

The expression \eqref{eq:marg1} may be further expanded as visualized by the factor graph in Fig.~\ref{fig:factor_graph_forward_mp}: $\Prob{}{y_1^n}$ can be written as
\begin{align}
    \sum\limits_{s_{n/2}}\sum\limits_{\substack{x_{n/2} \\ s_{n/2-1}}} \cdots
    \overbrace{\sum\limits_{\substack{x_2\\s_1}}
    \underbrace{\sum\limits_{ \substack{x_1 \\ s_0}}   p(s_0)  \cdot \Prob{}{x_1, \bm{y}_{1} s_1 \lvert s_{0}}}_{\vv{\mu}\left(s_1 \right)}\cdot 
    \Prob{}{x_2, \bm{y}_{2} s_2 \lvert s_{1}} }^{\vv{\mu}(s_2 )}
    \cdots
    \label{eq:sum_product_mp}
\end{align}
where the state metrics are $\vv{\mu}(s_\kappa)=p(\bm{y}_1^\kappa, s_\kappa)$.
Eq.~\eqref{eq:sum_product_mp} is computed on the forward (FW) path of the forward-backward algorithm~\cite{bcjr_1974} through the graph in Fig.~\ref{fig:factor_graph_forward_mp}. We will later also consider the backward (BW) path to compute \eqref{eq:fw_bw_message_passing} shown in Fig.~\ref{fig:factor_graph_forward_mp}.
The state metrics $\vv{\mu}(s_\kappa)$ quickly approach zero for large $\kappa$ and we thus normalize as in~\cite{arnoldsimulationmi}. Next, decompose 
\begin{align}
    \Prob{}{x_\kappa, \bm{y}_{\kappa}, s_\kappa \lvert s_{\kappa-1}} = 
    \underbrace{\Prob{}{\bm{y}_\kappa \lvert x_\kappa, s_{\kappa-1}}}_{(a)} 
    \underbrace{\Prob{}{s_\kappa \lvert x_\kappa, s_{\kappa-1}}}_{(b)} 
    \underbrace{\Prob{}{x_\kappa  \lvert s_{\kappa-1}}}_{(c)}
    \label{eq:decomposition_joint_input_state_obs}
\end{align}
where the term $(a)$ is
\begin{align}
    \Prob{}{\bm{y}_\kappa \lvert x_\kappa, s_{\kappa-1}} = \Prob{}{y^\diamondsuit_\kappa \lvert s_{\kappa-1}} \cdot \Prob{}{y^\spadesuit_\kappa \lvert x_\kappa, s_{\kappa-1} }  
    \label{eq:factored_channel_diamond_spade}
\end{align}
and where
\begin{align}
    & \Prob{}{y^\diamondsuit_\kappa \lvert s_{\kappa-1}} =  \Prob{N}{y^\diamondsuit_\kappa - \left\lvert 
    \begin{bmatrix}
        0,\, x_{\kappa-\widetilde{M}},\, 0, \ldots ,\,  x_{\kappa-1} ,\,
        0
    \end{bmatrix} \cdot \bm{\uppsi}
     \right\rvert^2 }
     \label{eq:diamond_channel} \\
    & \Prob{}{y^\spadesuit_\kappa \lvert x_\kappa, s_{\kappa-1}} =  \Prob{N}{y^\spadesuit_\kappa - \left\lvert 
    \begin{bmatrix}
     x_{\kappa-\widetilde{ M}},  \ldots,   \, x_{\kappa-1},\,   0, x_\kappa
    \end{bmatrix} \cdot \bm{\uppsi}
     \right\rvert^2 }
    \label{eq:spade_channel}.
\end{align}
The term $(b)$ in~\eqref{eq:decomposition_joint_input_state_obs} is 1 if, given $X_\kappa=x_\kappa$, there is a state transition $s_{\kappa-1} \rightarrow s_{\kappa}$ and is 0 otherwise. Since we use i.i.d. signaling, the term $(c)$ is
\begin{align}
\Prob{}{x_\kappa  \lvert s_{\kappa-1}} = \Prob{}{x_\kappa}. 
\label{eq:p_xk_given_past_state}
\end{align}

\subsection{Auxiliary Channel and Mutual Information Rates}
\begin{figure}[t]
\centering
\begin{tikzpicture}[scale=0.88,]
    \begin{axis}[width=11cm,height=6.4cm,yshift=0cm,
    grid=none,xticklabel=\empty,yticklabel=\empty,
    xtick={-180,180},
    ytick={1},yticklabels={},
    ymin=-1.3,ymax=2.8,xmin=-695,xmax=695,
    ticks=none,
    axis line style={draw=none}
    ]
    \addplot+[ycomb,mark=o,mark size=3pt,line width=1pt,blue,samples at={-630, -540,  -450,  -360,  -270,  -180,   -90,     0.01,    90,   180,   270,   360,   450, 540, 630},line width=1.0pt] {1.5*sin(x)/ ((x)*(pi)/180) }; 
    \addplot+[ycomb,red,densely dashed,mark=x,mark options={solid},mark size=4pt,samples at={   -270,  -180,   -90,     0.01,    90,   180,   270},line width=1.0pt] {1.5*sin(x)/ ((x)*(pi)/180) }; 
    \addplot+[ycomb,red,densely dashed,mark=x,mark options={solid},mark size=4pt,samples at={-630, -540,  -450,  -360,     360,   450, 540, 630},line width=1.0pt] {0*x) }; 
    \draw[-,line width=0.7pt]  (axis cs:-630,0) -- (axis cs:630,0); %
    \draw [decorate,thick,decoration={brace,amplitude=10pt},yshift=0pt]
    (axis cs:270,-0.55) -- (-270,-0.55) node [font=\footnotesize,midway, yshift=-0.20cm, below, align=center, text width=8cm] 
    {Memory of $\bm{\uppsi}'$ in auxiliary channel: $\widetilde{N}$ taps};
    \draw [decorate,thick,decoration={brace,mirror, amplitude=10pt},yshift=0pt]
    (axis cs:690,+1.6) -- (-690,+1.6) node [font=\footnotesize,midway, above, yshift=+0.22cm, align=center, text width=6cm]
    {Memory of $\bm{\uppsi}$ in true channel: $\widetilde{M}$ taps};
    \node at (axis cs:670,0) {$\bm{\ldots}$}; 
    \node at (axis cs:-670,0) {$\bm{\ldots}$}; 
    \end{axis}
    \end{tikzpicture}
\caption{Memory comparison of true and auxiliary channels.}
\label{fig:auxiliary_channel}
\end{figure}
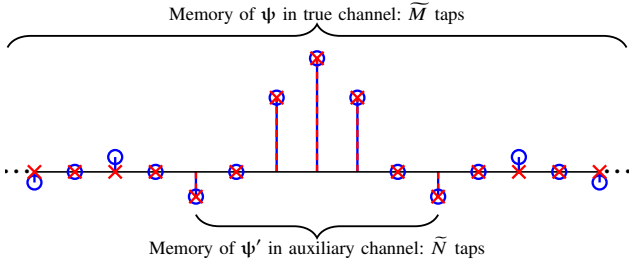

The complexity of computing $h_n(\bm{Y})$ grows with the number of states, which in turn grows exponentially in the memory length $\smash{\widetilde{M}}$. To limit receiver complexity, we must usually use a much smaller number $\smash{\widetilde{N}}$ of taps, see Fig.~\ref{fig:auxiliary_channel}. 

Consider the Gaussian auxiliary channel with mean $\boldsymbol{\upmu}_{\bm{Q}}$ and covariance matrix $\bm{C}_{\bm{Q}\bm{Q}}$:
\begin{align}
    q(\bm{y}\lvert\bm{x}) = \mathcal{N}\big( \bm{y}- \left\lvert \bm{\Psi}^\prime
    \tilde{\bm{x}}'
    \right\rvert^{\circ 2}\;; \boldsymbol{\upmu}_{\bm{Q}},\, \bm{C}_{\bm{Q}\bm{Q}}\big)
    \label{eq:auxiliary_channel}
\end{align}
where $\bm{\Psi}^\prime$ is a Toeplitz channel matrix that has entries corresponding to a time-reversed channel $\bm{\uppsi}' = \left[\psi_{N-1}, \ldots,  \psi_0  \right]^\mathrm{T}$ with memory $\smash{\widetilde{N}}=(N-1)/2$ rather than the channel $\bm{\uppsi} = \left[\psi_{M-1}, \ldots,  \psi_0  \right]^\mathrm{T}$ that has memory $\smash{\widetilde{M}}=(M-1)/2$. The corresponding channel output density is 
\begin{equation}
    \begin{aligned}
        q(\bm{y}) = \sum\limits_{\bm{x}}  p(\bm{x}) \cdot q(\bm{y}\lvert\bm{x})
        \label{eq:auxiliary_output_dist}
    \end{aligned}
\end{equation}
and we define the ``reverse'' channel as
\begin{equation}
    \begin{aligned}
    r(\bm{x}\lvert \bm{y}) = \frac{p(\bm{x}) \cdot q(\bm{y} \lvert  \bm{x})}{ q(\bm{y}) }.
    \label{eq:reverse_auxiliary_channel}
    \end{aligned}
\end{equation}

We now follow the steps in~\cite[p.~3503]{arnoldsimulationmi} and write
\begin{equation}
    I_n(\bm{X};\bm{Y}) = I_{q,n}(\bm{X};\bm{Y}) + \frac{2}{n}\Div{p_{\bm{XY}}}{p_{\bm{Y}} \cdot r_{\bm{X}\lvert \bm{Y}}}
    \label{eq:inequality_lb_mi}
\end{equation}
where
\begin{equation}
    I_{q,n}(\bm{X};\bm{Y})
    = \frac{2}{n} \E  \left[ \log_2 \!
        \frac{q(\bm{Y}\lvert\bm{X})}{q(\bm{Y})} \right]
     \label{eq:mi_lb_auxiliary_channel}
\end{equation}
and where the expectation is with respect to (w.r.t.) $p(\bm{x},\bm{y})$. The limiting rate is (see \eqref{eq:limiting})
\begin{align}
    \MIsub{\mathcal{X}}{\mathcal{Y}}{_q}
    = \lim_{n \rightarrow \infty} I_{q,n}(\bm{X};\bm{Y})
    \label{eq:limiting-q}
\end{align}
and this quantity exists~\cite[Thm.~4.2.1]{cover1991elementsofIT} if one uses stationary signaling. Since \eqref{eq:limiting} is an achievable rate for reliable communication, so is the rate \eqref{eq:limiting-q} by using \eqref{eq:inequality_lb_mi}. The expression $I_{q,n}(\bm{X};\bm{Y})$ can be approximated via simulation as
\begin{equation}
    I_{q,n}(\bm{X};\bm{Y}) \approx \frac{2}{n} \log_2
    \frac{q(\bm{y}\lvert\bm{x})}{q(\bm{y})}
    \label{eq:mi_lb_auxiliary_channel_sim}
\end{equation}
where $\bm{x}$ and $\bm{y}$ are realizations of $\bm{X}$ and $\bm{Y}$, respectively. The approximation becomes exact in the limit $n\rightarrow\infty$ since $\mathcal{X}$ is stationary and $\mathcal{Y}$ is cyclostationary with period $N_\text{os}$, and both processes are ergodic.

\subsection{Computing Achievable Rates}
\label{sec:computing-rates}
We use the auxiliary channel \eqref{eq:auxiliary_channel} with reduced memory $\widetilde{N}$ to simplify computation. Given $\smash{\widetilde{N}}$, we seek the $\bm{\upmu}_{\bm{Q}}$ and the \emph{diagonal} $\bm{C}_{\bm{QQ}}$ that maximize $I_{q,n}(\bm{X};\bm{Y})$, which is equivalent to minimizing the divergence in~\eqref{eq:inequality_lb_mi} for large $n$. Expanding this divergence, we have
\begin{align}
    &\Div{p_{\bm{X}\bm{Y}}}{p_{\bm{Y}} \cdot r_{\bm{X}\lvert\bm{Y}}} \nonumber \\
    &= \Div{p_{\bm{X}\bm{Y}}}{p_{\bm{X}} \cdot q_{\bm{Y}\lvert\bm{X}}} - \Div{p_{\bm{Y}}}{q_{\bm{Y}}} \ge 0.
    \label{eq:div_decomposition_subtraction}
\end{align}
Minimizing $\Div{p_{\bm{X}\bm{Y}}}{p_{\bm{X}} \cdot q_{\bm{Y}\lvert\bm{X}}}$ thus makes both the divergences $\Div{p_{\bm{X}\bm{Y}}}{p_{\bm{Y}} \cdot r_{\bm{X}\lvert\bm{Y}}}$ and $\Div{p_{\bm{Y}}}{q_{\bm{Y}}}$ small, and serves as a proxy for finding good $\bm{\upmu}_{\bm{Q}}$ and diagonal $\bm{C}_{\bm{QQ}}$. The Appendix formulates an optimization problem for this proxy and provides the solution for $\bm{\upmu}_{\bm{Q}}$ and general $\bm{C}_{\bm{QQ}}$. We will consider \textit{independent} noise that is identically distributed on the $\diamondsuit$-samples, and identically distributed on the $\spadesuit$-samples. The mean and covariance matrix thus have the form $\bm{\upmu}_{\bm{Q}}=\left([1,\ldots, 1]\otimes[\mu_1, \mu_2]\right)^\mathrm{T}$ and $\bm{C}_{\bm{QQ}}=\text{diag}\left([1,\ldots, 1]\otimes[\sigma_\mathrm{N1}^2, \sigma_\mathrm{N2}^2]\right)$, respectively, where $\mu_1$, $\mu_2$, $\sigma_\mathrm{N1}^2$, $\sigma_\mathrm{N2}^2$ can be computed analytically or estimated via Monte Carlo simulation.

We will also be interested in the individual contributions of the $\bm{Y}^\diamondsuit$ and $\bm{Y}^\spadesuit$ samples to the achievable rates. To compute these, consider the chain rule for mutual information:
\begin{equation}
    I_{q,n}(\bm{X};\bm{Y}) = I_{q,n}(\bm{X};\bm{Y}^\diamondsuit) + I_{q,n}(\bm{X};\bm{Y}^\spadesuit \,\lvert\, \bm{Y}^\diamondsuit).
   \label{eq:I_q_yspade_and_x_given_ydiamond}
\end{equation}
We compute $I_{q,n}(\bm{X};\bm{Y})$ via \eqref{eq:mi_lb_auxiliary_channel_sim} and by using the FW path of the graph in Fig.~\ref{fig:factor_graph_forward_mp} for the auxiliary channel $q(\bm{y}\lvert\bm{x})$ and input distribution $p(\bm{x})$. The term $I_{q,n}(\bm{X};\bm{Y}^\diamondsuit)$ is the usual IM-DD rate with Nyquist-rate sampling that can be found by numerical integration, e.g., Gauss-Hermite quadrature algorithms~\cite[A.3]{steiner_coding_for_higher_order}. The final term $I_{q,n}(\bm{X};\bm{Y}^\spadesuit \,\lvert\, \bm{Y}^\diamondsuit)$ is now obtained by~\eqref{eq:I_q_yspade_and_x_given_ydiamond}. The limiting versions of the individual expressions are
\begin{align}
    & \MIsub{\mathcal{X}}{{\mathcal{Y}^\diamondsuit}}{_q} = \lim_{n \rightarrow \infty} I_{q,n}(\bm{X};\bm{Y}^\diamondsuit) 
    \label{eq:I_q_limit} \\
    &
    \MIsub{\mathcal{X}}{{\mathcal{Y}^\spadesuit} \lvert \mathcal{Y}^\diamondsuit }{_q} = \lim_{n \rightarrow \infty} I_{q,n}(\bm{X};\bm{Y}^\spadesuit \,\lvert\, \bm{Y}^\diamondsuit)
   \label{eq:I_q_cond_limit}
\end{align}
which exist for stationary signaling.

\subsection{Upper Bounds on Mutual Information}
An \emph{upper} bound on mutual information is~\cite{topsoe1967information}: 
\begin{align}
    I(\bm{X};\bm{Y})
    &= \Div{p_{\bm{Y}\lvert \bm{X}}}{w_{\bm{Y}} \,\lvert\, p_{\bm{X}}} -  \Div{p_{\bm{Y}}}{w_{\bm{Y}}} \nonumber \\
    & \le \Div{p_{\bm{Y}\lvert \bm{X}}}{w_{\bm{Y}} \,\lvert\, p_{\bm{X}} }
    \label{eq:div-upper}
\end{align}
where $w(\bm{y})$ is any output density and $p(\bm{y})$ is absolutely continuous w.r.t. $w(\bm{y})$. For simplicity, we choose
\begin{align}
    w(\bm{y}) = \GaussDist{\bm{y}}{\bm{\upmu}_{\bm{Y}}}{ \bm{C}_{\bm{YY}} }.
    \label{eq:upper-Gauss}
\end{align}
Using $\bm{C}_{\bm{YY}}=\bm{C}_{\bm{ZZ}}+\sigma_\mathrm{N}^2\,\mident{n}$, the bound \eqref{eq:div-upper} is
\begin{align}
    I(\bm{X};\bm{Y}) \le \frac{1}{2}
    \log_2 \det \left(\mident{n} +  \frac{\bm{C}_{\bm{ZZ}}}{\sigma_\mathrm{N}^2 } \right)
    \label{eq:upper_bound_determinant_covariance}
\end{align}
where we applied standard identities~\cite[Ch.~8]{cover1991elementsofIT}. Intuitively, the Gaussian density \eqref{eq:upper-Gauss} is a good choice at low Signal-to-Noise Ratio (SNR) where the AWGN dominates the SLD output. At high SNR, however, a Gaussian density fails to accurately model the SLD output.

A simpler upper bound follows by the Hadamard~\cite[Eq. (8.64)]{cover1991elementsofIT} and Jensen inequalities~\cite[Eq. (2.75)]{cover1991elementsofIT}: 
\begin{align}
        \frac{1}{2} \log_2 \det \left(\mident{n} +  \frac{\bm{C}_{\bm{ZZ}}}{\sigma_\mathrm{N}^2 } \right) \le \frac{n}{2} \log_2 \left(1 +  \frac{\bar{\sigma}^2_{Z}}{\sigma_\mathrm{N}^2} \right)
       \label{eq:upper_bound_scalar_covariance}
\end{align}
where $\bar{\sigma}^2_{Z} = \frac{1}{n}  \trace \bm{C}_{\bm{ZZ}}$. An expression for $\mathbf{C}_{\mathbf{Z}\mathbf{Z}}$ for real-valued i.i.d. inputs that have symmetries is derived in~\cite[Eq. (23)]{wiener_filter_plabst2020}. For other input distributions, $\mathbf{C}_{\mathbf{Z}\mathbf{Z}}$ can be found by Monte-Carlo simulation. We remark that the bounds~\eqref{eq:upper_bound_determinant_covariance} and~\eqref{eq:upper_bound_scalar_covariance} depend on the pulse shape and higher-order moments of the transmit symbols $\bm{X}$ due to the SLD~\cite[Eq. (23)]{wiener_filter_plabst2020}.

\section{Symbol-Wise MAP Detection}
\label{sec:symbolwise_app_map}
In addition to computing achievable rates, we study decoding with the symbol \emph{a posteriori} probabilities (APPs) $p(x_\kappa \lvert \mathbf{y}_1^{n/2})$. We have
$p(x_\kappa \lvert \mathbf{y}_1^{n/2}) \propto  p(x_\kappa, \mathbf{y}_1^{n/2})$
because $p(\bm{y}_1^{n/2})$ is a constant w.r.t. the decisions. Taking a hard decision (HD) based on the APP gives the MAP symbol estimates
\begin{equation}
    \begin{aligned}
        \hat{x}_\kappa = \argmax_{x_\kappa}  \,p(x_\kappa, \mathbf{y}_1^{n/2})
        \label{eq:map_rule}
    \end{aligned}
    \end{equation}
where
\begin{equation}
\begin{aligned}
    p(x_\kappa, \mathbf{y}_1^{n/2}) 
&= \sum\limits_{s_0} \sum\limits_{x_1^{n/2}  \setminus x_\kappa  } p( s_0, x_1^{n/2},\mathbf{y}_1^{n/2})
\label{eq:app_marginal}
\end{aligned}
\end{equation}
and $\sum_{x_1^{n/2} \setminus x_\kappa} f(x_1^{n/2})$ denotes a sum over all strings $x_1^{n/2}$ but where the $\kappa^\text{th}$ entry of the strings remains fixed.

The value~\eqref{eq:app_marginal} can be efficiently computed for finite state channels using the forward-backward algorithm, as depicted in Fig.~\ref{fig:factor_graph_forward_mp} and \eqref{eq:fw_bw_message_passing}. The BW path metrics are defined as
\begin{align}
    \cvv{\mu}(s_\kappa) =  p(\bm{y}^{n/2}_{\kappa+1}\lvert s_\kappa).
\end{align}
The FW and BW path recursions thus read:
\begin{equation}
    \begin{aligned}
     \text{FW: }\; && \vv{\mu}{(s_{\kappa})} &= \sum\limits_{x_{\kappa} ,s_{\kappa-1} } p(x_\kappa, \bm{y}_\kappa, s_\kappa \lvert s_{\kappa-1}) \cdot  \vv{\mu}{(s_{\kappa-1})} \\ 
    \text{BW: }\; && \cvv{\mu}{(s_{\kappa-1})} &= \sum\limits_{ x_{\kappa} , s_{\kappa}  } p(x_\kappa, \bm{y}_\kappa, s_\kappa \lvert s_{\kappa-1}) \cdot \cvv{\mu}{(s_{\kappa})}
     \label{eq:fw_bw_recursions_bcjr}
    \end{aligned}
\end{equation}
where $\kappa \in \lbrace 1,\ldots,n/2 \rbrace$, $\vv{\mu}(s_0)=p(s_0)$ and $\cvv{\mu}(s_{n/2})=p(s_{n/2})$. We again normalize the FW and BW path metrics for every iteration to avoid numerical issues. The APPs can also be computed for the auxiliary channel~\eqref{eq:auxiliary_channel} rather than the true channel.

The SLD causes phase ambiguities in MAP detection for certain modulation alphabets with i.i.d. encoding~\cite{JaganathanSparsePR2013,qiao_sparse_phase_retrieval,schniter_gamp_2015}. These ambiguities can be avoided by differential encoding~\cite{WeberDiffEncPSK1978,HoeherTurobDPSK1999,SchmalenAdvancesDetectionCoptical2017,howard_differential_turbo,SecondiniDirectDetectionBPAM2020} or other precoders. We use differential encoding for the transmit symbol phases and include decoding in the forward-backward algorithm at the receiver. 

\section{Simulation Results and Discussion}
\label{sec:simulation_results}
We compute achievable rates and error probabilities  for a standard single-mode fiber (SSMF) at wavelength \SI{1550}{nm} with $\beta_2=\SI{-2.168e-23}{\second^2\per\kilo\meter}$, attenuation factor \SI{0.2}{dB \per\kilo\meter},  
link lengths $L=0$ and $L=30\,\text{km}$,
receiver oversampling rate $N_\text{os}=2$ and symbol rate $B=\SI{35}{\giga Baud}$. We use single-polarization transmission and neglect the Kerr nonlinearity. The transmit blocks have at least $n/2=\SI{20E3}{}$ symbols for all plots. The average transmit power is %
\begin{align}
    P_\text{tx} =  \frac{\E\big[ \lVert X(t) \rVert^2 \big]}{(n/2)\, T_\mathrm{s}}
\end{align} 
and the noise variance is $\sigma_\text{N}^2 = 1$ so that $\text{SNR} = P_\text{tx}$.

The constellations have one of $Q=2,4,8$. The PAM and ASK constellations have equidistant spacing for the $X_\kappa$, e.g., the 4-PAM and 4-ASK alphabets are $\mathcal{A}=\{0,1,2,3\}$ and $\mathcal{A}=\{-3,-1,1,3\}$, respectively. This means that the alphabets for the $|X_\kappa|^2$ after the SLD are $\{0,1,4,9\}$ and $\{1,9\}$, respectively. We remark that these constellations implement \emph{geometric shaping} after the SLD that is beneficial at intermediate SNRs. One could, of course, optimize the spacing for each SNR and this is interesting future work. We also consider quadrature amplitude modulation (QAM) with $Q=4$ (QPSK) and $\mathcal{A}=\lbrace \pm 1, \pm j\rbrace$, as well as the 2-ring/4-ary alphabet $\mathcal{A}=\lbrace \pm 1, \pm j, \pm 2, \pm 2j \rbrace$, denoted as 8-Star-QAM (SQAM). The corresponding $|X_\kappa|^2$ alphabets are $\{1\}$ and $\{1,4\}$, respectively. We use differential phase encoding for all constellations.

The FD-RC pulses have small roll-off factors $\alpha=0$ (sinc pulse) and $\alpha=0.2$, while the TD-RC pulse~\cite[Eq. (2)]{tasbihi2021direct} has a large roll-off factor $\alpha=0.9$. The SE of FD-RC is
\begin{align}
    \mathrm{SE} = \frac{R}{\left(1+ \alpha\right)\cdot B \cdot T_\text{s}} \;\; \mathrm{[bit/s/Hz]}.
    \label{eq:se_definition}
\end{align}
For TD-RC, we measure the bandwidth as the smallest frequency range with 95\% of the transmit power. This bandwidth turns out to be 15\% larger than the corresponding sinc pulse, see~\cite[Table I]{tasbihi2021direct}.

We remark that the expression \eqref{eq:z_time-discrete_sld_output} changes for FD-RC pulses with positive roll-off factor and for TD-RC pulses because the bandwidth of $Z^\prime(t)$ is larger than $2B$. The receiver filter $g_\text{rx}(t)$ thus cuts off part of the spectrum.
We simulated the system with oversampling factor $N_\text{sim} = 4$ and then downsampled to $N_\mathrm{os} = 2$. The result is
\begin{align}
    Z_k
     = \sum\nolimits_{\ell=-\infty}^{\infty} g_\text{rx}(\ell\,T_\mathrm{s}^{\prime}) \left| \, \sum\nolimits_{m = -\infty}^{\infty} \psi_m X^\prime_{(d\cdot k -\ell)-m} \, \right|^2
     \label{eq:z_time-discrete_sld_output_2}
\end{align}
where $\psi_m = \psi(m T_\mathrm{s}')$ and the $N_\text{sim}$-fold upsampled string is $(X_m')_{m\in \mathbb{Z}} = ((0,0,0, X_\kappa) )_{\kappa \in \mathbb{Z}}$ with the sampling time $T_\mathrm{s}^{\prime}=T_\mathrm{s}/N_\text{sim}$ and with $d = N_\mathrm{sim}/N_\mathrm{os}$.

\subsection{\texorpdfstring{$L=0$}{L=0} Rates}
\label{subsec:back-to-back}

\newcommand{\gridfigurewidth}{7.4cm}
\newcommand{\gridfigureheight}{4cm}
\definecolor{mycolor1}{HTML}{0072bd}%
\definecolor{mycolor1bright}{HTML}{0fa0ff}%
\definecolor{mycolor1dark}{HTML}{0b5b8f}%

\definecolor{mycolor2}{rgb}{1.25000,0.32500,0.09800}%
\definecolor{mycolor3}{rgb}{0.92900,0.69400,0.12500}%
\definecolor{mycolor4}{rgb}{0.49400,0.18400,0.55600}%

\definecolor{mycolor5}{HTML}{27ae60}%
\definecolor{mycolor5bright}{HTML}{11d966}%
\definecolor{mycolor5dark}{HTML}{166337}%
\definecolor{mycolor6}{HTML}{FF0000}%
\definecolor{mycolor6bright}{HTML}{f76565}%
\definecolor{mycolor6dark}{HTML}{961717}%

\definecolor{mycolor7}{HTML}{0ae4fc}

\pgfplotsset{
   compat=1.15,
    PAM/.style={
       mycolor1,
       line width=0.85pt,
       mark=triangle, 
       mark options={solid,mycolor1},
       mark size=2.9pt,
    },
    ASK/.style={
       mycolor6,
       line width=0.85pt,
       mark=square, 
       mark options={solid,mycolor6},
       mark size=2pt,
    },
    QAM/.style={
       mycolor5,
       line width=0.85pt,
       mark=o, 
       mark options={solid,mycolor5},
       mark size=2pt,
    },
    PAM_I_q_YoXgYe/.style={
       mark=triangle,
       mark options={solid},
       mark size=2.9pt,
       line width=1.4pt,
       densely dotted,
   },
    ASK_I_q_YoXgYe/.style={
        mark=square,
        mark options={solid},
        mark size=2pt,
       line width=1.4pt,
        densely dotted,
    },
    QAM_I_q_YoXgYe/.style={
       mark=o,
       mark options={solid},
       mark size=2pt,
       line width=1.4pt,
      densely dotted,
    },
    PAM_I_q_YeX/.style={
       line width=2pt,
       solid,
       mark=star,
       mark options={solid},
       mark size=2.9pt,
   },
    ASK_I_q_YeX/.style={
        line width=2pt,
        solid,
        mark=star,
        mark options={solid},
        mark size=2.9pt,
    },
    QAM_I_q_YeX/.style={
        line width=2pt,
       solid,
       mark=star,
       mark options={solid},
       mark size=2.9pt,
    },
    TDRC/.style={
       solid,mark=*,mark size=2pt,mark options={draw=white,solid,line width=0.5pt}
    },
    TDRC_ICI/.style={
       solid,mark=diamond*,mark size=2.7pt,mark options={draw=white,solid,line width=0.5pt}
    },
    UB/.style={
       line width=0.85pt,
       mark=none, 
       densely dotted
    },
     LB/.style={
       line width=0.85pt,
       mark=none, 
       densely dash dot
    },
    PAM_L30km/.style={
       mark=triangle*, 
       mark size=4pt,
       mark options={draw=white,solid,line width=0.4pt}
    },
    ASK_L30km/.style={
       mark=square*, 
       mark size=2.5pt,
       mark options={draw=white,solid,line width=0.4pt}
    },
    QAM_L30km/.style={
       mark=*, 
       mark size=2.4pt,
       mark options={draw=white,solid,line width=0.4pt}
    },
    SERGeneralStyle/.style={
      width=\gridfigurewidth,
      height=\gridfigureheight,
      scale only axis,
      xmin=-5,
      xmax=13,
      xlabel style={font=\color{white!15!black}},
      xlabel={SNR $\text{[dB]}$},
      yminorticks=true,
      ylabel style={font=\color{white!15!black}},
      ylabel={SER},
      axis background/.style={fill=white},
      xmajorgrids,
      ymajorgrids,
      yminorgrids,
      xminorgrids,
      xtick distance={2},
      legend style={legend cell align=left, font=\scriptsize, align=left, draw=white!15!black,legend pos=south west},
    },
    MIGeneralStyle/.style={
      width=\gridfigurewidth,
      height=\gridfigureheight,
      at={(0.968in,0.544in)},
      scale only axis,
      xmin=-5,
      xmax=13,
      xlabel style={font=\color{white!15!black}},
      xlabel={SNR [dB]},
      ymin=0,
      ylabel style={font=\color{white!15!black}},
      ylabel={bpcu},
      axis background/.style={fill=white},
      xmajorgrids,
      ymajorgrids,
      xminorgrids,
      xtick distance={2},
      legend style={legend cell align=left, font=\scriptsize, align=left, draw=white!15!black,legend pos=north west},
      }
}

\tikzset{
    arr/.style={stealth-stealth,very thick},
    shadowed/.style={
    preaction={transform canvas={shift={(0.2pt,-0.2pt)}},draw=white,ultra thick}},
}

\begin{figure*}[t!]
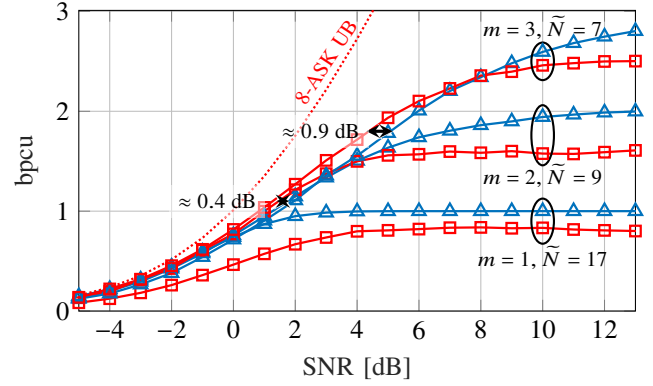
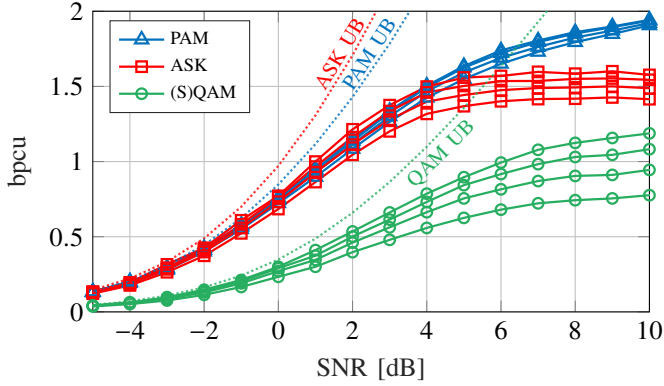
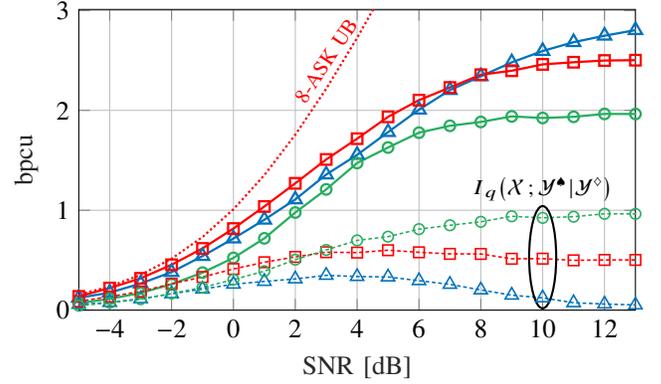
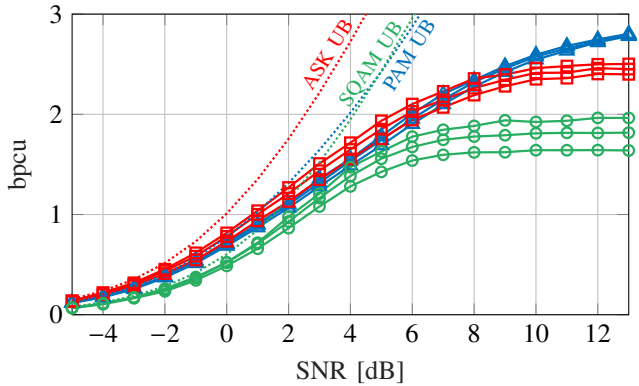
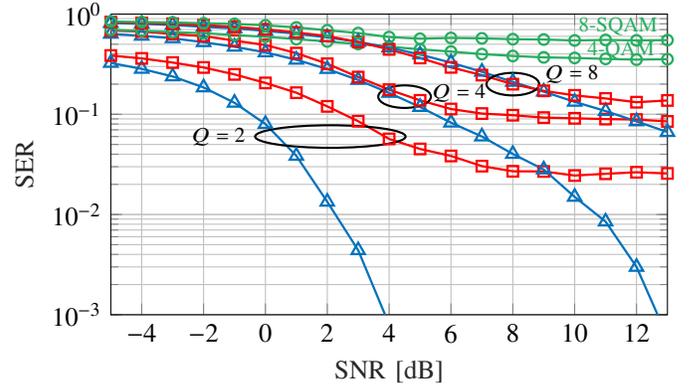
 %
\centering
\begin{subfigure}[b]{0.49\textwidth}
\centering
\begin{tikzpicture}

  \begin{axis}[%
  MIGeneralStyle,
  ymax=1.005,
  xmax=10]
  \input{Figures/standalone/Q2/dataQ2_MI_TXPOW.tex}
  
\addplot [PAM,opacity=1,forget plot] table {\PAMrcosspanseven};
\addplot [PAM,opacity=1,forget plot] table {\PAMrcosspannine};
\addplot [PAM,opacity=1,forget plot] table {\PAMrcosspaneleven};

\addplot [ASK,opacity=1,forget plot] table {\ASKrcosspanseven};
\addplot [ASK,opacity=1,forget plot] table {\ASKrcosspannine};
\addplot [ASK,opacity=1,forget plot] table {\ASKrcosspaneleven};
\addplot [ASK,opacity=1,forget plot] table {\ASKrcosspanthirten};
\addplot [ASK,opacity=1,forget plot] table {\ASKrcosspanfiften};
\addplot [ASK,opacity=1,forget plot] table {\ASKrcosspanseventen};

\addplot [PAM,UB,forget plot] table {\PAMCovUpperBound};
\addplot [ASK,UB,forget plot] table {\ASKCovUpperBound};

\node at (axis cs:-0.6,0.85)[mycolor1,rotate=64,font=\footnotesize]{PAM UB};
\node at (axis cs:1.3,0.75)[mycolor6,rotate=58,font=\footnotesize]{ASK UB};

\end{axis}
\end{tikzpicture}%
\captionsetup{width=.86\linewidth,margin={1cm,0cm}}
\caption{$2$-PAM/ASK, $ \smash{\widetilde{N}} = [7,9,11,13,15,17]$. \vspace*{0.4\baselineskip} }
\centering 
\begin{tikzpicture}

\begin{axis}[
  MIGeneralStyle,
  ymax=2.0,
  xmax=10]
\input{Figures/standalone/Q4/dataQ4_MI_TXPOW.tex}

\addplot [PAM,forget plot,opacity=1] table {\PAMrcosspanthree};
\addplot [PAM,forget plot,opacity=1] table {\PAMrcosspanfive};
\addplot [PAM,forget plot,opacity=1] table {\PAMrcosspanseven};
\addplot [PAM,opacity=1] table {\PAMrcosspannine};
\addlegendentry{PAM};

\addplot [ASK,forget plot,opacity=1] table {\ASKrcosspanthree};
\addplot [ASK,forget plot,opacity=1] table {\ASKrcosspanfive};
\addplot [ASK,opacity=1,forget plot] table {\ASKrcosspanseven};
\addplot [ASK,opacity=1] table {\ASKrcosspannine};
\addlegendentry{ASK};

\addplot [QAM,opacity=1] table {\QAMrcosspanthree};
\addplot [QAM,opacity=1] table {\QAMrcosspanfive};
\addplot [QAM,opacity=1] table {\QAMrcosspanseven};
\addplot [QAM,opacity=1] table {\QAMrcosspannine};
\addlegendentry{(S)QAM};

\addplot [PAM,UB,opacity=0.8,forget plot] table {\PAMCovUpperBound};
\addplot [ASK,UB,opacity=0.8,forget plot] table {\ASKCovUpperBound};
\addplot [QAM,UB,opacity=0.8,forget plot] table {\QAMCovUpperBound};

\node at (axis cs:1.65,1.72)[mycolor6,rotate=59,font=\footnotesize]{ASK UB};
\node at (axis cs:2.45,1.72)[mycolor1,rotate=57,font=\footnotesize]{PAM UB};
\node at (axis cs:4.4,1.05)[mycolor5,rotate=45,font=\footnotesize]{QAM UB};

\end{axis}
\end{tikzpicture}%
\captionsetup{width=.86\linewidth,margin={1cm,0cm}}
\caption{$4$-PAM/ASK/QAM, $\smash{\widetilde{N}} = [3,5,7,9]$. \vspace*{1.1\baselineskip}}
\centering
\begin{tikzpicture}

\begin{axis}[
  MIGeneralStyle,
  ymax=3.0]
\input{Figures/standalone/Q8/dataQ8_MI_TXPOW.tex}

\addplot [PAM,forget plot,opacity=1] table {\PAMrcosspanthree};
\addplot [PAM,opacity=1] table {\PAMrcosspanfive};
\addplot [PAM,forget plot,opacity=1] table {\PAMrcosspanseven};

\addplot [ASK,forget plot,opacity=1] table {\ASKrcosspanthree};
\addplot [ASK,opacity=1] table {\ASKrcosspanfive};
\addplot [ASK,opacity=1,forget plot] table {\ASKrcosspanseven};

\addplot [QAM,opacity=1,forget plot] table {\QAMrcosspanthree};
\addplot [QAM,opacity=1] table {\QAMrcosspanfive};
\addplot [QAM,opacity=1] table {\QAMrcosspanseven};

\addplot [PAM,UB,forget plot] table {\PAMCovUpperBound};
\addplot [ASK,UB,forget plot] table {\ASKCovUpperBound};
\addplot [QAM,UB,forget plot] table {\QAMCovUpperBound};

\node at (axis cs:3.25,2.6)[mycolor6,rotate=58,font=\footnotesize]{ASK UB};
\node at (axis cs:4.7,2.50)[mycolor5,rotate=56,font=\footnotesize]{SQAM UB};
\node at (axis cs:5.9,2.6)[mycolor1,rotate=58,font=\footnotesize]{PAM UB};

\end{axis}
\end{tikzpicture}%
\captionsetup{width=.86\linewidth,margin={1cm,0cm}}
\caption{$8$-PAM/ASK/SQAM, $\smash{\widetilde{N}} = [3,5,7]$. \vspace*{1.0\baselineskip}}
\end{subfigure}%
\hspace*{\fill}
\begin{subfigure}[b]{0.49\textwidth}
\begin{flushright}
\vspace*{20pt}
\begin{tikzpicture}

\pgfmathsetmacro\rcosrolloffa{0}    
\pgfmathsetmacro\rcosrolloffb{0}

\pgfdeclarelayer{fg}    %
\pgfdeclarelayer{bg}    %
\pgfsetlayers{bg,main,fg}  %

\begin{axis}[%
MIGeneralStyle,ymax=3+0.01,
ylabel={bpcu}
]
\input{Figures/standalone/Q2/dataQ2_MI_TXPOW.tex}
\addplot [PAM,forget plot] table[y expr=\thisrowno{1}*1/(1+\rcosrolloffa)] {\PAMrcosspaneleven};
\addplot [ASK,forget plot] table[y expr=\thisrowno{1}*1/(1+\rcosrolloffa)] {\ASKrcosspanseventen};

\input{Figures/standalone/Q4/dataQ4_MI_TXPOW.tex}
\addplot [PAM] table[y expr=\thisrowno{1}*1/(1+\rcosrolloffa)] {\PAMrcosspannine};
\addplot [ASK] table[y expr=\thisrowno{1}*1/(1+\rcosrolloffa)] {\ASKrcosspannine};

\input{Figures/standalone/Q8/dataQ8_MI_TXPOW.tex}
\addplot [PAM,forget plot] table[y expr=\thisrowno{1}*1/(1+\rcosrolloffa)] {\PAMrcosspanseven};
\addplot [ASK,forget plot] table[y expr=\thisrowno{1}*1/(1+\rcosrolloffa)] {\ASKrcosspanseven};

\addplot [ASK,UB,forget plot] table[y expr=\thisrowno{1}*1/(1+\rcosrolloffa)] {\ASKCovUpperBound};

\draw[thick]  (axis cs:10,0.9)  ellipse (0.15cm and 0.3cm)node[below,yshift=-0.2cm,font=\footnotesize]{$m=1, \widetilde{N} = 17$};
\draw[thick]  (axis cs:10,1.76)  ellipse (0.15cm and 0.4cm)node[below,yshift=-0.25cm,font=\footnotesize]{$m=2, \widetilde{N} = 9$};
\draw[thick]  (axis cs:10,2.50)  ellipse (0.15cm and 0.25cm)node[above,yshift=+0.17cm,font=\footnotesize]{$m=3, \widetilde{N} = 7$};

\begin{pgfonlayer}{fg}
\draw[arr,shadowed] (axis cs:1.4,1.1) -- (axis cs:1.8,1.1) node[font=\footnotesize,left,xshift=-0.3cm,yshift=0cm,fill=white,opacity=0.5,text opacity=1]{$\approx 0.4\;\text{dB}$};
\draw[arr,shadowed] (axis cs:4.3,1.8) -- (axis cs:5.15,1.8) node[font=\footnotesize,left,xshift=-0.3cm,yshift=0cm,fill=white,opacity=0.5,text opacity=1]{$\approx 0.9\;\text{dB}$};
\end{pgfonlayer}

\node at (axis cs:3.0,2.50)[red,rotate=56.5,font=\footnotesize]{$8$-ASK UB};

\end{axis}
\end{tikzpicture}%
\captionsetup{width=.86\linewidth,margin={1cm,0cm}}
\caption{$Q$-PAM/ASK for the largest $\smash{\widetilde{N}}$ studied here. }
\vspace{-1.5\baselineskip}
\vspace*{20pt}
\begin{tikzpicture}

\pgfmathsetmacro\rcosrolloffa{0}    
\pgfmathsetmacro\rcosrolloffb{0}    

\begin{axis}[%
MIGeneralStyle,ymax=3/1.0+0.01,
ylabel={bpcu}]

\input{Figures/standalone/Q8/dataQ8_MI_TXPOW.tex}
\addplot [PAM] table[,forget plot] {\PAMrcosspanseven};
\addplot [ASK] table[,forget plot] {\ASKrcosspanseven};
\addplot [QAM] table[,forget plot] {\QAMrcosspanseven};

\addplot [ASK,UB,forget plot] table[] {\ASKCovUpperBound};
\node at (axis cs:3.0,2.50)[red,rotate=56.5,font=\footnotesize]{$8$-ASK UB};

\addplot [PAM,PAM_I_q_YoXgYe, line width=0.6pt] table[] {\PAMIqYoXgYe};
\addplot [ASK,ASK_I_q_YoXgYe, line width=0.6pt,forget plot] table[] {\ASKIqYoXgYe};
\addplot [QAM,QAM_I_q_YoXgYe, line width=0.6pt,forget plot] table[] {\QAMIqYoXgYe};

\draw[thick]  (axis cs:10,0.53)  ellipse (0.18cm and 0.68cm)node[above,yshift=+0.62cm,font=\footnotesize]{$\MIsub{ \mathcal{X}}{\mathcal{Y}^\spadesuit \lvert \mathcal{Y}^\diamondsuit}{_q}$};

\end{axis}
\end{tikzpicture}%
\captionsetup{width=.86\linewidth,margin={1cm,0cm}}
\caption{$8$-PAM/ASK/SQAM for the largest studied $\smash{\widetilde{N}}$ and conditional information rate~\eqref{eq:I_q_cond_limit} (dotted with marks).}
\begin{tikzpicture}

\begin{axis}[%
SERGeneralStyle,
ymin=1E-3,
ymax=1,
ymode=log
]
\input{Figures/standalone/Q2/dataQ2_SER_TXPOW.tex}
\addplot [PAM,solid] table {\PAMrcosspanseventeen};
\addplot [ASK,solid] table {\ASKrcosspanseventeen};

\pgfplotstableread{
    X Y
-2	0.6913
-1	0.678
0	0.65896
1	0.63246
2	0.60408
3	0.57076
4	0.52456
5	0.48376
6	0.41558
7	0.35404
8	0.2928
9	0.22698
10	0.17422
11	0.13174
12	0.10042
13	0.07794
14	0.06544
15	0.05424
16	0.04842
17	0.04182
18	0.03838
19	0.75068
20	0.75084
}{\PAMrcosspanthree}

\pgfplotstableread{
    X Y
-2	0.69292
-1	0.6774
0	0.65882
1	0.63048
2	0.60178
3	0.56792
4	0.51984
5	0.48042
6	0.40804
7	0.34524
8	0.283
9	0.21158
10	0.15532
11	0.1118
12	0.0788
13	0.0561
14	0.0426
15	0.0322
16	0.02538
17	0.02094
18	0.01846
19	0.01662
20	0.01642
}{\PAMrcosspanfive}

\pgfplotstableread{
    X Y
-2	0.69044
-1	0.67602
0	0.65864
1	0.63
2	0.60126
3	0.56742
4	0.52006
5	0.47828
6	0.40586
7	0.3403
8	0.27656
9	0.2036
10	0.1436
11	0.09566
12	0.05988
13	0.03712
14	0.0214
15	0.0108
16	0.0068
17	0.00354
18	0.002
19	0.00196
20	0.00092
}{\PAMrcosspanseven}

\pgfplotstableread{
    X Y
  -13.0000    0.7309
  -12.0000    0.7286
  -11.0000    0.7230
  -10.0000    0.7145
   -9.0000    0.7049
   -8.0000    0.6926
   -7.0000    0.6736
   -6.0000    0.6620
   -5.0000    0.6353
   -4.0000    0.6069
   -3.0000    0.5740
   -2.0000    0.5240
   -1.0000    0.4720
         0    0.4147
    1.0000    0.3502
    2.0000    0.2830
    3.0000    0.2201
    4.0000    0.1635
    5.0000    0.1182
    6.0000    0.0825
    7.0000    0.0600
    8.0000    0.0402
    9.0000    0.0281
   10.0000    0.0150
   11.0000    0.0085
   12.0000    0.0030
   13.0000    0.0006
   14.0000    0.0002
   15.0000    0.0002
   16.0000    0.0001
   17.0000    0.0001
   18.0000    0.0002
}{\PAMrcosspannine}

\pgfplotstableread{
    X Y
-2	0.69416
-1	0.67822
0	0.65636
1	0.63588
2	0.60278
3	0.56346
4	0.51794
5	0.47032
6	0.4134
7	0.34664
8	0.28708
9	0.22484
10	0.18266
11	0.15196
12	0.1353
13	0.12294
14	0.11232
15	0.10672
16	0.10312
17	0.74812
18	0.75236
19	0.75068
20	0.75084
}{\ASKrcosspanthree}

\pgfplotstableread{
    X Y
-2	0.6945
-1	0.67774
0	0.65684
1	0.63598
2	0.60298
3	0.56218
4	0.5139
5	0.46754
6	0.408
7	0.33742
8	0.2692
9	0.19888
10	0.15066
11	0.11394
12	0.09942
13	0.08558
14	0.07528
15	0.07034
16	0.06158
17	0.06556
18	0.06032
19	0.75068
20	0.75084
}{\ASKrcosspanfive}

\pgfplotstableread{
    X Y
-2	0.69402
-1	0.67706
0	0.65738
1	0.6366
2	0.60306
3	0.56206
4	0.5136
5	0.46664
6	0.40604
7	0.33388
8	0.26154
9	0.18458
10	0.13082
11	0.0924
12	0.0733
13	0.05744
14	0.04606
15	0.04114
16	0.03342
17	0.03298
18	0.03074
19	0.02932
20	0.02946
}{\ASKrcosspanseven}

\pgfplotstableread{
    X Y
  -13.0000    0.7418
  -12.0000    0.7415
  -11.0000    0.7292
  -10.0000    0.7313
   -9.0000    0.7228
   -8.0000    0.7191
   -7.0000    0.7069
   -6.0000    0.6999
   -5.0000    0.6861
   -4.0000    0.6695
   -3.0000    0.6328
   -2.0000    0.6035
   -1.0000    0.5431
         0    0.4914
    1.0000    0.4076
    2.0000    0.3194
    3.0000    0.2352
    4.0000    0.1762
    5.0000    0.1378
    6.0000    0.1129
    7.0000    0.1017
    8.0000    0.0979
    9.0000    0.0927
   10.0000    0.0912
   11.0000    0.0896
   12.0000    0.0886
   13.0000    0.0849
   14.0000    0.0828
   15.0000    0.0827
   16.0000    0.0840
   17.0000    0.0848
   18.0000    0.0815
}{\ASKrcosspannine}

\pgfplotstableread{
    X Y
  -13.0000    0.7418
  -12.0000    0.7406
  -11.0000    0.7335
  -10.0000    0.7323
   -9.0000    0.7277
   -8.0000    0.7203
   -7.0000    0.7111
   -6.0000    0.7077
   -5.0000    0.6970
   -4.0000    0.6797
   -3.0000    0.6608
   -2.0000    0.6445
   -1.0000    0.6184
         0    0.5936
    1.0000    0.5622
    2.0000    0.5334
    3.0000    0.5021
    4.0000    0.4729
    5.0000    0.4448
    6.0000    0.4228
    7.0000    0.3989
    8.0000    0.3820
    9.0000    0.3701
   10.0000    0.3652
   11.0000    0.3573
   12.0000    0.3513
   13.0000    0.3549
   14.0000    0.3541
   15.0000    0.3502
   16.0000    0.3500
   17.0000    0.3504
   18.0000    0.3489
}{\QAMrcosspannine}
\addplot [PAM,solid] table {\PAMrcosspannine};
\addplot [ASK,solid] table {\ASKrcosspannine};
\addplot [QAM,solid] table {\QAMrcosspannine};

\pgfplotstableread{
    X Y
  -13.0000    0.8657
  -12.0000    0.8618
  -11.0000    0.8604
  -10.0000    0.8556
   -9.0000    0.8505
   -8.0000    0.8486
   -7.0000    0.8343
   -6.0000    0.8304
   -5.0000    0.8137
   -4.0000    0.7996
   -3.0000    0.7779
   -2.0000    0.7477
   -1.0000    0.7203
         0    0.6795
    1.0000    0.6378
    2.0000    0.5853
    3.0000    0.5282
    4.0000    0.4630
    5.0000    0.3951
    6.0000    0.3278
    7.0000    0.2656
    8.0000    0.2106
    9.0000    0.1682
   10.0000    0.1332
   11.0000    0.1077
   12.0000    0.0850
   13.0000    0.0665
   14.0000    0.0534
   15.0000    0.0417
   16.0000    0.0297
   17.0000    0.0177
   18.0000    0.0089
}{\PAMrcosspanseven}

\pgfplotstableread{
    X Y
   -13.0000    0.8686
  -12.0000    0.8650
  -11.0000    0.8649
  -10.0000    0.8578
   -9.0000    0.8597
   -8.0000    0.8525
   -7.0000    0.8481
   -6.0000    0.8381
   -5.0000    0.8331
   -4.0000    0.8143
   -3.0000    0.8013
   -2.0000    0.7765
   -1.0000    0.7468
         0    0.7051
    1.0000    0.6556
    2.0000    0.6040
    3.0000    0.5262
    4.0000    0.4444
    5.0000    0.3646
    6.0000    0.2940
    7.0000    0.2461
    8.0000    0.1998
    9.0000    0.1714
   10.0000    0.1545
   11.0000    0.1439
   12.0000    0.1320
   13.0000    0.1375
   14.0000    0.1364
   15.0000    0.1325
   16.0000    0.1374
   17.0000    0.1338
   18.0000    0.1326
}{\ASKrcosspanseven}

\pgfplotstableread{
    X Y
  -13.0000    0.8669
  -12.0000    0.8691
  -11.0000    0.8727
  -10.0000    0.8636
   -9.0000    0.8637
   -8.0000    0.8598
   -7.0000    0.8582
   -6.0000    0.8559
   -5.0000    0.8431
   -4.0000    0.8324
   -3.0000    0.8252
   -2.0000    0.8098
   -1.0000    0.7977
         0    0.7726
    1.0000    0.7362
    2.0000    0.6897
    3.0000    0.6469
    4.0000    0.5902
    5.0000    0.5676
    6.0000    0.5781
    7.0000    0.5728
    8.0000    0.5617
    9.0000    0.5575
   10.0000    0.5457
   11.0000    0.5523
   12.0000    0.5467
   13.0000    0.5502
   14.0000    0.5465
   15.0000    0.5482
   16.0000    0.5501
   17.0000    0.5502
   18.0000    0.5455
}{\QAMrcosspanseven}
\addplot [PAM,solid] table {\PAMrcosspanseven};
\addplot [ASK,solid] table {\ASKrcosspanseven};
\addplot [QAM,solid] table {\QAMrcosspanseven};

\draw[thick]  (axis cs:2.1,6E-2)  ellipse (1.0cm and 0.15cm)node[left,xshift=-1.0cm,font=\footnotesize,align=center]{$Q=2$};
\draw[thick]   (axis cs:4.5,1.5E-1)  ellipse (0.35cm and 0.15cm)node[right,xshift=0.24cm,yshift=0.05cm,font=\footnotesize]{$Q=4$};
\draw[thick]   (axis cs:8,2E-1)  ellipse (0.35cm and 0.15cm)node[right,xshift=+0.3cm,yshift=0.14cm,font=\footnotesize]{$Q=8$};

\node[] at (axis cs:11.5,4.4E-1)[mycolor5,font=\footnotesize]{$4$-QAM};
\node[] at (axis cs:11.4,7.5E-1)[mycolor5,font=\footnotesize]{$8$-SQAM};

\end{axis}
\end{tikzpicture}%
\captionsetup{width=.86\linewidth,margin={1cm,0cm}}
\caption{Uncoded HD SERs of $Q$-PAM/ASK/SQAM according to (a)-(c) for largest  studied $\smash{\widetilde{N}}$.}
\end{flushright}
\end{subfigure}%
\caption{Achievable rates, upper bounds and HD SERs for $L=0$. The colors in subplot (b) are used for all plots in the figure.}
\label{fig:2x3_plot_L=0km}
\end{figure*}

Fig.~\ref{fig:2x3_plot_L=0km} plots the rates~\eqref{eq:mi_lb_auxiliary_channel_sim}, upper bounds~\eqref{eq:upper_bound_determinant_covariance}, and HD symbol error rates (SERs) for a sinc pulse and $L=0$. Within the subplots, we vary the memory $\smash{\widetilde{N}}$ of the auxiliary channel. We remark that the simulation complexity grows rapidly with $Q$ and $\smash{\widetilde{N}}$ so that we had to use a smaller $\smash{\widetilde{N}}$ for larger $Q$.

Subplots (a)-(b) show that ASK and QAM benefit considerably from increased $\smash{\widetilde{N}}$ because the phase transitions must be recovered from the $\spadesuit$-samples that experience significant ISI. PAM is less sensitive to model mismatch and subplot (a) suggests that $2$-PAM should be preferred over $2$-ASK. However, from subplot (d) we see that the bipolar $4$-ASK and $8$-ASK alphabets gain approximately \SI{0.4}{dB} and \SI{0.9}{dB} over $4$-PAM and $8$-PAM, respectively, at intermediate SNRs.

\begin{figure*}[t!]
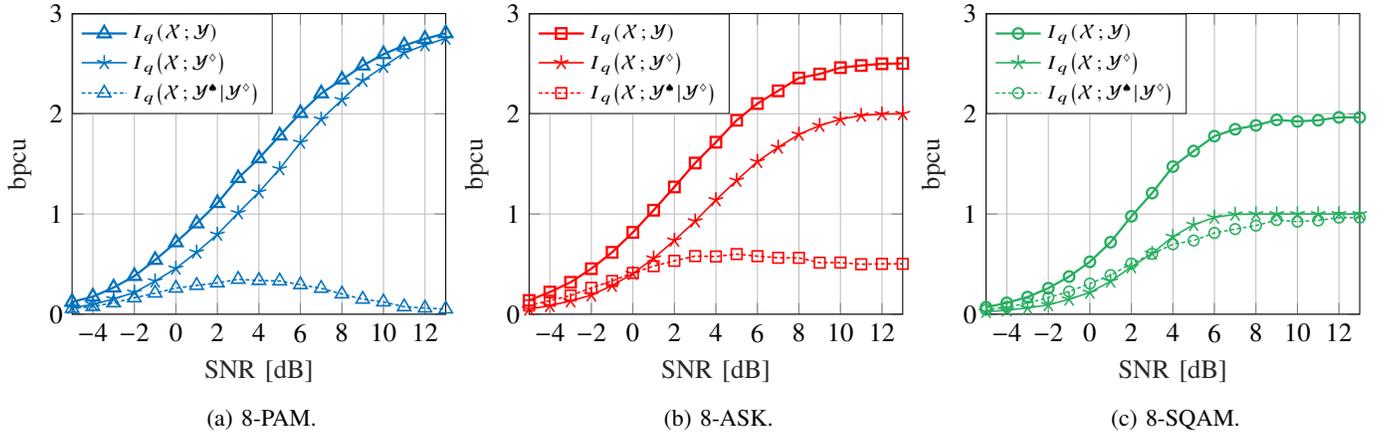
 %
\centering
\begin{subfigure}[b]{0.33\textwidth}
\centering
    \begin{tikzpicture}
    \begin{axis}[
      MIGeneralStyle,
      ymax=3.0,
      width=0.83\textwidth,
      legend style={at={(0.00,1.0)},anchor=north west}]
    \input{Figures/standalone/Q8/dataQ8_MI_TXPOW.tex}
    
    \addplot [PAM,opacity=1] table {\PAMrcosspanseven};
    \addplot [PAM,PAM_I_q_YeX, line width=0.6pt] table[] {\PAMIqYeX};
    \addplot [PAM,PAM_I_q_YoXgYe, line width=0.6pt] table[] {\PAMIqYoXgYe};
    
    \addlegendentry{$\MIsub{\mathcal{X}}{{\mathcal{Y}}}{_q}$}
    \addlegendentry{$\MIsub{\mathcal{X}}{{\mathcal{Y}^\diamondsuit}}{_q}$}
    \addlegendentry{$\MIsub{\mathcal{X}}{{\mathcal{Y}^\spadesuit \lvert \mathcal{Y}^\diamondsuit}}{_q}$}
    \end{axis}
    \end{tikzpicture}%
    \captionsetup{width=.86\linewidth,margin={1cm,0cm}}
    \caption{$8$-PAM.}

\end{subfigure}%
\hspace*{\fill}
\begin{subfigure}[b]{0.33\textwidth}
\centering
    \begin{tikzpicture}
    \begin{axis}[
      MIGeneralStyle,
      ymax=3.0,
      width=0.83\textwidth,
      legend style={at={(0.00,1.0)},anchor=north west},
      ]]
    \input{Figures/standalone/Q8/dataQ8_MI_TXPOW.tex}
    
    \addplot [ASK,opacity=1] table {\ASKrcosspanseven};
    \addplot [ASK,ASK_I_q_YeX, line width=0.6pt] table[] {\ASKIqYeX};
    \addplot [ASK,ASK_I_q_YoXgYe, line width=0.6pt] table[] {\ASKIqYoXgYe};
    
    \addlegendentry{$\MIsub{\mathcal{X}}{{\mathcal{Y}}}{_q}$}
    \addlegendentry{$\MIsub{\mathcal{X}}{{\mathcal{Y}^\diamondsuit}}{_q}$}
    \addlegendentry{$\MIsub{\mathcal{X}}{{\mathcal{Y}^\spadesuit \lvert \mathcal{Y}^\diamondsuit}}{_q}$}
    \end{axis}
    \end{tikzpicture}%
    \captionsetup{width=.86\linewidth,margin={1cm,0cm}}
    \caption{$8$-ASK.}

\end{subfigure}%
\hspace*{\fill}
\begin{subfigure}[b]{0.33\textwidth}
\centering
    \begin{tikzpicture}
    \begin{axis}[
      MIGeneralStyle,
      ymax=3.0,
      width=0.83\textwidth,
            legend style={at={(0.00,1.0)},anchor=north west}]]
    \input{Figures/standalone/Q8/dataQ8_MI_TXPOW.tex}
    
    \addplot [QAM,opacity=1] table {\QAMrcosspanseven};
    \addplot [QAM,QAM_I_q_YeX, line width=0.6pt] table[] {\QAMIqYeX};
    \addplot [QAM,QAM_I_q_YoXgYe, line width=0.6pt] table[] {\QAMIqYoXgYe};
    \addlegendentry{$\MIsub{\mathcal{X}}{{\mathcal{Y}}}{_q}$}
    \addlegendentry{$\MIsub{\mathcal{X}}{{\mathcal{Y}^\diamondsuit}}{_q}$}
    \addlegendentry{$\MIsub{\mathcal{X}}{{\mathcal{Y}^\spadesuit \lvert \mathcal{Y}^\diamondsuit}}{_q}$}
    \end{axis}
    \end{tikzpicture}%
    \captionsetup{width=.86\linewidth,margin={1cm,0cm}}
    \caption{$8$-SQAM.}

\end{subfigure}
\caption{Achievable rates for $L=0$, $Q=8$, and $\widetilde{N} = 7$. }
\label{fig:comparision_even_odd_samples_information_L=0}
\end{figure*}
Subplot (e) shows~\eqref{eq:mi_lb_auxiliary_channel_sim} for $8$-PAM/ASK/SQAM as well as the information in the $\spadesuit$-samples given the $\diamondsuit$-samples. As expected, $\MIsub{\mathcal{X}}{{\mathcal{Y}^\spadesuit} \lvert \mathcal{Y}^\diamondsuit }{_q}$ for $8$-PAM approaches zero at high SNR because the $\diamondsuit$-samples suffice to reconstruct the transmitted symbols. However, the $\spadesuit$-samples are useful at intermediate SNRs. For $8$-ASK and $8$-SQAM, the $\spadesuit$-samples contribute approximately \SI{0.5}{bpcu} and \SI{1}{bpcu}, respectively, at high SNR. One would expect the phase modulations to give \SI{1}{bpcu} and \SI{2}{bpcu}, respectively, which shows that there are substantial losses due to the auxiliary channel mismatch. 

Observe that the upper bounds are tight at low SNR and loose at high SNR. The upper bounds for $4$-QAM and $8$-SQAM are lower than for $Q$-ASK. The information in the $\diamondsuit$-samples is zero for $4$-QAM.

Finally, subplot (f) depicts the uncoded HD SERs with MAP decisions. The SERs of PAM decrease with SNR whereas the SERs of ASK and (S)QAM saturate due to the auxiliary channel mismatch. Moreover, (S)QAM with a sinc filter exhibits phase ambiguities despite differential encoding at the transmitter, and this causes an error floor.

Fig.~\ref{fig:comparision_even_odd_samples_information_L=0} plots the $L=0$ rates~\eqref{eq:limiting-q}, \eqref{eq:I_q_limit} and \eqref{eq:I_q_cond_limit} with oversampling, Nyquist-rate sampling~and conditional mutual information, respectively, for $Q=8$ and the largest studied auxiliary memory. Observe that there is considerable information in the $\spadesuit$-samples. As expected, $I_q(\mathcal{X}\,;\, \mathcal{Y}^\diamondsuit)$ saturates at the logarithm of the number of intensities of the modulation, i.e., 3, 2, 1 bits for 8-PAM, 8-ASK, 8-SQAM, respectively. 

\subsection{CD Rates}
\label{subsec:CDrates}

\begin{figure*}[!t]
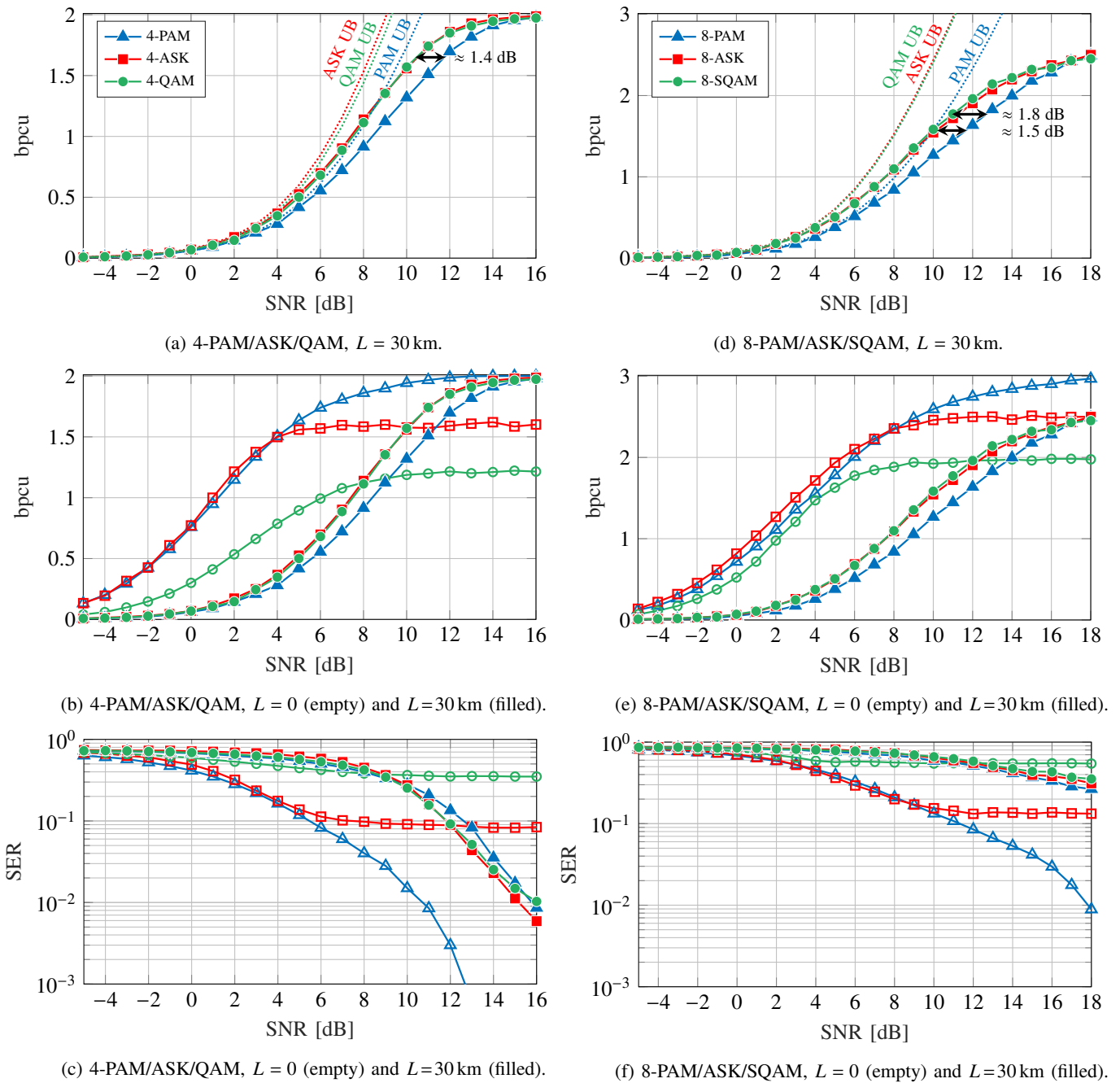
 %
        \begin{subfigure}[t]{0.5\textwidth}
        \begin{flushright}
        \begin{tikzpicture}

\pgfmathsetmacro\rcosrolloffa{0}    
\pgfmathsetmacro\rcosrolloffb{0} 
\pgfmathsetmacro\attshift{6}

\begin{axis}[%
MIGeneralStyle,ymax=2+0.01,
xmax=16,
ylabel={bpcu}
]

\input{Figures/standalone/Q4/dataQ4_MI_TXPOW}

\pgfplotstableread{
    X Y
   -2.0000    0.0222
   -1.0000    0.0347
         0    0.0528
    1.0000    0.0862
    2.0000    0.1314
    3.0000    0.1900
    4.0000    0.2773
    5.0000    0.3724
    6.0000    0.5196
    7.0000    0.6718
    8.0000    0.8300
    9.0000    1.0148
   10.0000    1.1881
   11.0000    1.3574
   12.0000    1.5058
   13.0000    1.6370
   14.0000    1.7429
   15.0000    1.8248
   16.0000    1.8698
   17.0000    1.8840
   18.0000    1.8809
   19.0000    1.8548
   20.0000    1.8231
}{\PAMrcosspanfive}

\pgfplotstableread{
    X Y
 -2.0000    0.0233
   -1.0000    0.0365
         0    0.0575
    1.0000    0.0861
    2.0000    0.1397
    3.0000    0.2144
    4.0000    0.3076
    5.0000    0.4475
    6.0000    0.6333
    7.0000    0.8232
    8.0000    1.0510
    9.0000    1.2795
   10.0000    1.4934
   11.0000    1.6580
   12.0000    1.7685
   13.0000    1.8402
   14.0000    1.8596
   15.0000    1.8740
   16.0000    1.8706
   17.0000    1.8088
   18.0000    1.7296
   19.0000    1.6315
   20.0000   -1.0000
}{\ASKrcosspanfive}

\pgfplotstableread{
    X Y
   -2.0000    0.0237
   -1.0000    0.0335
         0    0.0537
    1.0000    0.0813
    2.0000    0.1254
    3.0000    0.2012
    4.0000    0.2947
    5.0000    0.4124
    6.0000    0.5888
    7.0000    0.7880
    8.0000    1.0010
    9.0000    1.2097
   10.0000    1.3913
   11.0000    1.5561
   12.0000    1.6544
   13.0000    1.7033
   14.0000    1.7197
   15.0000    1.7024
   16.0000    1.6556
   17.0000    1.5085
   18.0000    1.3625
   19.0000    1.0889
   20.0000   -1.0000
}{\QAMrcosspanfive}

\pgfplotstableread{
    X Y
   -13.0000    0.0017
  -12.0000    0.0031
  -11.0000    0.0066
  -10.0000    0.0101
   -9.0000    0.0165
   -8.0000    0.0229
   -7.0000    0.0378
   -6.0000    0.0613
   -5.0000    0.0910
   -4.0000    0.1442
   -3.0000    0.2087
   -2.0000    0.2797
   -1.0000    0.4179
         0    0.5555
    1.0000    0.7218
    2.0000    0.9149
    3.0000    1.1239
    4.0000    1.3194
    5.0000    1.5106
    6.0000    1.6983
    7.0000    1.8187
    8.0000    1.9124
    9.0000    1.9540
   10.0000    1.9787
   11.0000    1.9902
   12.0000    1.9947
   13.0000    1.9956
   14.0000    1.9986
   15.0000    1.9969
   16.0000    1.9972
   17.0000    1.9981
   18.0000    1.9986
}{\PAMrcosspannine}

\pgfplotstableread{
    X Y
  -13.0000    0.0031
  -12.0000    0.0054
  -11.0000    0.0069
  -10.0000    0.0147
   -9.0000    0.0208
   -8.0000    0.0309
   -7.0000    0.0453
   -6.0000    0.0704
   -5.0000    0.1126
   -4.0000    0.1731
   -3.0000    0.2506
   -2.0000    0.3669
   -1.0000    0.5260
         0    0.6985
    1.0000    0.9021
    2.0000    1.1400
    3.0000    1.3574
    4.0000    1.5586
    5.0000    1.7414
    6.0000    1.8592
    7.0000    1.9292
    8.0000    1.9624
    9.0000    1.9792
   10.0000    1.9874
   11.0000    1.9925
   12.0000    1.9962
   13.0000    1.9971
   14.0000    1.9962
   15.0000    1.9965
   16.0000    1.9975
   17.0000    1.9970
   18.0000    1.9950
}{\ASKrcosspannine}

\pgfplotstableread{
    X Y
  -13.0000    0.0025
  -12.0000    0.0046
  -11.0000    0.0082
  -10.0000    0.0114
   -9.0000    0.0172
   -8.0000    0.0275
   -7.0000    0.0441
   -6.0000    0.0683
   -5.0000    0.1064
   -4.0000    0.1463
   -3.0000    0.2448
   -2.0000    0.3474
   -1.0000    0.5008
         0    0.6810
    1.0000    0.8860
    2.0000    1.1131
    3.0000    1.3532
    4.0000    1.5704
    5.0000    1.7415
    6.0000    1.8502
    7.0000    1.9079
    8.0000    1.9457
    9.0000    1.9655
   10.0000    1.9729
   11.0000    1.9795
   12.0000    1.9840
   13.0000    1.9850
   14.0000    1.9888
   15.0000    1.9875
   16.0000    1.9866
   17.0000    1.9878
   18.0000    1.9865
}{\QAMrcosspannine}

\pgfplotstableread{
    X Y
  -13.0000    0.0032
  -12.0000    0.0050
  -11.0000    0.0079
  -10.0000    0.0125
   -9.0000    0.0198
   -8.0000    0.0312
   -7.0000    0.0490
   -6.0000    0.0766
   -5.0000    0.1190
   -4.0000    0.1830
   -3.0000    0.2774
   -2.0000    0.4125
   -1.0000    0.5988
         0    0.8449
    1.0000    1.1555
    2.0000    1.5309
    3.0000    1.9667
    4.0000    2.4557
    5.0000       Inf
    6.0000       Inf
    7.0000       Inf
    8.0000       Inf
    9.0000       Inf
   10.0000       Inf
   11.0000       Inf
   12.0000       Inf
   13.0000       Inf
   14.0000       Inf
   15.0000       Inf
   16.0000       Inf
   17.0000       Inf
   18.0000       Inf
}{\ASKCovUpperBound}

\pgfplotstableread{
    X Y
 -13.0000    0.0025
  -12.0000    0.0040
  -11.0000    0.0063
  -10.0000    0.0100
   -9.0000    0.0157
   -8.0000    0.0248
   -7.0000    0.0388
   -6.0000    0.0604
   -5.0000    0.0932
   -4.0000    0.1421
   -3.0000    0.2129
   -2.0000    0.3119
   -1.0000    0.4451
         0    0.6172
    1.0000    0.8311
    2.0000    1.0881
    3.0000    1.3890
    4.0000    1.7339
    5.0000    2.1230
    6.0000       Inf
    7.0000       Inf
    8.0000       Inf
    9.0000       Inf
   10.0000       Inf
   11.0000       Inf
   12.0000       Inf
   13.0000       Inf
   14.0000       Inf
   15.0000       Inf
   16.0000       Inf
   17.0000       Inf
   18.0000       Inf
}{\PAMCovUpperBound}

\pgfplotstableread{
    X Y
  -13.0000    0.0029
  -12.0000    0.0046
  -11.0000    0.0073
  -10.0000    0.0116
   -9.0000    0.0183
   -8.0000    0.0289
   -7.0000    0.0454
   -6.0000    0.0711
   -5.0000    0.1104
   -4.0000    0.1699
   -3.0000    0.2578
   -2.0000    0.3838
   -1.0000    0.5580
         0    0.7890
    1.0000    1.0821
    2.0000    1.4381
    3.0000    1.8535
    4.0000    2.3219
    5.0000       Inf
    6.0000       Inf
    7.0000       Inf
    8.0000       Inf
    9.0000       Inf
   10.0000       Inf
   11.0000       Inf
   12.0000       Inf
   13.0000       Inf
   14.0000       Inf
   15.0000       Inf
   16.0000       Inf
   17.0000       Inf
   18.0000       Inf
}{\QAMCovUpperBound}
\addplot [PAM,PAM_L30km] table[y expr=\thisrowno{1}*1/(1+\rcosrolloffa),x expr=\thisrowno{0}+\attshift] {\PAMrcosspannine}; 
\addplot [ASK,ASK_L30km] table[y expr=\thisrowno{1}*1/(1+\rcosrolloffa),x expr=\thisrowno{0}+\attshift] {\ASKrcosspannine}; 
\addplot [QAM,QAM_L30km] table[y expr=\thisrowno{1}*1/(1+\rcosrolloffa),x expr=\thisrowno{0}+\attshift] {\QAMrcosspannine};

\addplot [PAM,UB,forget plot] table[y expr=\thisrowno{1}*1/(1+\rcosrolloffa),x expr=\thisrowno{0}+\attshift] {\PAMCovUpperBound};
\addplot [ASK,UB,forget plot] table[y expr=\thisrowno{1}*1/(1+\rcosrolloffa),x expr=\thisrowno{0}+\attshift] {\ASKCovUpperBound};
\addplot [QAM,UB,forget plot] table[y expr=\thisrowno{1}*1/(1+\rcosrolloffa),x expr=\thisrowno{0}+\attshift] {\QAMCovUpperBound};

\draw[arr,shadowed] (axis cs:10.4,1.65) -- (axis cs:11.8,1.65) node[font=\footnotesize,right]{$\approx 1.4\;\text{dB}$};

\addlegendentry{$4$-PAM}
\addlegendentry{$4$-ASK}
\addlegendentry{$4$-QAM}

\node at (axis cs:6.9,1.72)[mycolor6,rotate=65,font=\footnotesize]{ASK UB};
\node at (axis cs:7.80,1.67)[mycolor5,rotate=65,font=\footnotesize]{QAM UB};
\node at (axis cs:9.35,1.72)[mycolor1,rotate=64,font=\footnotesize]{PAM UB};

\end{axis}
\end{tikzpicture}%
        \captionsetup{width=.90\linewidth,margin={1cm,0cm}}
        \caption{$4$-PAM/ASK/QAM, $L=\SI{30}{\kilo\meter}$. \vspace*{0.3\baselineskip} }
        \begin{tikzpicture}

\pgfmathsetmacro\rcosrolloffa{0}    
\pgfmathsetmacro\rcosrolloffb{0}    
\pgfmathsetmacro\attshift{6}

\begin{axis}[%
MIGeneralStyle,ymax=2+0.01,
xmax=16,
ylabel={bpcu}
]

\input{Figures/standalone/Q4/dataQ4_MI_TXPOW}
\addplot [PAM,forget plot] table[y expr=\thisrowno{1}*1/(1+\rcosrolloffa)] {\PAMrcosspannine};
\addplot [ASK,forget plot] table[y expr=\thisrowno{1}*1/(1+\rcosrolloffa)] {\ASKrcosspannine};
\addplot [QAM,forget plot] table[y expr=\thisrowno{1}*1/(1+\rcosrolloffa)] {\QAMrcosspannine};

\pgfplotstableread{
    X Y
   -2.0000    0.0222
   -1.0000    0.0347
         0    0.0528
    1.0000    0.0862
    2.0000    0.1314
    3.0000    0.1900
    4.0000    0.2773
    5.0000    0.3724
    6.0000    0.5196
    7.0000    0.6718
    8.0000    0.8300
    9.0000    1.0148
   10.0000    1.1881
   11.0000    1.3574
   12.0000    1.5058
   13.0000    1.6370
   14.0000    1.7429
   15.0000    1.8248
   16.0000    1.8698
   17.0000    1.8840
   18.0000    1.8809
   19.0000    1.8548
   20.0000    1.8231
}{\PAMrcosspanfive}

\pgfplotstableread{
    X Y
 -2.0000    0.0233
   -1.0000    0.0365
         0    0.0575
    1.0000    0.0861
    2.0000    0.1397
    3.0000    0.2144
    4.0000    0.3076
    5.0000    0.4475
    6.0000    0.6333
    7.0000    0.8232
    8.0000    1.0510
    9.0000    1.2795
   10.0000    1.4934
   11.0000    1.6580
   12.0000    1.7685
   13.0000    1.8402
   14.0000    1.8596
   15.0000    1.8740
   16.0000    1.8706
   17.0000    1.8088
   18.0000    1.7296
   19.0000    1.6315
   20.0000   -1.0000
}{\ASKrcosspanfive}

\pgfplotstableread{
    X Y
   -2.0000    0.0237
   -1.0000    0.0335
         0    0.0537
    1.0000    0.0813
    2.0000    0.1254
    3.0000    0.2012
    4.0000    0.2947
    5.0000    0.4124
    6.0000    0.5888
    7.0000    0.7880
    8.0000    1.0010
    9.0000    1.2097
   10.0000    1.3913
   11.0000    1.5561
   12.0000    1.6544
   13.0000    1.7033
   14.0000    1.7197
   15.0000    1.7024
   16.0000    1.6556
   17.0000    1.5085
   18.0000    1.3625
   19.0000    1.0889
   20.0000   -1.0000
}{\QAMrcosspanfive}

\pgfplotstableread{
    X Y
   -13.0000    0.0017
  -12.0000    0.0031
  -11.0000    0.0066
  -10.0000    0.0101
   -9.0000    0.0165
   -8.0000    0.0229
   -7.0000    0.0378
   -6.0000    0.0613
   -5.0000    0.0910
   -4.0000    0.1442
   -3.0000    0.2087
   -2.0000    0.2797
   -1.0000    0.4179
         0    0.5555
    1.0000    0.7218
    2.0000    0.9149
    3.0000    1.1239
    4.0000    1.3194
    5.0000    1.5106
    6.0000    1.6983
    7.0000    1.8187
    8.0000    1.9124
    9.0000    1.9540
   10.0000    1.9787
   11.0000    1.9902
   12.0000    1.9947
   13.0000    1.9956
   14.0000    1.9986
   15.0000    1.9969
   16.0000    1.9972
   17.0000    1.9981
   18.0000    1.9986
}{\PAMrcosspannine}

\pgfplotstableread{
    X Y
  -13.0000    0.0031
  -12.0000    0.0054
  -11.0000    0.0069
  -10.0000    0.0147
   -9.0000    0.0208
   -8.0000    0.0309
   -7.0000    0.0453
   -6.0000    0.0704
   -5.0000    0.1126
   -4.0000    0.1731
   -3.0000    0.2506
   -2.0000    0.3669
   -1.0000    0.5260
         0    0.6985
    1.0000    0.9021
    2.0000    1.1400
    3.0000    1.3574
    4.0000    1.5586
    5.0000    1.7414
    6.0000    1.8592
    7.0000    1.9292
    8.0000    1.9624
    9.0000    1.9792
   10.0000    1.9874
   11.0000    1.9925
   12.0000    1.9962
   13.0000    1.9971
   14.0000    1.9962
   15.0000    1.9965
   16.0000    1.9975
   17.0000    1.9970
   18.0000    1.9950
}{\ASKrcosspannine}

\pgfplotstableread{
    X Y
  -13.0000    0.0025
  -12.0000    0.0046
  -11.0000    0.0082
  -10.0000    0.0114
   -9.0000    0.0172
   -8.0000    0.0275
   -7.0000    0.0441
   -6.0000    0.0683
   -5.0000    0.1064
   -4.0000    0.1463
   -3.0000    0.2448
   -2.0000    0.3474
   -1.0000    0.5008
         0    0.6810
    1.0000    0.8860
    2.0000    1.1131
    3.0000    1.3532
    4.0000    1.5704
    5.0000    1.7415
    6.0000    1.8502
    7.0000    1.9079
    8.0000    1.9457
    9.0000    1.9655
   10.0000    1.9729
   11.0000    1.9795
   12.0000    1.9840
   13.0000    1.9850
   14.0000    1.9888
   15.0000    1.9875
   16.0000    1.9866
   17.0000    1.9878
   18.0000    1.9865
}{\QAMrcosspannine}

\pgfplotstableread{
    X Y
  -13.0000    0.0032
  -12.0000    0.0050
  -11.0000    0.0079
  -10.0000    0.0125
   -9.0000    0.0198
   -8.0000    0.0312
   -7.0000    0.0490
   -6.0000    0.0766
   -5.0000    0.1190
   -4.0000    0.1830
   -3.0000    0.2774
   -2.0000    0.4125
   -1.0000    0.5988
         0    0.8449
    1.0000    1.1555
    2.0000    1.5309
    3.0000    1.9667
    4.0000    2.4557
    5.0000       Inf
    6.0000       Inf
    7.0000       Inf
    8.0000       Inf
    9.0000       Inf
   10.0000       Inf
   11.0000       Inf
   12.0000       Inf
   13.0000       Inf
   14.0000       Inf
   15.0000       Inf
   16.0000       Inf
   17.0000       Inf
   18.0000       Inf
}{\ASKCovUpperBound}

\pgfplotstableread{
    X Y
 -13.0000    0.0025
  -12.0000    0.0040
  -11.0000    0.0063
  -10.0000    0.0100
   -9.0000    0.0157
   -8.0000    0.0248
   -7.0000    0.0388
   -6.0000    0.0604
   -5.0000    0.0932
   -4.0000    0.1421
   -3.0000    0.2129
   -2.0000    0.3119
   -1.0000    0.4451
         0    0.6172
    1.0000    0.8311
    2.0000    1.0881
    3.0000    1.3890
    4.0000    1.7339
    5.0000    2.1230
    6.0000       Inf
    7.0000       Inf
    8.0000       Inf
    9.0000       Inf
   10.0000       Inf
   11.0000       Inf
   12.0000       Inf
   13.0000       Inf
   14.0000       Inf
   15.0000       Inf
   16.0000       Inf
   17.0000       Inf
   18.0000       Inf
}{\PAMCovUpperBound}

\pgfplotstableread{
    X Y
  -13.0000    0.0029
  -12.0000    0.0046
  -11.0000    0.0073
  -10.0000    0.0116
   -9.0000    0.0183
   -8.0000    0.0289
   -7.0000    0.0454
   -6.0000    0.0711
   -5.0000    0.1104
   -4.0000    0.1699
   -3.0000    0.2578
   -2.0000    0.3838
   -1.0000    0.5580
         0    0.7890
    1.0000    1.0821
    2.0000    1.4381
    3.0000    1.8535
    4.0000    2.3219
    5.0000       Inf
    6.0000       Inf
    7.0000       Inf
    8.0000       Inf
    9.0000       Inf
   10.0000       Inf
   11.0000       Inf
   12.0000       Inf
   13.0000       Inf
   14.0000       Inf
   15.0000       Inf
   16.0000       Inf
   17.0000       Inf
   18.0000       Inf
}{\QAMCovUpperBound}
\addplot [PAM,PAM_L30km] table[y expr=\thisrowno{1}*1/(1+\rcosrolloffa),x expr=\thisrowno{0}+\attshift] {\PAMrcosspannine}; 
\addplot [ASK,ASK_L30km] table[y expr=\thisrowno{1}*1/(1+\rcosrolloffa),x expr=\thisrowno{0}+\attshift] {\ASKrcosspannine}; 
\addplot [QAM,QAM_L30km] table[y expr=\thisrowno{1}*1/(1+\rcosrolloffa),x expr=\thisrowno{0}+\attshift] {\QAMrcosspannine};

\end{axis}
\end{tikzpicture}%
        \captionsetup{width=.90\linewidth,margin={1cm,0cm}}
        \caption{$4$-PAM/ASK/QAM, $L=0$ (empty) and $L\!=\!\SI{30}{\kilo\meter}$ (filled). \vspace*{0.3\baselineskip}   }
        \begin{tikzpicture}
\pgfmathsetmacro\attshift{6}    
\begin{axis}[%
SERGeneralStyle,
ymin=1E-3,
ymax=1,
xmax=16,
ymode=log
]
\pgfplotstableread{
    X Y
-2	0.6913
-1	0.678
0	0.65896
1	0.63246
2	0.60408
3	0.57076
4	0.52456
5	0.48376
6	0.41558
7	0.35404
8	0.2928
9	0.22698
10	0.17422
11	0.13174
12	0.10042
13	0.07794
14	0.06544
15	0.05424
16	0.04842
17	0.04182
18	0.03838
19	0.75068
20	0.75084
}{\PAMrcosspanthree}

\pgfplotstableread{
    X Y
-2	0.69292
-1	0.6774
0	0.65882
1	0.63048
2	0.60178
3	0.56792
4	0.51984
5	0.48042
6	0.40804
7	0.34524
8	0.283
9	0.21158
10	0.15532
11	0.1118
12	0.0788
13	0.0561
14	0.0426
15	0.0322
16	0.02538
17	0.02094
18	0.01846
19	0.01662
20	0.01642
}{\PAMrcosspanfive}

\pgfplotstableread{
    X Y
-2	0.69044
-1	0.67602
0	0.65864
1	0.63
2	0.60126
3	0.56742
4	0.52006
5	0.47828
6	0.40586
7	0.3403
8	0.27656
9	0.2036
10	0.1436
11	0.09566
12	0.05988
13	0.03712
14	0.0214
15	0.0108
16	0.0068
17	0.00354
18	0.002
19	0.00196
20	0.00092
}{\PAMrcosspanseven}

\pgfplotstableread{
    X Y
  -13.0000    0.7309
  -12.0000    0.7286
  -11.0000    0.7230
  -10.0000    0.7145
   -9.0000    0.7049
   -8.0000    0.6926
   -7.0000    0.6736
   -6.0000    0.6620
   -5.0000    0.6353
   -4.0000    0.6069
   -3.0000    0.5740
   -2.0000    0.5240
   -1.0000    0.4720
         0    0.4147
    1.0000    0.3502
    2.0000    0.2830
    3.0000    0.2201
    4.0000    0.1635
    5.0000    0.1182
    6.0000    0.0825
    7.0000    0.0600
    8.0000    0.0402
    9.0000    0.0281
   10.0000    0.0150
   11.0000    0.0085
   12.0000    0.0030
   13.0000    0.0006
   14.0000    0.0002
   15.0000    0.0002
   16.0000    0.0001
   17.0000    0.0001
   18.0000    0.0002
}{\PAMrcosspannine}

\pgfplotstableread{
    X Y
-2	0.69416
-1	0.67822
0	0.65636
1	0.63588
2	0.60278
3	0.56346
4	0.51794
5	0.47032
6	0.4134
7	0.34664
8	0.28708
9	0.22484
10	0.18266
11	0.15196
12	0.1353
13	0.12294
14	0.11232
15	0.10672
16	0.10312
17	0.74812
18	0.75236
19	0.75068
20	0.75084
}{\ASKrcosspanthree}

\pgfplotstableread{
    X Y
-2	0.6945
-1	0.67774
0	0.65684
1	0.63598
2	0.60298
3	0.56218
4	0.5139
5	0.46754
6	0.408
7	0.33742
8	0.2692
9	0.19888
10	0.15066
11	0.11394
12	0.09942
13	0.08558
14	0.07528
15	0.07034
16	0.06158
17	0.06556
18	0.06032
19	0.75068
20	0.75084
}{\ASKrcosspanfive}

\pgfplotstableread{
    X Y
-2	0.69402
-1	0.67706
0	0.65738
1	0.6366
2	0.60306
3	0.56206
4	0.5136
5	0.46664
6	0.40604
7	0.33388
8	0.26154
9	0.18458
10	0.13082
11	0.0924
12	0.0733
13	0.05744
14	0.04606
15	0.04114
16	0.03342
17	0.03298
18	0.03074
19	0.02932
20	0.02946
}{\ASKrcosspanseven}

\pgfplotstableread{
    X Y
  -13.0000    0.7418
  -12.0000    0.7415
  -11.0000    0.7292
  -10.0000    0.7313
   -9.0000    0.7228
   -8.0000    0.7191
   -7.0000    0.7069
   -6.0000    0.6999
   -5.0000    0.6861
   -4.0000    0.6695
   -3.0000    0.6328
   -2.0000    0.6035
   -1.0000    0.5431
         0    0.4914
    1.0000    0.4076
    2.0000    0.3194
    3.0000    0.2352
    4.0000    0.1762
    5.0000    0.1378
    6.0000    0.1129
    7.0000    0.1017
    8.0000    0.0979
    9.0000    0.0927
   10.0000    0.0912
   11.0000    0.0896
   12.0000    0.0886
   13.0000    0.0849
   14.0000    0.0828
   15.0000    0.0827
   16.0000    0.0840
   17.0000    0.0848
   18.0000    0.0815
}{\ASKrcosspannine}

\pgfplotstableread{
    X Y
  -13.0000    0.7418
  -12.0000    0.7406
  -11.0000    0.7335
  -10.0000    0.7323
   -9.0000    0.7277
   -8.0000    0.7203
   -7.0000    0.7111
   -6.0000    0.7077
   -5.0000    0.6970
   -4.0000    0.6797
   -3.0000    0.6608
   -2.0000    0.6445
   -1.0000    0.6184
         0    0.5936
    1.0000    0.5622
    2.0000    0.5334
    3.0000    0.5021
    4.0000    0.4729
    5.0000    0.4448
    6.0000    0.4228
    7.0000    0.3989
    8.0000    0.3820
    9.0000    0.3701
   10.0000    0.3652
   11.0000    0.3573
   12.0000    0.3513
   13.0000    0.3549
   14.0000    0.3541
   15.0000    0.3502
   16.0000    0.3500
   17.0000    0.3504
   18.0000    0.3489
}{\QAMrcosspannine}
\addplot [PAM] table {\PAMrcosspannine};
\addplot [ASK] table {\ASKrcosspannine};
\addplot [QAM] table {\QAMrcosspannine};

\pgfplotstableread{
    X Y
  -13.0000    0.7467
  -12.0000    0.7403
  -11.0000    0.7418
  -10.0000    0.7362
   -9.0000    0.7306
   -8.0000    0.7304
   -7.0000    0.7292
   -6.0000    0.7152
   -5.0000    0.7090
   -4.0000    0.6929
   -3.0000    0.6786
   -2.0000    0.6570
   -1.0000    0.6181
         0    0.5808
    1.0000    0.5289
    2.0000    0.4499
    3.0000    0.3654
    4.0000    0.2743
    5.0000    0.1607
    6.0000    0.0895
    7.0000    0.0438
    8.0000    0.0229
    9.0000    0.0112
   10.0000    0.0059
   11.0000    0.0043
   12.0000    0.0031
   13.0000    0.0017
   14.0000    0.0027
   15.0000    0.0022
   16.0000    0.0018
   17.0000    0.0020
   18.0000    0.0029
}{\ASKrcosspannine}

 \pgfplotstableread{
     X Y
   -13.0000    0.7373
  -12.0000    0.7308
  -11.0000    0.7277
  -10.0000    0.7218
   -9.0000    0.7152
   -8.0000    0.7062
   -7.0000    0.6897
   -6.0000    0.6772
   -5.0000    0.6564
   -4.0000    0.6315
   -3.0000    0.6049
   -2.0000    0.5849
   -1.0000    0.5420
         0    0.5020
    1.0000    0.4553
    2.0000    0.3989
    3.0000    0.3323
    4.0000    0.2804
    5.0000    0.2079
    6.0000    0.1351
    7.0000    0.0829
    8.0000    0.0355
    9.0000    0.0174
   10.0000    0.0086
   11.0000    0.0049
   12.0000    0.0022
   13.0000    0.0018
   14.0000    0.0011
   15.0000    0.0014
   16.0000    0.0009
   17.0000    0.0011
   18.0000    0.0012
 }{\PAMrcosspannine}

\pgfplotstableread{
    X Y
  -13.0000    0.7357
  -12.0000    0.7339
  -11.0000    0.7255
  -10.0000    0.7285
   -9.0000    0.7215
   -8.0000    0.7100
   -7.0000    0.7035
   -6.0000    0.6889
   -5.0000    0.6735
   -4.0000    0.6575
   -3.0000    0.6250
   -2.0000    0.6046
   -1.0000    0.5681
         0    0.5310
    1.0000    0.4827
    2.0000    0.4219
    3.0000    0.3439
    4.0000    0.2531
    5.0000    0.1563
    6.0000    0.0918
    7.0000    0.0515
    8.0000    0.0253
    9.0000    0.0149
   10.0000    0.0103
   11.0000    0.0083
   12.0000    0.0059
   13.0000    0.0050
   14.0000    0.0045
   15.0000    0.0047
   16.0000    0.0046
   17.0000    0.0046
   18.0000    0.0055
}{\QAMrcosspannine}
\addplot [PAM,PAM_L30km] table[x expr=\thisrowno{0}+\attshift] {\PAMrcosspannine};
\addplot [ASK,ASK_L30km] table[x expr=\thisrowno{0}+\attshift] {\ASKrcosspannine};
\addplot [QAM,QAM_L30km] table[x expr=\thisrowno{0}+\attshift] {\QAMrcosspannine};

\end{axis}
\end{tikzpicture}%
        \captionsetup{width=.90\linewidth,margin={1cm,0cm}}
        \caption{$4$-PAM/ASK/QAM, $L=0$ (empty) and $L\!=\!\SI{30}{\kilo\meter}$ (filled). \vspace*{0.3\baselineskip}   }
        \end{flushright}
        \end{subfigure}%
    \begin{subfigure}[t]{0.5\textwidth}
        \begin{flushright}
        \begin{tikzpicture}

\pgfmathsetmacro\rcosrolloffa{0}    
\pgfmathsetmacro\rcosrolloffb{0} 
\pgfmathsetmacro\attshift{6}    

\pgfdeclarelayer{fg}    %
\pgfdeclarelayer{bg}    %
\pgfsetlayers{bg,main,fg}  %

\begin{axis}[%
MIGeneralStyle,ymax=3+0.01,
xmax=18,
ylabel={bpcu}
]

\input{Figures/standalone/Q8/dataQ8_MI_TXPOW}

 \pgfplotstableread{
     X Y
-13.0000    0.0028
  -12.0000   -0.0003
  -11.0000    0.0055
  -10.0000    0.0085
   -9.0000    0.0125
   -8.0000    0.0189
   -7.0000    0.0258
   -6.0000    0.0494
   -5.0000    0.0783
   -4.0000    0.1154
   -3.0000    0.1743
   -2.0000    0.2581
   -1.0000    0.3805
         0    0.5147
    1.0000    0.6809
    2.0000    0.8389
    3.0000    1.0534
    4.0000    1.2683
    5.0000    1.4468
    6.0000    1.6373
    7.0000    1.8302
    8.0000    1.9982
    9.0000    2.1806
   10.0000    2.2823
   11.0000    2.4208
   12.0000    2.4909
   13.0000    2.5755
   14.0000    2.6121
   15.0000    2.6120
   16.0000    2.6408
   17.0000    2.6648
   18.0000    2.6921
 }{\PAMrcosspanseven}

\pgfplotstableread{
    X Y 
  -13.0000    0.0023
  -12.0000    0.0048
  -11.0000    0.0093
  -10.0000    0.0135
   -9.0000    0.0187
   -8.0000    0.0296
   -7.0000    0.0449
   -6.0000    0.0629
   -5.0000    0.1020
   -4.0000    0.1736
   -3.0000    0.2587
   -2.0000    0.3616
   -1.0000    0.5017
         0    0.6865
    1.0000    0.8753
    2.0000    1.0897
    3.0000    1.3322
    4.0000    1.5440
    5.0000    1.7212
    6.0000    1.9047
    7.0000    2.0746
    8.0000    2.1975
    9.0000    2.2985
   10.0000    2.3750
   11.0000    2.4311
   12.0000    2.5008
   13.0000    2.5342
   14.0000    2.5186
   15.0000    2.5176
   16.0000    2.5284
   17.0000    2.5311
   18.0000    2.5340
}{\ASKrcosspanseven}

\pgfplotstableread{
    X Y
  -13.0000    0.0022
  -12.0000    0.0035
  -11.0000    0.0056
  -10.0000    0.0088
   -9.0000    0.0139
   -8.0000    0.0219
   -7.0000    0.0344
   -6.0000    0.0536
   -5.0000    0.0828
   -4.0000    0.1264
   -3.0000    0.1897
   -2.0000    0.2787
   -1.0000    0.3991
         0    0.5553
    1.0000    0.7503
    2.0000    0.9853
    3.0000    1.2606
    4.0000    1.5767
    5.0000    1.9338
    6.0000    2.3323
    7.0000    2.7721
    8.0000    3.2517
    9.0000    3.7685
   10.0000    4.3184
   11.0000    4.8963
   12.0000    5.4970
   13.0000    6.1154
   14.0000    6.7470
   15.0000    7.3882
   16.0000    8.0364
   17.0000    8.6896
   18.0000    9.3463
}{\PAMCovUpperBound}

\pgfplotstableread{
    X Y
 -13.0000    0.0032
  -12.0000    0.0051
  -11.0000    0.0081
  -10.0000    0.0128
   -9.0000    0.0203
   -8.0000    0.0319
   -7.0000    0.0502
   -6.0000    0.0784
   -5.0000    0.1215
   -4.0000    0.1864
   -3.0000    0.2817
   -2.0000    0.4172
   -1.0000    0.6029
         0    0.8469
    1.0000    1.1538
    2.0000    1.5239
    3.0000    1.9535
    4.0000    2.4360
    5.0000    2.9634
    6.0000    3.5270
    7.0000    4.1188
    8.0000    4.7317
    9.0000    5.3600
   10.0000    5.9993
   11.0000    6.6464
   12.0000    7.2990
   13.0000    7.9554
   14.0000    8.6144
   15.0000    9.2752
   16.0000    9.9372
   17.0000   10.6001
   18.0000   11.2635
   }{\ASKCovUpperBound}
   
\pgfplotstableread{
    X Y
   -13.0000    0.0032
  -12.0000    0.0050
  -11.0000    0.0080
  -10.0000    0.0126
   -9.0000    0.0199
   -8.0000    0.0314
   -7.0000    0.0493
   -6.0000    0.0770
   -5.0000    0.1194
   -4.0000    0.1831
   -3.0000    0.2768
   -2.0000    0.4102
   -1.0000    0.5932
         0    0.8340
    1.0000    1.1375
    2.0000    1.5040
    3.0000    1.9299
    4.0000    2.4084
    5.0000    2.9314
    6.0000    3.4903
    7.0000    4.0772
    8.0000    4.6853
    9.0000    5.3092
   10.0000    5.9446
   11.0000    6.5884
   12.0000    7.2382
   13.0000    7.8924
   14.0000    8.5497
   15.0000    9.2093
   16.0000    9.8704
   17.0000   10.5327
   18.0000   11.1956
   }{\QAMCovUpperBound}

\pgfplotstableread{
    X Y
  -13.0000    0.0033
  -12.0000    0.0032
  -11.0000    0.0100
  -10.0000    0.0130
   -9.0000    0.0179
   -8.0000    0.0323
   -7.0000    0.0371
   -6.0000    0.0698
   -5.0000    0.1081
   -4.0000    0.1795
   -3.0000    0.2461
   -2.0000    0.3748
   -1.0000    0.5063
         0    0.6711
    1.0000    0.8790
    2.0000    1.0980
    3.0000    1.3553
    4.0000    1.5845
    5.0000    1.7736
    6.0000    1.9614
    7.0000    2.1424
    8.0000    2.2210
    9.0000    2.3216
   10.0000    2.3403
   11.0000    2.4296
   12.0000    2.4525
   13.0000    2.4773
   14.0000    2.4453
   15.0000    2.4648
   16.0000    2.5088
   17.0000    2.4970
   18.0000    2.5111
 }{\QAMrcosspanseven}

\pgfplotstableread{
    X Y
   -13.0000    0.0022
  -12.0000    0.0037
  -11.0000    0.0051
  -10.0000    0.0072
   -9.0000    0.0099
   -8.0000    0.0148
   -7.0000    0.0298
   -6.0000    0.0337
   -5.0000    0.0574
   -4.0000    0.0771
   -3.0000    0.1137
   -2.0000    0.1635
   -1.0000    0.2104
         0    0.2597
    1.0000    0.2872
    2.0000    0.3123
    3.0000    0.3482
    4.0000    0.3363
    5.0000    0.3318
    6.0000    0.2947
    7.0000    0.2580
    8.0000    0.2021
    9.0000    0.1497
   10.0000    0.1225
   11.0000    0.0741
   12.0000    0.0611
   13.0000    0.0517
   14.0000    0.0411
   15.0000    0.0266
   16.0000    0.0237
   17.0000    0.0118
   18.0000    0.0080
      }{\PAMIqYoXgYe}

\pgfplotstableread{
    X Y
  -13.0000    0.0032
  -12.0000    0.0050
  -11.0000    0.0072
  -10.0000    0.0106
   -9.0000    0.0179
   -8.0000    0.0233
   -7.0000    0.0381
   -6.0000    0.0577
   -5.0000    0.0875
   -4.0000    0.1364
   -3.0000    0.1901
   -2.0000    0.2644
   -1.0000    0.3348
         0    0.4095
    1.0000    0.4756
    2.0000    0.5327
    3.0000    0.5773
    4.0000    0.5754
    5.0000    0.6060
    6.0000    0.5765
    7.0000    0.5530
    8.0000    0.5575
    9.0000    0.5210
   10.0000    0.5185
   11.0000    0.4861
   12.0000    0.5097
   13.0000    0.5030
   14.0000    0.4675
   15.0000    0.5197
   16.0000    0.4757
   17.0000    0.4917
   18.0000    0.5025
    }{\ASKIqYoXgYe}

\pgfplotstableread{
    X Y
  -13.0000    0.0021
  -12.0000    0.0039
  -11.0000    0.0049
  -10.0000    0.0076
   -9.0000    0.0109
   -8.0000    0.0210
   -7.0000    0.0213
   -6.0000    0.0404
   -5.0000    0.0695
   -4.0000    0.0956
   -3.0000    0.1376
   -2.0000    0.2073
   -1.0000    0.2791
         0    0.3604
    1.0000    0.4327
    2.0000    0.5081
    3.0000    0.5690
    4.0000    0.5829
    5.0000    0.5860
    6.0000    0.6125
    7.0000    0.6311
    8.0000    0.6246
    9.0000    0.6559
   10.0000    0.6728
   11.0000    0.6974
   12.0000    0.6680
   13.0000    0.6779
   14.0000    0.6772
   15.0000    0.6894
   16.0000    0.6997
   17.0000    0.7016
   18.0000    0.6899
    }{\QAMIqYoXgYe}
\addplot [PAM,PAM_L30km] table[y expr=\thisrowno{1}*1/(1+\rcosrolloffa),x expr=\thisrowno{0}+\attshift] {\PAMrcosspanseven}; 
\addplot [ASK,ASK_L30km] table[y expr=\thisrowno{1}*1/(1+\rcosrolloffa),x expr=\thisrowno{0}+\attshift] {\ASKrcosspanseven}; 
\addplot [QAM,QAM_L30km] table[y expr=\thisrowno{1}*1/(1+\rcosrolloffa),x expr=\thisrowno{0}+\attshift] {\QAMrcosspanseven};

\addplot [PAM,UB,forget plot] table[y expr=\thisrowno{1}*1/(1+\rcosrolloffa),x expr=\thisrowno{0}+\attshift] {\PAMCovUpperBound};
\addplot [ASK,UB,forget plot] table[y expr=\thisrowno{1}*1/(1+\rcosrolloffa),x expr=\thisrowno{0}+\attshift] {\ASKCovUpperBound};
\addplot [QAM,UB,forget plot] table[y expr=\thisrowno{1}*1/(1+\rcosrolloffa),x expr=\thisrowno{0}+\attshift] {\QAMCovUpperBound};

\begin{pgfonlayer}{fg}  
    \draw[arr,shadowed] (axis cs:10.2,1.57) -- (axis cs:11.7,1.57) node[xshift=0.43cm,font=\footnotesize,right]{$\approx 1.5\;\text{dB}$};
    
    \draw[arr,shadowed] (axis cs:10.95,1.77) -- (axis cs:12.75,1.77) node[xshift=0.1cm,font=\footnotesize,right]{$\approx 1.8\;\text{dB}$};
\end{pgfonlayer}

\addlegendentry{$8$-PAM}
\addlegendentry{$8$-ASK}
\addlegendentry{$8$-SQAM}

\node at (axis cs:9.5,2.52)[mycolor6,rotate=62,font=\footnotesize]{ASK UB};
\node at (axis cs:8.5,2.47)[mycolor5,rotate=62,font=\footnotesize]{QAM UB};
\node at (axis cs:11.85,2.52)[mycolor1,rotate=60,font=\footnotesize]{PAM UB};

\end{axis}
\end{tikzpicture}%
        \captionsetup{width=.90\linewidth,margin={1cm,0cm}}
        \caption{$8$-PAM/ASK/SQAM, $L=\SI{30}{\kilo\meter}$. \vspace*{0.3\baselineskip} }
        \begin{tikzpicture}

\pgfmathsetmacro\rcosrolloffa{0}    
\pgfmathsetmacro\rcosrolloffb{0}    
\pgfmathsetmacro\attshift{6}    

\begin{axis}[%
MIGeneralStyle,ymax=3+0.01,
xmax=18,
ylabel={bpcu}
]

\input{Figures/standalone/Q8/dataQ8_MI_TXPOW}
\addplot [PAM,forget plot] table[y expr=\thisrowno{1}*1/(1+\rcosrolloffa)] {\PAMrcosspanseven}; 
\addplot [ASK,forget plot] table[y expr=\thisrowno{1}*1/(1+\rcosrolloffa)] {\ASKrcosspanseven};
\addplot [QAM,forget plot] table[y expr=\thisrowno{1}*1/(1+\rcosrolloffa)] {\QAMrcosspanseven};

 \pgfplotstableread{
     X Y
-13.0000    0.0028
  -12.0000   -0.0003
  -11.0000    0.0055
  -10.0000    0.0085
   -9.0000    0.0125
   -8.0000    0.0189
   -7.0000    0.0258
   -6.0000    0.0494
   -5.0000    0.0783
   -4.0000    0.1154
   -3.0000    0.1743
   -2.0000    0.2581
   -1.0000    0.3805
         0    0.5147
    1.0000    0.6809
    2.0000    0.8389
    3.0000    1.0534
    4.0000    1.2683
    5.0000    1.4468
    6.0000    1.6373
    7.0000    1.8302
    8.0000    1.9982
    9.0000    2.1806
   10.0000    2.2823
   11.0000    2.4208
   12.0000    2.4909
   13.0000    2.5755
   14.0000    2.6121
   15.0000    2.6120
   16.0000    2.6408
   17.0000    2.6648
   18.0000    2.6921
 }{\PAMrcosspanseven}

\pgfplotstableread{
    X Y 
  -13.0000    0.0023
  -12.0000    0.0048
  -11.0000    0.0093
  -10.0000    0.0135
   -9.0000    0.0187
   -8.0000    0.0296
   -7.0000    0.0449
   -6.0000    0.0629
   -5.0000    0.1020
   -4.0000    0.1736
   -3.0000    0.2587
   -2.0000    0.3616
   -1.0000    0.5017
         0    0.6865
    1.0000    0.8753
    2.0000    1.0897
    3.0000    1.3322
    4.0000    1.5440
    5.0000    1.7212
    6.0000    1.9047
    7.0000    2.0746
    8.0000    2.1975
    9.0000    2.2985
   10.0000    2.3750
   11.0000    2.4311
   12.0000    2.5008
   13.0000    2.5342
   14.0000    2.5186
   15.0000    2.5176
   16.0000    2.5284
   17.0000    2.5311
   18.0000    2.5340
}{\ASKrcosspanseven}

\pgfplotstableread{
    X Y
  -13.0000    0.0022
  -12.0000    0.0035
  -11.0000    0.0056
  -10.0000    0.0088
   -9.0000    0.0139
   -8.0000    0.0219
   -7.0000    0.0344
   -6.0000    0.0536
   -5.0000    0.0828
   -4.0000    0.1264
   -3.0000    0.1897
   -2.0000    0.2787
   -1.0000    0.3991
         0    0.5553
    1.0000    0.7503
    2.0000    0.9853
    3.0000    1.2606
    4.0000    1.5767
    5.0000    1.9338
    6.0000    2.3323
    7.0000    2.7721
    8.0000    3.2517
    9.0000    3.7685
   10.0000    4.3184
   11.0000    4.8963
   12.0000    5.4970
   13.0000    6.1154
   14.0000    6.7470
   15.0000    7.3882
   16.0000    8.0364
   17.0000    8.6896
   18.0000    9.3463
}{\PAMCovUpperBound}

\pgfplotstableread{
    X Y
 -13.0000    0.0032
  -12.0000    0.0051
  -11.0000    0.0081
  -10.0000    0.0128
   -9.0000    0.0203
   -8.0000    0.0319
   -7.0000    0.0502
   -6.0000    0.0784
   -5.0000    0.1215
   -4.0000    0.1864
   -3.0000    0.2817
   -2.0000    0.4172
   -1.0000    0.6029
         0    0.8469
    1.0000    1.1538
    2.0000    1.5239
    3.0000    1.9535
    4.0000    2.4360
    5.0000    2.9634
    6.0000    3.5270
    7.0000    4.1188
    8.0000    4.7317
    9.0000    5.3600
   10.0000    5.9993
   11.0000    6.6464
   12.0000    7.2990
   13.0000    7.9554
   14.0000    8.6144
   15.0000    9.2752
   16.0000    9.9372
   17.0000   10.6001
   18.0000   11.2635
   }{\ASKCovUpperBound}
   
\pgfplotstableread{
    X Y
   -13.0000    0.0032
  -12.0000    0.0050
  -11.0000    0.0080
  -10.0000    0.0126
   -9.0000    0.0199
   -8.0000    0.0314
   -7.0000    0.0493
   -6.0000    0.0770
   -5.0000    0.1194
   -4.0000    0.1831
   -3.0000    0.2768
   -2.0000    0.4102
   -1.0000    0.5932
         0    0.8340
    1.0000    1.1375
    2.0000    1.5040
    3.0000    1.9299
    4.0000    2.4084
    5.0000    2.9314
    6.0000    3.4903
    7.0000    4.0772
    8.0000    4.6853
    9.0000    5.3092
   10.0000    5.9446
   11.0000    6.5884
   12.0000    7.2382
   13.0000    7.8924
   14.0000    8.5497
   15.0000    9.2093
   16.0000    9.8704
   17.0000   10.5327
   18.0000   11.1956
   }{\QAMCovUpperBound}

\pgfplotstableread{
    X Y
  -13.0000    0.0033
  -12.0000    0.0032
  -11.0000    0.0100
  -10.0000    0.0130
   -9.0000    0.0179
   -8.0000    0.0323
   -7.0000    0.0371
   -6.0000    0.0698
   -5.0000    0.1081
   -4.0000    0.1795
   -3.0000    0.2461
   -2.0000    0.3748
   -1.0000    0.5063
         0    0.6711
    1.0000    0.8790
    2.0000    1.0980
    3.0000    1.3553
    4.0000    1.5845
    5.0000    1.7736
    6.0000    1.9614
    7.0000    2.1424
    8.0000    2.2210
    9.0000    2.3216
   10.0000    2.3403
   11.0000    2.4296
   12.0000    2.4525
   13.0000    2.4773
   14.0000    2.4453
   15.0000    2.4648
   16.0000    2.5088
   17.0000    2.4970
   18.0000    2.5111
 }{\QAMrcosspanseven}

\pgfplotstableread{
    X Y
   -13.0000    0.0022
  -12.0000    0.0037
  -11.0000    0.0051
  -10.0000    0.0072
   -9.0000    0.0099
   -8.0000    0.0148
   -7.0000    0.0298
   -6.0000    0.0337
   -5.0000    0.0574
   -4.0000    0.0771
   -3.0000    0.1137
   -2.0000    0.1635
   -1.0000    0.2104
         0    0.2597
    1.0000    0.2872
    2.0000    0.3123
    3.0000    0.3482
    4.0000    0.3363
    5.0000    0.3318
    6.0000    0.2947
    7.0000    0.2580
    8.0000    0.2021
    9.0000    0.1497
   10.0000    0.1225
   11.0000    0.0741
   12.0000    0.0611
   13.0000    0.0517
   14.0000    0.0411
   15.0000    0.0266
   16.0000    0.0237
   17.0000    0.0118
   18.0000    0.0080
      }{\PAMIqYoXgYe}

\pgfplotstableread{
    X Y
  -13.0000    0.0032
  -12.0000    0.0050
  -11.0000    0.0072
  -10.0000    0.0106
   -9.0000    0.0179
   -8.0000    0.0233
   -7.0000    0.0381
   -6.0000    0.0577
   -5.0000    0.0875
   -4.0000    0.1364
   -3.0000    0.1901
   -2.0000    0.2644
   -1.0000    0.3348
         0    0.4095
    1.0000    0.4756
    2.0000    0.5327
    3.0000    0.5773
    4.0000    0.5754
    5.0000    0.6060
    6.0000    0.5765
    7.0000    0.5530
    8.0000    0.5575
    9.0000    0.5210
   10.0000    0.5185
   11.0000    0.4861
   12.0000    0.5097
   13.0000    0.5030
   14.0000    0.4675
   15.0000    0.5197
   16.0000    0.4757
   17.0000    0.4917
   18.0000    0.5025
    }{\ASKIqYoXgYe}

\pgfplotstableread{
    X Y
  -13.0000    0.0021
  -12.0000    0.0039
  -11.0000    0.0049
  -10.0000    0.0076
   -9.0000    0.0109
   -8.0000    0.0210
   -7.0000    0.0213
   -6.0000    0.0404
   -5.0000    0.0695
   -4.0000    0.0956
   -3.0000    0.1376
   -2.0000    0.2073
   -1.0000    0.2791
         0    0.3604
    1.0000    0.4327
    2.0000    0.5081
    3.0000    0.5690
    4.0000    0.5829
    5.0000    0.5860
    6.0000    0.6125
    7.0000    0.6311
    8.0000    0.6246
    9.0000    0.6559
   10.0000    0.6728
   11.0000    0.6974
   12.0000    0.6680
   13.0000    0.6779
   14.0000    0.6772
   15.0000    0.6894
   16.0000    0.6997
   17.0000    0.7016
   18.0000    0.6899
    }{\QAMIqYoXgYe}
\addplot [PAM,PAM_L30km] table[y expr=\thisrowno{1}*1/(1+\rcosrolloffa),x expr=\thisrowno{0}+\attshift] {\PAMrcosspanseven}; 
\addplot [ASK,ASK_L30km] table[y expr=\thisrowno{1}*1/(1+\rcosrolloffa),x expr=\thisrowno{0}+\attshift] {\ASKrcosspanseven}; 
\addplot [QAM,QAM_L30km] table[y expr=\thisrowno{1}*1/(1+\rcosrolloffa),x expr=\thisrowno{0}+\attshift] {\QAMrcosspanseven};

\end{axis}
\end{tikzpicture}%
        \captionsetup{width=.90\linewidth,margin={1cm,0cm}}
        \caption{$8$-PAM/ASK/SQAM, $L=0$ (empty) and $L\!=\!\SI{30}{\kilo\meter}$ (filled). \vspace*{-0.6\baselineskip}   }
        \begin{tikzpicture}
\pgfmathsetmacro\attshift{6}    

\begin{axis}[%
SERGeneralStyle,
ymin=1E-3,
ymax=1,
xmax=18,
ymode=log
]
\pgfplotstableread{
    X Y
  -13.0000    0.8657
  -12.0000    0.8618
  -11.0000    0.8604
  -10.0000    0.8556
   -9.0000    0.8505
   -8.0000    0.8486
   -7.0000    0.8343
   -6.0000    0.8304
   -5.0000    0.8137
   -4.0000    0.7996
   -3.0000    0.7779
   -2.0000    0.7477
   -1.0000    0.7203
         0    0.6795
    1.0000    0.6378
    2.0000    0.5853
    3.0000    0.5282
    4.0000    0.4630
    5.0000    0.3951
    6.0000    0.3278
    7.0000    0.2656
    8.0000    0.2106
    9.0000    0.1682
   10.0000    0.1332
   11.0000    0.1077
   12.0000    0.0850
   13.0000    0.0665
   14.0000    0.0534
   15.0000    0.0417
   16.0000    0.0297
   17.0000    0.0177
   18.0000    0.0089
}{\PAMrcosspanseven}

\pgfplotstableread{
    X Y
   -13.0000    0.8686
  -12.0000    0.8650
  -11.0000    0.8649
  -10.0000    0.8578
   -9.0000    0.8597
   -8.0000    0.8525
   -7.0000    0.8481
   -6.0000    0.8381
   -5.0000    0.8331
   -4.0000    0.8143
   -3.0000    0.8013
   -2.0000    0.7765
   -1.0000    0.7468
         0    0.7051
    1.0000    0.6556
    2.0000    0.6040
    3.0000    0.5262
    4.0000    0.4444
    5.0000    0.3646
    6.0000    0.2940
    7.0000    0.2461
    8.0000    0.1998
    9.0000    0.1714
   10.0000    0.1545
   11.0000    0.1439
   12.0000    0.1320
   13.0000    0.1375
   14.0000    0.1364
   15.0000    0.1325
   16.0000    0.1374
   17.0000    0.1338
   18.0000    0.1326
}{\ASKrcosspanseven}

\pgfplotstableread{
    X Y
  -13.0000    0.8669
  -12.0000    0.8691
  -11.0000    0.8727
  -10.0000    0.8636
   -9.0000    0.8637
   -8.0000    0.8598
   -7.0000    0.8582
   -6.0000    0.8559
   -5.0000    0.8431
   -4.0000    0.8324
   -3.0000    0.8252
   -2.0000    0.8098
   -1.0000    0.7977
         0    0.7726
    1.0000    0.7362
    2.0000    0.6897
    3.0000    0.6469
    4.0000    0.5902
    5.0000    0.5676
    6.0000    0.5781
    7.0000    0.5728
    8.0000    0.5617
    9.0000    0.5575
   10.0000    0.5457
   11.0000    0.5523
   12.0000    0.5467
   13.0000    0.5502
   14.0000    0.5465
   15.0000    0.5482
   16.0000    0.5501
   17.0000    0.5502
   18.0000    0.5455
}{\QAMrcosspanseven}
\addplot [PAM,solid] table {\PAMrcosspanseven};
\addplot [ASK,solid] table {\ASKrcosspanseven};
\addplot [QAM,solid] table {\QAMrcosspanseven};

\pgfplotstableread{
    X Y
  -13.0000    0.8667
  -12.0000    0.8698
  -11.0000    0.8638
  -10.0000    0.8525
   -9.0000    0.8537
   -8.0000    0.8478
   -7.0000    0.8508
   -6.0000    0.8370
   -5.0000    0.8280
   -4.0000    0.8128
   -3.0000    0.8062
   -2.0000    0.7753
   -1.0000    0.7595
         0    0.7333
    1.0000    0.7067
    2.0000    0.6852
    3.0000    0.6475
    4.0000    0.5878
    5.0000    0.5625
    6.0000    0.5135
    7.0000    0.4678
    8.0000    0.4140
    9.0000    0.3688
   10.0000    0.3342
   11.0000    0.2852
   12.0000    0.2655
   13.0000    0.2258
   14.0000    0.2173
   15.0000    0.2043
   16.0000    0.1898
   17.0000    0.1788
   18.0000    0.1627
}{\PAMrcosspanseven}

\pgfplotstableread{
    X Y
   -13.0000    0.8719
  -12.0000    0.8680
  -11.0000    0.8674
  -10.0000    0.8685
   -9.0000    0.8656
   -8.0000    0.8621
   -7.0000    0.8556
   -6.0000    0.8605
   -5.0000    0.8484
   -4.0000    0.8364
   -3.0000    0.8340
   -2.0000    0.8181
   -1.0000    0.8108
         0    0.7827
    1.0000    0.7607
    2.0000    0.7323
    3.0000    0.6851
    4.0000    0.6406
    5.0000    0.5976
    6.0000    0.5549
    7.0000    0.4948
    8.0000    0.4527
    9.0000    0.3957
   10.0000    0.3822
   11.0000    0.3534
   12.0000    0.3122
   13.0000    0.2930
   14.0000    0.2780
   15.0000    0.2872
   16.0000    0.2882
   17.0000    0.2876
   18.0000    0.2677
}{\ASKrcosspanseven}

\pgfplotstableread{
    X Y
  -13.0000    0.8612
  -12.0000    0.8713
  -11.0000    0.8627
  -10.0000    0.8612
   -9.0000    0.8648
   -8.0000    0.8536
   -7.0000    0.8481
   -6.0000    0.8474
   -5.0000    0.8406
   -4.0000    0.8220
   -3.0000    0.8162
   -2.0000    0.8018
   -1.0000    0.7861
         0    0.7755
    1.0000    0.7511
    2.0000    0.7308
    3.0000    0.6953
    4.0000    0.6609
    5.0000    0.6257
    6.0000    0.5831
    7.0000    0.5143
    8.0000    0.4724
    9.0000    0.4361
   10.0000    0.4323
   11.0000    0.3708
   12.0000    0.3538
   13.0000    0.3564
   14.0000    0.3483
   15.0000    0.3440
   16.0000    0.3265
   17.0000    0.3406
   18.0000    0.3285
}{\QAMrcosspanseven}
\addplot [PAM,PAM_L30km,solid] table[x expr=\thisrowno{0}+\attshift] {\PAMrcosspanseven};
\addplot [ASK,ASK_L30km,solid] table[x expr=\thisrowno{0}+\attshift]  {\ASKrcosspanseven};
\addplot [QAM,QAM_L30km,solid] table[x expr=\thisrowno{0}+\attshift] {\QAMrcosspanseven};

\end{axis}
\end{tikzpicture}%
        \captionsetup{width=.90\linewidth,margin={1cm,0cm}}
        \caption{$8$-PAM/ASK/SQAM, $L=0$ (empty) and $L\!=\!\SI{30}{\kilo\meter}$ (filled). \vspace*{0.3\baselineskip}   }
        \end{flushright}
        \end{subfigure}  
        \caption{Achievable rates, upper bounds and HD SERs for $L=0$ and $L=\SI{30}{\kilo\meter}$. The auxiliary channel memory is $\smash{\widetilde{N}} = 9$ and $\smash{\widetilde{N}} = 7$ for $Q=4$ and $Q=8$, respectively.}
        \label{fig:se_hdser_Q4_Q8_L0km_and_L30km}
\end{figure*}

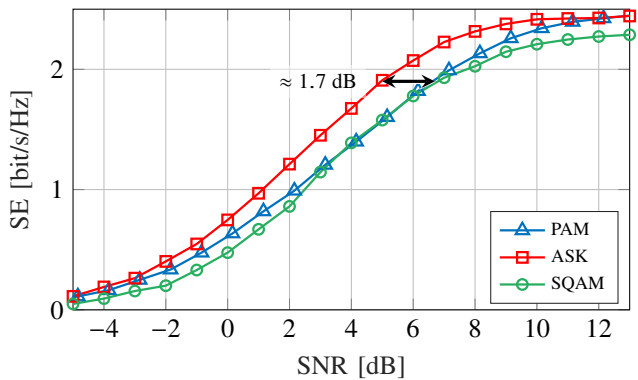
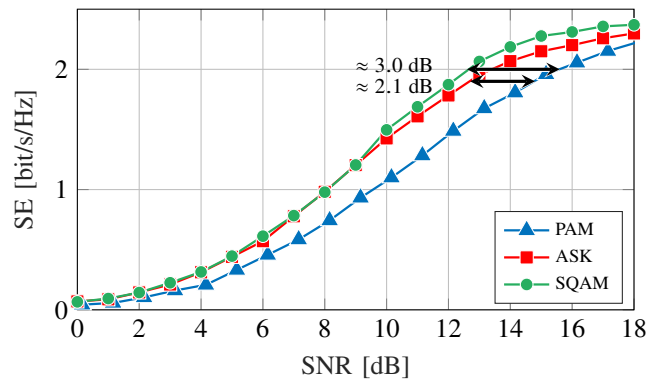
\begin{figure*}[!t] %
\begin{subfigure}[t]{0.49\textwidth}
    \centering
    \begin{tikzpicture}

\pgfdeclarelayer{fg}    %
\pgfdeclarelayer{bg}    %
\pgfsetlayers{bg,main,fg}  %

\pgfmathsetmacro\rcosrolloffb{0.2}    

\begin{axis}[%
MIGeneralStyle,
ymax=2.5,
xmin=-5,
ylabel={SE [bit/s/Hz]},
width=\gridfigurewidth,
height=\gridfigureheight,
legend style={at={(0.98,0.02)},anchor=south east}
]

 \pgfplotstableread{
     X Y
  -12.8432    0.0031
  -11.8432    0.0028
  -10.8432    0.0092
   -9.8432    0.0088
   -8.8432    0.0258
   -7.8432    0.0347
   -6.8432    0.0563
   -5.8432    0.0713
   -4.8432    0.1339
   -3.8432    0.1961
   -2.8432    0.2992
   -1.8432    0.4021
   -0.8432    0.5708
    0.1568    0.7625
    1.1568    0.9826
    2.1568    1.1890
    3.1568    1.4472
    4.1568    1.6783
    5.1568    1.9228
    6.1568    2.1834
    7.1568    2.3894
    8.1568    2.5596
    9.1568    2.7100
   10.1568    2.8102
   11.1568    2.8738
   12.1568    2.9090
   13.1568    2.9413
   14.1568    2.9523
   15.1568    2.9539
   16.1568    2.9723
   17.1568    2.9733
   18.1568    2.9668
 }{\PAMrcosspanseven}

 \pgfplotstableread{
     X Y
       -13.0000    0.0037
  -12.0000    0.0058
  -11.0000    0.0105
  -10.0000    0.0149
   -9.0000    0.0261
   -8.0000    0.0407
   -7.0000    0.0604
   -6.0000    0.0925
   -5.0000    0.1344
   -4.0000    0.2283
   -3.0000    0.3161
   -2.0000    0.4830
   -1.0000    0.6573
         0    0.8973
    1.0000    1.1632
    2.0000    1.4541
    3.0000    1.7418
    4.0000    2.0098
    5.0000    2.2908
    6.0000    2.4875
    7.0000    2.6724
    8.0000    2.7784
    9.0000    2.8531
   10.0000    2.8991
   11.0000    2.9070
   12.0000    2.9121
   13.0000    2.9323
   14.0000    2.9354
   15.0000    2.9457
   16.0000    2.9321
   17.0000    2.9270
   18.0000    2.9256 
   }{\ASKrcosspanseven}

 \pgfplotstableread{
     X Y
-13.0000    0.0019
  -12.0000    0.0011
  -11.0000    0.0053
  -10.0000    0.0087
   -9.0000    0.0116
   -8.0000    0.0172
   -7.0000    0.0331
   -6.0000    0.0480
   -5.0000    0.0587
   -4.0000    0.1123
   -3.0000    0.1877
   -2.0000    0.2410
   -1.0000    0.3972
         0    0.5708
    1.0000    0.8034
    2.0000    1.0320
    3.0000    1.3744
    4.0000    1.6663
    5.0000    1.8934
    6.0000    2.1340
    7.0000    2.3162
    8.0000    2.4305
    9.0000    2.5755
   10.0000    2.6507
   11.0000    2.6973
   12.0000    2.7274
   13.0000    2.7449
   14.0000    2.7811
   15.0000    2.7687
   16.0000    2.7356
   17.0000    2.7792
   18.0000    2.7780
}{\QAMrcosspanseven}
 
\addplot [PAM] table[y expr=\thisrowno{1}*1/(1+\rcosrolloffb)] {\PAMrcosspanseven}; 
\addplot [ASK] table[y expr=\thisrowno{1}*1/(1+\rcosrolloffb)] {\ASKrcosspanseven}; 
\addplot [QAM] table[y expr=\thisrowno{1}*1/(1+\rcosrolloffb)] {\QAMrcosspanseven};

\begin{pgfonlayer}{fg}  
    \draw[arr,shadowed] (axis cs:5.0,1.9)node[font=\footnotesize,left,xshift=-0.2cm,yshift=0cm,fill=white,opacity=0.5,text opacity=1]{$\approx 1.7 \;\text{dB}$} --  (axis cs:6.7,1.9) ;
\end{pgfonlayer}

\addlegendentry{PAM}
\addlegendentry{ASK}
\addlegendentry{SQAM}

\end{axis}
\end{tikzpicture}%
    \captionsetup{width=.90\linewidth,margin={0.0cm,0cm}}
    \caption{SE for $L=0$.}
    \label{fig:se_Q8_rc_pulse_rolloff_0.2_L0km}
\end{subfigure}%
\hspace*{\fill}
\begin{subfigure}[t]{0.49\textwidth}
    \centering
    \begin{tikzpicture}

\pgfdeclarelayer{fg}    %
\pgfdeclarelayer{bg}    %
\pgfsetlayers{bg,main,fg}  %

\pgfmathsetmacro\rcosrolloffb{0.2}    
\pgfmathsetmacro\attshift{6}    

\begin{axis}[%
MIGeneralStyle,
ymax=2.5,
xmin=0,
xmax=18,
ylabel={SE [bit/s/Hz]},
width=\gridfigurewidth,
height=\gridfigureheight,
legend style={at={(0.98,0.02)},anchor=south east}
]

\pgfplotstableread{
     X Y
  -12.8432    0.0029
  -11.8432    0.0026
  -10.8432    0.0072
   -9.8432    0.0086
   -8.8432    0.0105
   -7.8432    0.0273
   -6.8432    0.0315
   -5.8432    0.0515
   -4.8432    0.0691
   -3.8432    0.1244
   -2.8432    0.1929
   -1.8432    0.2496
   -0.8432    0.3986
    0.1568    0.5487
    1.1568    0.7019
    2.1568    0.8920
    3.1568    1.1186
    4.1568    1.3199
    5.1568    1.5401
    6.1568    1.7849
    7.1568    2.0098
    8.1568    2.1678
    9.1568    2.3507
   10.1568    2.4664
   11.1568    2.5829
   12.1568    2.6746
   13.1568    2.7403
   14.1568    2.7685
   15.1568    2.8020
   16.1568    2.8047
   17.1568    2.7854
   18.1568    2.8234
}{\PAMrcosspanseven}

 \pgfplotstableread{
 X Y
   -13.0000    0.0029
  -12.0000    0.0062
  -11.0000    0.0055
  -10.0000    0.0136
   -9.0000    0.0159
   -8.0000    0.0289
   -7.0000    0.0457
   -6.0000    0.0839
   -5.0000    0.1050
   -4.0000    0.1737
   -3.0000    0.2528
   -2.0000    0.3746
   -1.0000    0.5289
         0    0.6851
    1.0000    0.9336
    2.0000    1.1779
    3.0000    1.4439
    4.0000    1.7109
    5.0000    1.9308
    6.0000    2.1394
    7.0000    2.3455
    8.0000    2.4817
    9.0000    2.5801
   10.0000    2.6412
   11.0000    2.7101
   12.0000    2.7583
   13.0000    2.7586
   14.0000    2.7568
   15.0000    2.7540
   16.0000    2.7777
   17.0000    2.7924
   18.0000    2.7602
 }{\ASKrcosspanseven}

 \pgfplotstableread{
     X Y
  -13.0000    0.0037
  -12.0000    0.0048
  -11.0000    0.0038
  -10.0000    0.0107
   -9.0000    0.0173
   -8.0000    0.0352
   -7.0000    0.0404
   -6.0000    0.0795
   -5.0000    0.1127
   -4.0000    0.1725
   -3.0000    0.2702
   -2.0000    0.3782
   -1.0000    0.5361
         0    0.7352
    1.0000    0.9404
    2.0000    1.1745
    3.0000    1.4456
    4.0000    1.7961
    5.0000    2.0264
    6.0000    2.2476
    7.0000    2.4785
    8.0000    2.6236
    9.0000    2.7324
   10.0000    2.7730
   11.0000    2.8285
   12.0000    2.8445
   13.0000    2.8601
   14.0000    2.8850
   15.0000    2.8516
   16.0000    2.8691
   17.0000    2.8674
   18.0000    2.9031
    }{\QAMrcosspanseven}
 
\addplot [PAM,PAM_L30km] table[y expr=\thisrowno{1}*1/(1+\rcosrolloffb),x expr=\thisrowno{0}+\attshift] {\PAMrcosspanseven}; 
\addplot [ASK,ASK_L30km] table[y expr=\thisrowno{1}*1/(1+\rcosrolloffb),x expr=\thisrowno{0}+\attshift] {\ASKrcosspanseven}; 
\addplot [QAM,QAM_L30km] table[y expr=\thisrowno{1}*1/(1+\rcosrolloffb),x expr=\thisrowno{0}+\attshift] {\QAMrcosspanseven};

\begin{pgfonlayer}{fg}  
    \draw[arr,shadowed] (axis cs:12.6,2)node[font=\footnotesize,left,xshift=-0.31cm,yshift=0.06cm]{$\approx 3.0\;\text{dB}$} -- (axis cs:15.6,2) ;
    \draw[arr,shadowed] (axis cs:12.7,1.9)node[font=\footnotesize,left,xshift=-0.35cm,yshift=-0.06cm]{$\approx 2.1\;\text{dB}$} -- (axis cs:14.8,1.9) ;
\end{pgfonlayer}

\addlegendentry{PAM}
\addlegendentry{ASK}
\addlegendentry{SQAM}

\end{axis}
\end{tikzpicture}%
    \captionsetup{width=.90\linewidth,margin={0.2cm,0cm}}
    \caption{SE for $L=\SI{30}{\kilo\meter}$.}
    \label{fig:se_Q8_rc_pulse_rolloff_0.2_L30km}
\end{subfigure}
\caption{SEs for $Q=8$ and FD-RC pulses with roll-off $\alpha = 0.2$. The auxiliary channel memory is $\smash{\widetilde{N}} = 7$. }
\label{fig:se_Q8_rc_pulse_rolloff_0.2_L0km_and_L30km}

\end{figure*}
The next plots show that CD inherently resolves phase ambiguities. The reason is that CD spreads symbols over multiple receive samples, and consequently the $\diamondsuit$-samples have both intensity and phase information when using bipolar and complex alphabets. CD can thus be interpreted as a precoding for the receiver, i.e., even for short link lengths near $L=0$ one can avoid phase ambiguities by appropriate precoding.

Fig.~\ref{fig:se_hdser_Q4_Q8_L0km_and_L30km} shows rate bounds for a sinc pulse and a link with $L=\SI{30}{\kilo\meter}$ that has \SI{6}{dB} of attenuation. The first and second columns are for $Q=4$ and $Q=8$, respectively. The first column shows that 4-ASK and 4-QAM perform equally well, achieve the maximal $\log_2 Q=2$ bits at high SNR and gain approximately \SI{1.4}{dB} over 4-PAM. The SER plot (c) shows that differential coding gives a waterfall behaviour for ASK and QAM (here QPSK).
The plots in the second column show similar behaviour for $Q=8$, where 8-ASK (a bipolar constellation) and $8$-SQAM (a complex-alphabet constellation) gain approximately \SI{1.5}{dB} and \SI{1.8}{dB} over PAM, respectively. The achievable rates for both modulation formats saturate due to the auxiliary model mismatch. The resulting uncoded HD SERs decay slowly and eventually saturate.

To reduce the model mismatch, we replace the sinc pulse by an FD-RC pulse with roll-off factor $\alpha=0.2$. The maximum SE is thus reduced by a factor of 5/6. The remaining system is described in Fig.~\ref{fig:continuous_detailed_system_model}. The SLD again doubles the bandwidth and the excess bandwidth is removed by the antialiasing filter $g_\text{rx}(t)$ in \eqref{eq:z_time-discrete_sld_output_sampled_convolution}. 
Fig.~\ref{fig:se_Q8_rc_pulse_rolloff_0.2_L0km_and_L30km} plots the SEs for $Q=8$. Observe that at high SNR the SEs for all constellations are closer to the maximum SE (here \SI{2.5}{bit/s/Hz}) as compared to Fig.~\ref{fig:2x3_plot_L=0km}.
For SEs near \SI{2}{bit/s/Hz} and $L=\SI{0}{}$, 8-ASK  gains approximately \SI{1.7}{dB} over 8-PAM, whereas 8-SQAM and 8-ASK perform equally well. 
For SEs near \SI{2}{bit/s/Hz} and $L=\SI{30}{\kilo\meter}$, 8-SQAM and 8-ASK gain approximately \SI{3}{dB} and \SI{2.1}{dB} over 8-PAM, respectively.

\subsection{Comparison of sinc and TD-RC Pulses for \texorpdfstring{$L=0$}{L=0}}
\label{subsec:fdrc-tdrc}
TD-RC pulses are shorter than FD-RC pulses and this reduces the detection complexity when using a receiver filter with finite time duration\footnote{The paper~\cite{tasbihi2021direct} uses an integrate-and-dump filter. An optimal detector instead projects $Y'(t)$ in Fig.~\ref{fig:continuous_detailed_system_model} onto the space spanned by all possible signals $Z'(t)$. For example, for FD-RC pulses one may project onto the space spanned by the sinc pulses $B_\alpha\sinc(B_\alpha t-\kappa)$ where $B_\alpha=2B(1+\alpha)$ and $\kappa\in\mathbb{Z}$, see~\eqref{eq:rx_ps_filter}.}. On the other hand, TD-RC leaks energy outside the FD-RC frequency band and this causes interference to neighboring bands and also distortions with devices that have limited bandwidth.

\begin{figure}[!ht]
    \centering
    \begin{subfigure}[t]{\columnwidth}
    \usetikzlibrary{decorations.markings}
\tikzset{node distance=2.3cm}

\pgfdeclarelayer{background}
\pgfdeclarelayer{foreground}
\pgfsetlayers{background,main,foreground}
\tikzset{boxlines/.style = {draw=black!20!white,}}
\tikzset{boxlinesred/.style = {densely dashed,draw=red!50!white,thick}}

\pgfmathsetmacro{\samplerwidth}{30}

\tikzset{midnodes/.style = {midway,above,text width=1.5cm,align=center,yshift=-0.2em}}
\tikzset{midnodesRP/.style = {midway,above,text width=1.5cm,align=center,yshift=-1.4em}}
\begin{tikzpicture}[]
    
    \node [input, name=input] {Input};
    \node [comblock,right of=input,node distance=2.1cm] (txfilter) {$g_\text{tx}(t)$};
    \node [comblock,above of=txfilter,node distance=1.5cm] (txfilter_left) {$g_\text{tx}(t)$};
    \node [comblock,below of=txfilter,node distance=1.5cm] (txfilter_right) {$g_\text{tx}(t)$};

    \node [sum,right of=txfilter,node distance=1.5cm] (sumnode) {$+$};
    \node [comblock,right of=sumnode,node distance=1.4cm] (channel_select) {$g_\text{c}(t)$};
    \node [comblock,right of=channel_select,node distance=1.4cm] (sld) {$\left\lvert \,\cdot\, \right\rvert^2$};
    \node [comblock,right of=sld,node distance=1.4cm] (rxfilter) {$g_\text{rx}(t)$};
    \node [node distance=0.9cm,right of=rxfilter] (ldots) {$\ldots$};
   
   \node[left of=txfilter,node distance=1.4cm,font=\footnotesize,text width=1cm,align=left]{Center\\ Channel};
   \node[left of=txfilter_left,node distance=1.3cm,font=\footnotesize]{Channel 1};
   \node[left of=txfilter_right,node distance=1.3cm,font=\footnotesize]{Channel 3};
   
   \draw[-latex,thick](txfilter) --  (sumnode);
   \draw[-latex,thick](txfilter_left) -- node[sum,inner sep=-0.7pt,fill=white](mult1){$+$} (sumnode);
   \draw[-latex,thick](txfilter_right) -- node[sum,inner sep=-0.7pt,fill=white](mult2){$+$} (sumnode);
   
   \node[above,yshift=0.25cm,xshift=0.5cm,font=\footnotesize](mult_desc1) at (mult1){$e^{-\mathrm{j} 2 \pi B_\text{c} t }$};
   \node[below,yshift=-0.25cm,xshift=0.5cm,font=\footnotesize](mult_desc2) at (mult2){$e^{+\mathrm{j} 2 \pi B_\text{c} t }$};

   \node[below of=channel_select,node distance=0.8cm,font=\footnotesize,text width=1cm,align=center]{Channel\\ selection};
   
   \draw[-latex,thick] (sumnode) -- (channel_select);
   \draw[-latex,thick] (channel_select) -- (sld);
   \draw[-latex,thick] (sld) -- (rxfilter); 
   
    \draw[-,thick] (mult_desc1) -- (mult1); 
    \draw[-,thick] (mult_desc2) -- (mult2); 
    
   \node[above of=rxfilter,xshift=0.15cm,yshift=0.1cm,node distance=1.3cm]{
   \begin{tikzpicture}
        \begin{axis}[
          axis lines=middle,
          width=3cm,
          ytick={},
          xtick={-2,2},
          ymax=1.2,
          ymajorticks=false,
          xlabel={$f$},
          xmin = -3.2,
          xmax = 3.2,
          ylabel={$G_\text{rx}(f)$},
          xticklabels={\footnotesize{$-B$},\footnotesize{$B$}},
          every axis x label/.style={at={(current axis.right of origin)},anchor=west},
          every axis y label/.style={
            at={(ticklabel* cs:1.05)},
            anchor=south,
        },]
        ]
        \addplot[mycolor1,line width=1.5pt,mark=none] table[]
         {
          -3 0
          -2 0
          -1.99 1
          1.99 1
          2 0
          3 0
         };
         \node[right,font=\footnotesize] at(axis cs:2,1){$1$}; 
        \end{axis}
        \end{tikzpicture}%
   };
   
   \node[above of=channel_select,yshift=0.30cm,xshift=0.25cm,node distance=1.2cm]{
   \begin{tikzpicture}
        \begin{axis}[
          clip=false,
          axis lines=middle,
          width=3cm,
          ytick={},
          xlabel={$f$},
          xtick={-1.15,1.15},
          xmin = -3.2,
          xmax = 3.2,
          ymax=1.2,
          ymajorticks=false,
          xticklabels={},
          ylabel={$G_\text{c}(f)$},
          every axis x label/.style={at={(current axis.right of origin)},anchor=west},
          every axis y label/.style={
            at={(ticklabel* cs:1.05)},
            anchor=south,
        },]
        \addplot[mycolor6,line width=1.5pt,mark=none] table[]
         {
          -3 0
          -1.14 0
          -1.15 1
          1.15 1
          1.14 0
          3 0
         };
         \draw[latex-latex] (axis cs: -1.15,0.5) --  (axis cs: +1.15,0.5) node[right,font=\footnotesize]{$B_\text{c}$}; 
         \node[right,font=\footnotesize] at(axis cs:1.15,1){$1$}; 
        \end{axis}
        \end{tikzpicture}%
   };
   
\end{tikzpicture}
    \caption{Three channels with ICI.}
    \end{subfigure}\\[1em]
    \begin{subfigure}[t]{\columnwidth}
    \usetikzlibrary{decorations.markings}
\tikzset{node distance=2.3cm}

\pgfdeclarelayer{background}
\pgfdeclarelayer{foreground}
\pgfsetlayers{background,main,foreground}
\tikzset{boxlines/.style = {draw=black!20!white,}}
\tikzset{boxlinesred/.style = {densely dashed,draw=red!50!white,thick}}

\pgfmathsetmacro{\samplerwidth}{30}

\tikzset{midnodes/.style = {midway,above,text width=1.5cm,align=center,yshift=-0.2em}}
\tikzset{midnodesRP/.style = {midway,above,text width=1.5cm,align=center,yshift=-1.4em}}
\begin{tikzpicture}[]
    
    \node [input, name=input] {Input};
    \node [comblock,right of=input,node distance=3.6cm] (txfilter) {$g_\text{tx}(t) * g_\text{c}(t)$};

    \node [comblock,right of=txfilter,node distance=2.7cm] (channel_select) {$g_\text{c}(t)$};
    \node [comblock,right of=channel_select,node distance=1.4cm] (sld) {$\left\lvert \,\cdot\, \right\rvert^2$};
    \node [comblock,right of=sld,node distance=1.4cm] (rxfilter) {$g_\text{rx}(t)$};
    \node [node distance=0.9cm,right of=rxfilter] (ldots) {$\ldots$};
   
   \node[left of=txfilter,node distance=1.6cm,font=\footnotesize,text width=1cm,align=left]{Center\\ Channel};

   \draw[-latex,thick] (txfilter) -- (channel_select);
   \draw[-latex,thick] (channel_select) -- (sld);
   \draw[-latex,thick] (sld) -- (rxfilter);

\end{tikzpicture}
    \caption{Equivalent single channel without ICI. }
    \end{subfigure}
    \caption{WDM for $L=0$. The noise is neglected here.}
    \label{fig:system_model_wdm}
\end{figure}
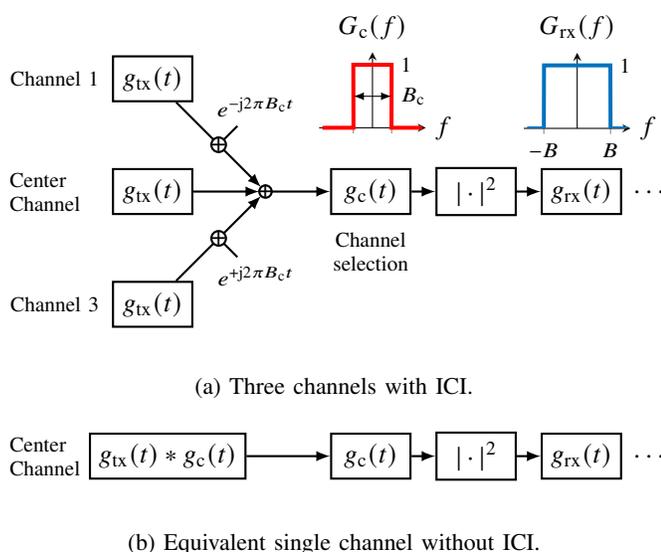

We consider two scenarios to compare the schemes, see Fig.~\ref{fig:system_model_wdm}. First, the case where one uses wavelength division multiplexing (WDM) to share the spectrum. We consider two neighbouring WDM channels centered at $\pm B_\text{c}$ from the channel of interest, see Fig.~\ref{fig:system_model_wdm} (a). TD-RC signals thus cause inter-channel interference (ICI) because their spectrum has infinite support. Second, the case where one uses an additional sinc pulse filter $g_\text{c}(t)$ with bandwidth $B_\text{c}$ at the transmitter to avoid ICI, see Fig.~\ref{fig:system_model_wdm} (b) where the new transmit filter is $g_\text{tx}(t)*g_\text{c}(t)$. For both cases, the filter $g_\text{c}(t)$ is also placed before the SLD to select the center channel, and there is an antialiasing filter $g_\text{rx}(t)$ of bandwidth $2B$ after the SLD, see~\eqref{eq:rx_ps_filter}. The receiver oversampling rate is $N_\text{os} = 2$.

\begin{figure*}[t!]
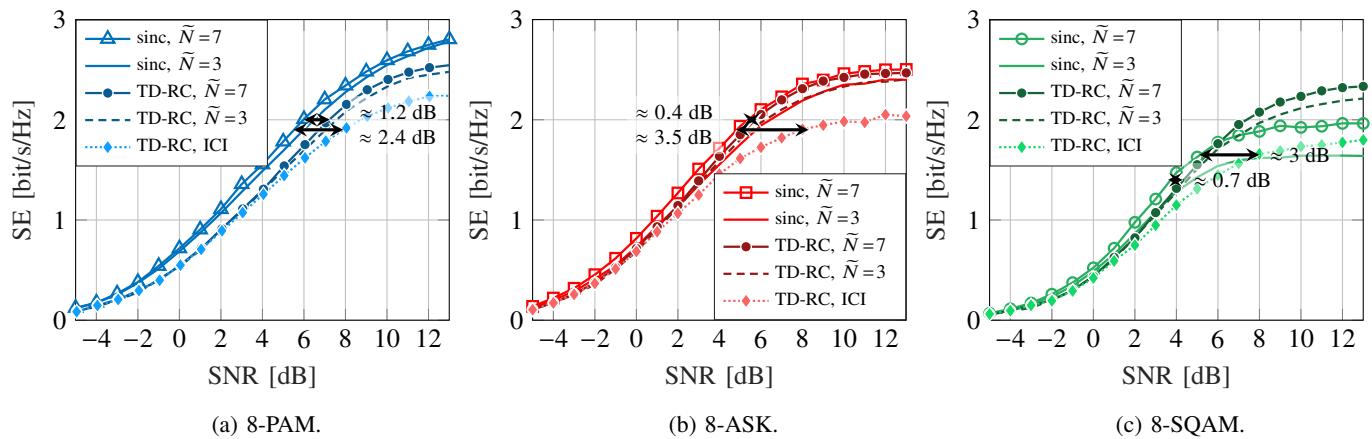
 %
\centering
\begin{subfigure}[b]{0.33\textwidth}
\pgfmathsetmacro\scaleTUKEYspeceff{0.15}    
\pgfdeclarelayer{fg}    %
\pgfdeclarelayer{bg}    %
\pgfsetlayers{bg,main,fg}  %
\centering
    \begin{tikzpicture}
    \begin{axis}[
      MIGeneralStyle,
      ymax=3.0,
      ylabel={SE [bit/s/Hz]},
      width=0.83\textwidth,
      legend style={at={(0,1)},anchor=north west}]
      
    \input{Figures/standalone/Q8/dataQ8_MI_TXPOW.tex}
    \addplot [PAM] table {\PAMrcosspanseven};
    \addplot [PAM,mark=none] table {\PAMrcosspanthree};
    \pgfplotstableread{
     X Y
   -11.9504    0.0034
  -10.9504    0.0076
   -9.9504    0.0114
   -8.9504    0.0181
   -7.9504    0.0352
   -6.9504    0.0381
   -5.9504    0.0678
   -4.9504    0.0977
   -3.9504    0.1696
   -2.9504    0.2379
   -1.9504    0.3410
   -0.9504    0.4589
    0.0496    0.6319
    1.0496    0.8130
    2.0496    1.0264
    3.0496    1.2351
    4.0496    1.4463
    5.0496    1.6606
    6.0496    1.8631
    7.0496    2.0538
    8.0496    2.2065
    9.0496    2.3381
   10.0496    2.4357
   11.0496    2.5116
   12.0496    2.5682
   13.0496    2.5758
   14.0496    2.6216
   15.0496    2.6225
   16.0496    2.6247
   17.0496    2.6338
   18.0496    2.6318
   19.0496    2.6584
 }{\PAMrcosspansevenICI}
 
 \pgfplotstableread{
     X Y
  -13.0000    0.0020
  -12.0000    0.0054
  -11.0000    0.0072
  -10.0000    0.0171
   -9.0000    0.0226
   -8.0000    0.0368
   -7.0000    0.0572
   -6.0000    0.0859
   -5.0000    0.1208
   -4.0000    0.1982
   -3.0000    0.2964
   -2.0000    0.4219
   -1.0000    0.5880
         0    0.7893
    1.0000    1.0156
    2.0000    1.2257
    3.0000    1.4371
    4.0000    1.6761
    5.0000    1.8555
    6.0000    1.9844
    7.0000    2.0937
    8.0000    2.1906
    9.0000    2.2385
   10.0000    2.2806
   11.0000    2.2709
   12.0000    2.3582
   13.0000    2.3416
   14.0000    2.3192
   15.0000    2.3615
   16.0000    2.3575
   17.0000    2.3588
   18.0000    2.3345
   }{\ASKrcosspansevenICI}

 \pgfplotstableread{
     X Y
   -13.0000    0.0009
  -12.0000    0.0030
  -11.0000    0.0027
  -10.0000    0.0092
   -9.0000    0.0106
   -8.0000    0.0167
   -7.0000    0.0257
   -6.0000    0.0447
   -5.0000    0.0709
   -4.0000    0.1090
   -3.0000    0.1713
   -2.0000    0.2231
   -1.0000    0.3363
         0    0.4843
    1.0000    0.6796
    2.0000    0.8618
    3.0000    1.0900
    4.0000    1.3210
    5.0000    1.5071
    6.0000    1.6647
    7.0000    1.7987
    8.0000    1.9090
    9.0000    1.9419
   10.0000    1.9938
   11.0000    2.0120
   12.0000    2.0401
   13.0000    2.0681
   14.0000    2.0542
   15.0000    2.0538
   16.0000    2.0723
   17.0000    2.0578
   18.0000    2.0566
}{\QAMrcosspansevenICI}

 \pgfplotstableread{
     X Y
  -13.0000    0.0020
  -12.0000    0.0054
  -11.0000    0.0072
  -10.0000    0.0172
   -9.0000    0.0226
   -8.0000    0.0368
   -7.0000    0.0573
   -6.0000    0.0857
   -5.0000    0.1221
   -4.0000    0.2006
   -3.0000    0.3004
   -2.0000    0.4312
   -1.0000    0.6057
         0    0.8225
    1.0000    1.0719
    2.0000    1.3162
    3.0000    1.5994
    4.0000    1.8810
    5.0000    2.1289
    6.0000    2.3611
    7.0000    2.5286
    8.0000    2.6567
    9.0000    2.7443
   10.0000    2.7861
   11.0000    2.8236
   12.0000    2.8317
   13.0000    2.8381
   14.0000    2.8642
   15.0000    2.8506
   16.0000    2.8597
   17.0000    2.8763
   18.0000    2.8787  
   }{\ASKrcosspanseven}
   
    \pgfplotstableread{
     X Y
     -13.0000    0.0029
  -12.0000    0.0048
  -11.0000    0.0096
  -10.0000    0.0135
   -9.0000    0.0208
   -8.0000    0.0371
   -7.0000    0.0542
   -6.0000    0.0884
   -5.0000    0.1350
   -4.0000    0.1981
   -3.0000    0.2959
   -2.0000    0.4093
   -1.0000    0.5884
         0    0.7851
    1.0000    1.0471
    2.0000    1.2934
    3.0000    1.5664
    4.0000    1.8249
    5.0000    2.0701
    6.0000    2.2685
    7.0000    2.4176
    8.0000    2.5387
    9.0000    2.6167
   10.0000    2.6811
   11.0000    2.7316
   12.0000    2.7374
   13.0000    2.7643
   14.0000    2.7669
   15.0000    2.7659
   16.0000    2.7718
   17.0000    2.7753
   18.0000    2.7797
    }{\ASKrcosspanthree}
 
  \pgfplotstableread{
     X Y
  -13.0000    0.0009
  -12.0000    0.0030
  -11.0000    0.0027
  -10.0000    0.0093
   -9.0000    0.0107
   -8.0000    0.0168
   -7.0000    0.0260
   -6.0000    0.0449
   -5.0000    0.0717
   -4.0000    0.1100
   -3.0000    0.1746
   -2.0000    0.2300
   -1.0000    0.3454
         0    0.5034
    1.0000    0.7211
    2.0000    0.9449
    3.0000    1.2346
    4.0000    1.5130
    5.0000    1.7846
    6.0000    2.0260
    7.0000    2.2457
    8.0000    2.3877
    9.0000    2.5040
   10.0000    2.5686
   11.0000    2.6322
   12.0000    2.6688
   13.0000    2.6836
   14.0000    2.6991
   15.0000    2.7166
   16.0000    2.7044
   17.0000    2.7099
   18.0000    2.7196
   }{\QAMrcosspanseven}

   \pgfplotstableread{
     X Y
      -13.0000    0.0020
  -12.0000    0.0012
  -11.0000    0.0038
  -10.0000    0.0061
   -9.0000    0.0113
   -8.0000    0.0185
   -7.0000    0.0274
   -6.0000    0.0399
   -5.0000    0.0647
   -4.0000    0.1036
   -3.0000    0.1417
   -2.0000    0.2321
   -1.0000    0.3400
         0    0.5041
    1.0000    0.7007
    2.0000    0.9533
    3.0000    1.2099
    4.0000    1.4689
    5.0000    1.7354
    6.0000    1.9679
    7.0000    2.1199
    8.0000    2.2644
    9.0000    2.3598
   10.0000    2.4329
   11.0000    2.4797
   12.0000    2.5193
   13.0000    2.5438
   14.0000    2.5524
   15.0000    2.5630
   16.0000    2.5730
   17.0000    2.5745
   18.0000    2.5732
    }{\QAMrcosspanthree}

 \pgfplotstableread{
     X Y
    -11.9504    0.0034
  -10.9504    0.0076
   -9.9504    0.0115
   -8.9504    0.0181
   -7.9504    0.0353
   -6.9504    0.0381
   -5.9504    0.0678
   -4.9504    0.0981
   -3.9504    0.1693
   -2.9504    0.2383
   -1.9504    0.3431
   -0.9504    0.4624
    0.0496    0.6363
    1.0496    0.8248
    2.0496    1.0499
    3.0496    1.2774
    4.0496    1.5035
    5.0496    1.7703
    6.0496    2.0150
    7.0496    2.2659
    8.0496    2.4767
    9.0496    2.6444
   10.0496    2.7628
   11.0496    2.8465
   12.0496    2.8984
   13.0496    2.9278
   14.0496    2.9423
   15.0496    2.9541
   16.0496    2.9640
   17.0496    2.9662
   18.0496    2.9711
   19.0496    2.9713
 }{\PAMrcosspanseven}
 
  \pgfplotstableread{
     X Y
 -11.9504    0.0046
  -10.9504    0.0052
   -9.9504    0.0109
   -8.9504    0.0175
   -7.9504    0.0271
   -6.9504    0.0414
   -5.9504    0.0654
   -4.9504    0.0994
   -3.9504    0.1594
   -2.9504    0.2328
   -1.9504    0.3227
   -0.9504    0.4755
    0.0496    0.6249
    1.0496    0.8121
    2.0496    1.0533
    3.0496    1.2594
    4.0496    1.4955
    5.0496    1.7336
    6.0496    1.9575
    7.0496    2.1871
    8.0496    2.4005
    9.0496    2.5536
   10.0496    2.6856
   11.0496    2.7795
   12.0496    2.8183
   13.0496    2.8522
   14.0496    2.8760
   15.0496    2.8910
   16.0496    2.8960
   17.0496    2.8997
   18.0496    2.9045
   19.0496    2.9044
   }{\PAMrcosspanthree}
 
    \addplot [PAM,TDRC,mycolor1dark] table[y expr=\thisrowno{1}*1/(1+\scaleTUKEYspeceff)] {\PAMrcosspanseven};
    \addplot [PAM,densely dashed,mark=none,mycolor1dark] table[y expr=\thisrowno{1}*1/(1+\scaleTUKEYspeceff)] {\PAMrcosspanthree};

    \addplot [PAM,TDRC_ICI,mycolor1bright,densely dotted] table[y expr=\thisrowno{1}*1/(1+\scaleTUKEYspeceff)] {\PAMrcosspansevenICI};

    \addlegendentry{sinc, \smash{$\widetilde{N} \!=\! 7$}}
    \addlegendentry{sinc, \smash{$\widetilde{N} \!=\! 3$}}
    \addlegendentry{TD-RC, \smash{$\widetilde{N} \!=\! 7$} }
    \addlegendentry{TD-RC, \smash{$\widetilde{N} \!=\! 3$}}
    \addlegendentry{TD-RC, ICI}
    
    \begin{pgfonlayer}{fg}  
        \draw[arr,shadowed] (axis cs:6.0,2)node[font=\footnotesize,right,xshift=+0.60cm,yshift=+0.1cm,fill=white,opacity=0.5,text opacity=1]{$\approx 1.2 \;\text{dB}$} --  (axis cs:7.25,2) ;
        \draw[arr,shadowed] (axis cs:5.5,1.9) node[font=\footnotesize,right,xshift=+0.73cm,yshift=-0.1cm,fill=white,opacity=0.5,text opacity=1]{$\approx 2.4 \;\text{dB}$}
        -- (axis cs:7.9,1.9) ;
    \end{pgfonlayer}
    
    \end{axis}
    \end{tikzpicture}%
    \captionsetup{width=.86\linewidth,margin={1cm,0cm}}
    \caption{$8$-PAM.}

\end{subfigure}%
\hspace*{\fill}
\begin{subfigure}[b]{0.33\textwidth}
\pgfmathsetmacro\scaleTUKEYspeceff{0.15}    
\pgfdeclarelayer{fg}    %
\pgfdeclarelayer{bg}    %
\pgfsetlayers{bg,main,fg}  %
\centering
    \begin{tikzpicture}
       \begin{axis}[
      MIGeneralStyle,
      ymax=3.0,
      ylabel={SE [bit/s/Hz]},
      width=0.83\textwidth,
      legend style={at={(1,0)},anchor=south east}]
      
    \input{Figures/standalone/Q8/dataQ8_MI_TXPOW.tex}
    \addplot [ASK] table {\ASKrcosspanseven};
    \addplot [ASK,mark=none] table {\ASKrcosspanthree};
    \pgfplotstableread{
     X Y
   -11.9504    0.0034
  -10.9504    0.0076
   -9.9504    0.0114
   -8.9504    0.0181
   -7.9504    0.0352
   -6.9504    0.0381
   -5.9504    0.0678
   -4.9504    0.0977
   -3.9504    0.1696
   -2.9504    0.2379
   -1.9504    0.3410
   -0.9504    0.4589
    0.0496    0.6319
    1.0496    0.8130
    2.0496    1.0264
    3.0496    1.2351
    4.0496    1.4463
    5.0496    1.6606
    6.0496    1.8631
    7.0496    2.0538
    8.0496    2.2065
    9.0496    2.3381
   10.0496    2.4357
   11.0496    2.5116
   12.0496    2.5682
   13.0496    2.5758
   14.0496    2.6216
   15.0496    2.6225
   16.0496    2.6247
   17.0496    2.6338
   18.0496    2.6318
   19.0496    2.6584
 }{\PAMrcosspansevenICI}
 
 \pgfplotstableread{
     X Y
  -13.0000    0.0020
  -12.0000    0.0054
  -11.0000    0.0072
  -10.0000    0.0171
   -9.0000    0.0226
   -8.0000    0.0368
   -7.0000    0.0572
   -6.0000    0.0859
   -5.0000    0.1208
   -4.0000    0.1982
   -3.0000    0.2964
   -2.0000    0.4219
   -1.0000    0.5880
         0    0.7893
    1.0000    1.0156
    2.0000    1.2257
    3.0000    1.4371
    4.0000    1.6761
    5.0000    1.8555
    6.0000    1.9844
    7.0000    2.0937
    8.0000    2.1906
    9.0000    2.2385
   10.0000    2.2806
   11.0000    2.2709
   12.0000    2.3582
   13.0000    2.3416
   14.0000    2.3192
   15.0000    2.3615
   16.0000    2.3575
   17.0000    2.3588
   18.0000    2.3345
   }{\ASKrcosspansevenICI}

 \pgfplotstableread{
     X Y
   -13.0000    0.0009
  -12.0000    0.0030
  -11.0000    0.0027
  -10.0000    0.0092
   -9.0000    0.0106
   -8.0000    0.0167
   -7.0000    0.0257
   -6.0000    0.0447
   -5.0000    0.0709
   -4.0000    0.1090
   -3.0000    0.1713
   -2.0000    0.2231
   -1.0000    0.3363
         0    0.4843
    1.0000    0.6796
    2.0000    0.8618
    3.0000    1.0900
    4.0000    1.3210
    5.0000    1.5071
    6.0000    1.6647
    7.0000    1.7987
    8.0000    1.9090
    9.0000    1.9419
   10.0000    1.9938
   11.0000    2.0120
   12.0000    2.0401
   13.0000    2.0681
   14.0000    2.0542
   15.0000    2.0538
   16.0000    2.0723
   17.0000    2.0578
   18.0000    2.0566
}{\QAMrcosspansevenICI}

 \pgfplotstableread{
     X Y
  -13.0000    0.0020
  -12.0000    0.0054
  -11.0000    0.0072
  -10.0000    0.0172
   -9.0000    0.0226
   -8.0000    0.0368
   -7.0000    0.0573
   -6.0000    0.0857
   -5.0000    0.1221
   -4.0000    0.2006
   -3.0000    0.3004
   -2.0000    0.4312
   -1.0000    0.6057
         0    0.8225
    1.0000    1.0719
    2.0000    1.3162
    3.0000    1.5994
    4.0000    1.8810
    5.0000    2.1289
    6.0000    2.3611
    7.0000    2.5286
    8.0000    2.6567
    9.0000    2.7443
   10.0000    2.7861
   11.0000    2.8236
   12.0000    2.8317
   13.0000    2.8381
   14.0000    2.8642
   15.0000    2.8506
   16.0000    2.8597
   17.0000    2.8763
   18.0000    2.8787  
   }{\ASKrcosspanseven}
   
    \pgfplotstableread{
     X Y
     -13.0000    0.0029
  -12.0000    0.0048
  -11.0000    0.0096
  -10.0000    0.0135
   -9.0000    0.0208
   -8.0000    0.0371
   -7.0000    0.0542
   -6.0000    0.0884
   -5.0000    0.1350
   -4.0000    0.1981
   -3.0000    0.2959
   -2.0000    0.4093
   -1.0000    0.5884
         0    0.7851
    1.0000    1.0471
    2.0000    1.2934
    3.0000    1.5664
    4.0000    1.8249
    5.0000    2.0701
    6.0000    2.2685
    7.0000    2.4176
    8.0000    2.5387
    9.0000    2.6167
   10.0000    2.6811
   11.0000    2.7316
   12.0000    2.7374
   13.0000    2.7643
   14.0000    2.7669
   15.0000    2.7659
   16.0000    2.7718
   17.0000    2.7753
   18.0000    2.7797
    }{\ASKrcosspanthree}
 
  \pgfplotstableread{
     X Y
  -13.0000    0.0009
  -12.0000    0.0030
  -11.0000    0.0027
  -10.0000    0.0093
   -9.0000    0.0107
   -8.0000    0.0168
   -7.0000    0.0260
   -6.0000    0.0449
   -5.0000    0.0717
   -4.0000    0.1100
   -3.0000    0.1746
   -2.0000    0.2300
   -1.0000    0.3454
         0    0.5034
    1.0000    0.7211
    2.0000    0.9449
    3.0000    1.2346
    4.0000    1.5130
    5.0000    1.7846
    6.0000    2.0260
    7.0000    2.2457
    8.0000    2.3877
    9.0000    2.5040
   10.0000    2.5686
   11.0000    2.6322
   12.0000    2.6688
   13.0000    2.6836
   14.0000    2.6991
   15.0000    2.7166
   16.0000    2.7044
   17.0000    2.7099
   18.0000    2.7196
   }{\QAMrcosspanseven}

   \pgfplotstableread{
     X Y
      -13.0000    0.0020
  -12.0000    0.0012
  -11.0000    0.0038
  -10.0000    0.0061
   -9.0000    0.0113
   -8.0000    0.0185
   -7.0000    0.0274
   -6.0000    0.0399
   -5.0000    0.0647
   -4.0000    0.1036
   -3.0000    0.1417
   -2.0000    0.2321
   -1.0000    0.3400
         0    0.5041
    1.0000    0.7007
    2.0000    0.9533
    3.0000    1.2099
    4.0000    1.4689
    5.0000    1.7354
    6.0000    1.9679
    7.0000    2.1199
    8.0000    2.2644
    9.0000    2.3598
   10.0000    2.4329
   11.0000    2.4797
   12.0000    2.5193
   13.0000    2.5438
   14.0000    2.5524
   15.0000    2.5630
   16.0000    2.5730
   17.0000    2.5745
   18.0000    2.5732
    }{\QAMrcosspanthree}

 \pgfplotstableread{
     X Y
    -11.9504    0.0034
  -10.9504    0.0076
   -9.9504    0.0115
   -8.9504    0.0181
   -7.9504    0.0353
   -6.9504    0.0381
   -5.9504    0.0678
   -4.9504    0.0981
   -3.9504    0.1693
   -2.9504    0.2383
   -1.9504    0.3431
   -0.9504    0.4624
    0.0496    0.6363
    1.0496    0.8248
    2.0496    1.0499
    3.0496    1.2774
    4.0496    1.5035
    5.0496    1.7703
    6.0496    2.0150
    7.0496    2.2659
    8.0496    2.4767
    9.0496    2.6444
   10.0496    2.7628
   11.0496    2.8465
   12.0496    2.8984
   13.0496    2.9278
   14.0496    2.9423
   15.0496    2.9541
   16.0496    2.9640
   17.0496    2.9662
   18.0496    2.9711
   19.0496    2.9713
 }{\PAMrcosspanseven}
 
  \pgfplotstableread{
     X Y
 -11.9504    0.0046
  -10.9504    0.0052
   -9.9504    0.0109
   -8.9504    0.0175
   -7.9504    0.0271
   -6.9504    0.0414
   -5.9504    0.0654
   -4.9504    0.0994
   -3.9504    0.1594
   -2.9504    0.2328
   -1.9504    0.3227
   -0.9504    0.4755
    0.0496    0.6249
    1.0496    0.8121
    2.0496    1.0533
    3.0496    1.2594
    4.0496    1.4955
    5.0496    1.7336
    6.0496    1.9575
    7.0496    2.1871
    8.0496    2.4005
    9.0496    2.5536
   10.0496    2.6856
   11.0496    2.7795
   12.0496    2.8183
   13.0496    2.8522
   14.0496    2.8760
   15.0496    2.8910
   16.0496    2.8960
   17.0496    2.8997
   18.0496    2.9045
   19.0496    2.9044
   }{\PAMrcosspanthree}
 
    \addplot [ASK,TDRC,mycolor6dark] table[y expr=\thisrowno{1}*1/(1+\scaleTUKEYspeceff)] {\ASKrcosspanseven};
    \addplot [ASK,densely dashed,mark=none,mycolor6dark] table[y expr=\thisrowno{1}*1/(1+\scaleTUKEYspeceff)] {\ASKrcosspanthree};
    \addplot [ASK,TDRC_ICI,densely dotted,mycolor6bright] table[y expr=\thisrowno{1}*1/(1+\scaleTUKEYspeceff)] {\ASKrcosspansevenICI};
    \addlegendentry{sinc, \smash{$\widetilde{N} \!=\! 7$}}
    \addlegendentry{sinc, \smash{$\widetilde{N} \!=\! 3$}}
    \addlegendentry{TD-RC, \smash{$\widetilde{N} \!=\! 7$} }
    \addlegendentry{TD-RC, \smash{$\widetilde{N} \!=\! 3$}}
    \addlegendentry{TD-RC, ICI}
    
    \begin{pgfonlayer}{fg}  
        \draw[arr,shadowed] (axis cs:5.3,2)node[font=\footnotesize,left,xshift=-0.3cm,yshift=+0.1cm,fill=white,opacity=0.5,text opacity=1]{$\approx 0.4 \;\text{dB}$} --  (axis cs:5.8,2) ;
        \draw[arr,shadowed] (axis cs:4.8,1.9) node[font=\footnotesize,left,xshift=-0.17cm,yshift=-0.1cm,fill=white,opacity=0.5,text opacity=1]{$\approx 3.5 \;\text{dB}$}
        -- (axis cs:8.3,1.9) ;
    \end{pgfonlayer}
    
    \end{axis}
    \end{tikzpicture}%
    \captionsetup{width=.86\linewidth,margin={1cm,0cm}}
    \caption{$8$-ASK.}

\end{subfigure}%
\hspace*{\fill}
\begin{subfigure}[b]{0.33\textwidth}
\pgfmathsetmacro\scaleTUKEYspeceff{0.15}    
\pgfdeclarelayer{fg}    %
\pgfdeclarelayer{bg}    %
\pgfsetlayers{bg,main,fg}  %
\centering
    \begin{tikzpicture}
        \begin{axis}[
      MIGeneralStyle,
      ymax=3.0,
      ylabel={SE [bit/s/Hz]},
      width=0.83\textwidth,
      legend style={at={(0,1)},anchor=north west}]
      
    \input{Figures/standalone/Q8/dataQ8_MI_TXPOW.tex}
    \addplot [QAM] table {\QAMrcosspanseven};
    \addplot [QAM,mark=none] table {\QAMrcosspanthree};
    
    \pgfplotstableread{
     X Y
   -11.9504    0.0034
  -10.9504    0.0076
   -9.9504    0.0114
   -8.9504    0.0181
   -7.9504    0.0352
   -6.9504    0.0381
   -5.9504    0.0678
   -4.9504    0.0977
   -3.9504    0.1696
   -2.9504    0.2379
   -1.9504    0.3410
   -0.9504    0.4589
    0.0496    0.6319
    1.0496    0.8130
    2.0496    1.0264
    3.0496    1.2351
    4.0496    1.4463
    5.0496    1.6606
    6.0496    1.8631
    7.0496    2.0538
    8.0496    2.2065
    9.0496    2.3381
   10.0496    2.4357
   11.0496    2.5116
   12.0496    2.5682
   13.0496    2.5758
   14.0496    2.6216
   15.0496    2.6225
   16.0496    2.6247
   17.0496    2.6338
   18.0496    2.6318
   19.0496    2.6584
 }{\PAMrcosspansevenICI}
 
 \pgfplotstableread{
     X Y
  -13.0000    0.0020
  -12.0000    0.0054
  -11.0000    0.0072
  -10.0000    0.0171
   -9.0000    0.0226
   -8.0000    0.0368
   -7.0000    0.0572
   -6.0000    0.0859
   -5.0000    0.1208
   -4.0000    0.1982
   -3.0000    0.2964
   -2.0000    0.4219
   -1.0000    0.5880
         0    0.7893
    1.0000    1.0156
    2.0000    1.2257
    3.0000    1.4371
    4.0000    1.6761
    5.0000    1.8555
    6.0000    1.9844
    7.0000    2.0937
    8.0000    2.1906
    9.0000    2.2385
   10.0000    2.2806
   11.0000    2.2709
   12.0000    2.3582
   13.0000    2.3416
   14.0000    2.3192
   15.0000    2.3615
   16.0000    2.3575
   17.0000    2.3588
   18.0000    2.3345
   }{\ASKrcosspansevenICI}

 \pgfplotstableread{
     X Y
   -13.0000    0.0009
  -12.0000    0.0030
  -11.0000    0.0027
  -10.0000    0.0092
   -9.0000    0.0106
   -8.0000    0.0167
   -7.0000    0.0257
   -6.0000    0.0447
   -5.0000    0.0709
   -4.0000    0.1090
   -3.0000    0.1713
   -2.0000    0.2231
   -1.0000    0.3363
         0    0.4843
    1.0000    0.6796
    2.0000    0.8618
    3.0000    1.0900
    4.0000    1.3210
    5.0000    1.5071
    6.0000    1.6647
    7.0000    1.7987
    8.0000    1.9090
    9.0000    1.9419
   10.0000    1.9938
   11.0000    2.0120
   12.0000    2.0401
   13.0000    2.0681
   14.0000    2.0542
   15.0000    2.0538
   16.0000    2.0723
   17.0000    2.0578
   18.0000    2.0566
}{\QAMrcosspansevenICI}

 \pgfplotstableread{
     X Y
  -13.0000    0.0020
  -12.0000    0.0054
  -11.0000    0.0072
  -10.0000    0.0172
   -9.0000    0.0226
   -8.0000    0.0368
   -7.0000    0.0573
   -6.0000    0.0857
   -5.0000    0.1221
   -4.0000    0.2006
   -3.0000    0.3004
   -2.0000    0.4312
   -1.0000    0.6057
         0    0.8225
    1.0000    1.0719
    2.0000    1.3162
    3.0000    1.5994
    4.0000    1.8810
    5.0000    2.1289
    6.0000    2.3611
    7.0000    2.5286
    8.0000    2.6567
    9.0000    2.7443
   10.0000    2.7861
   11.0000    2.8236
   12.0000    2.8317
   13.0000    2.8381
   14.0000    2.8642
   15.0000    2.8506
   16.0000    2.8597
   17.0000    2.8763
   18.0000    2.8787  
   }{\ASKrcosspanseven}
   
    \pgfplotstableread{
     X Y
     -13.0000    0.0029
  -12.0000    0.0048
  -11.0000    0.0096
  -10.0000    0.0135
   -9.0000    0.0208
   -8.0000    0.0371
   -7.0000    0.0542
   -6.0000    0.0884
   -5.0000    0.1350
   -4.0000    0.1981
   -3.0000    0.2959
   -2.0000    0.4093
   -1.0000    0.5884
         0    0.7851
    1.0000    1.0471
    2.0000    1.2934
    3.0000    1.5664
    4.0000    1.8249
    5.0000    2.0701
    6.0000    2.2685
    7.0000    2.4176
    8.0000    2.5387
    9.0000    2.6167
   10.0000    2.6811
   11.0000    2.7316
   12.0000    2.7374
   13.0000    2.7643
   14.0000    2.7669
   15.0000    2.7659
   16.0000    2.7718
   17.0000    2.7753
   18.0000    2.7797
    }{\ASKrcosspanthree}
 
  \pgfplotstableread{
     X Y
  -13.0000    0.0009
  -12.0000    0.0030
  -11.0000    0.0027
  -10.0000    0.0093
   -9.0000    0.0107
   -8.0000    0.0168
   -7.0000    0.0260
   -6.0000    0.0449
   -5.0000    0.0717
   -4.0000    0.1100
   -3.0000    0.1746
   -2.0000    0.2300
   -1.0000    0.3454
         0    0.5034
    1.0000    0.7211
    2.0000    0.9449
    3.0000    1.2346
    4.0000    1.5130
    5.0000    1.7846
    6.0000    2.0260
    7.0000    2.2457
    8.0000    2.3877
    9.0000    2.5040
   10.0000    2.5686
   11.0000    2.6322
   12.0000    2.6688
   13.0000    2.6836
   14.0000    2.6991
   15.0000    2.7166
   16.0000    2.7044
   17.0000    2.7099
   18.0000    2.7196
   }{\QAMrcosspanseven}

   \pgfplotstableread{
     X Y
      -13.0000    0.0020
  -12.0000    0.0012
  -11.0000    0.0038
  -10.0000    0.0061
   -9.0000    0.0113
   -8.0000    0.0185
   -7.0000    0.0274
   -6.0000    0.0399
   -5.0000    0.0647
   -4.0000    0.1036
   -3.0000    0.1417
   -2.0000    0.2321
   -1.0000    0.3400
         0    0.5041
    1.0000    0.7007
    2.0000    0.9533
    3.0000    1.2099
    4.0000    1.4689
    5.0000    1.7354
    6.0000    1.9679
    7.0000    2.1199
    8.0000    2.2644
    9.0000    2.3598
   10.0000    2.4329
   11.0000    2.4797
   12.0000    2.5193
   13.0000    2.5438
   14.0000    2.5524
   15.0000    2.5630
   16.0000    2.5730
   17.0000    2.5745
   18.0000    2.5732
    }{\QAMrcosspanthree}

 \pgfplotstableread{
     X Y
    -11.9504    0.0034
  -10.9504    0.0076
   -9.9504    0.0115
   -8.9504    0.0181
   -7.9504    0.0353
   -6.9504    0.0381
   -5.9504    0.0678
   -4.9504    0.0981
   -3.9504    0.1693
   -2.9504    0.2383
   -1.9504    0.3431
   -0.9504    0.4624
    0.0496    0.6363
    1.0496    0.8248
    2.0496    1.0499
    3.0496    1.2774
    4.0496    1.5035
    5.0496    1.7703
    6.0496    2.0150
    7.0496    2.2659
    8.0496    2.4767
    9.0496    2.6444
   10.0496    2.7628
   11.0496    2.8465
   12.0496    2.8984
   13.0496    2.9278
   14.0496    2.9423
   15.0496    2.9541
   16.0496    2.9640
   17.0496    2.9662
   18.0496    2.9711
   19.0496    2.9713
 }{\PAMrcosspanseven}
 
  \pgfplotstableread{
     X Y
 -11.9504    0.0046
  -10.9504    0.0052
   -9.9504    0.0109
   -8.9504    0.0175
   -7.9504    0.0271
   -6.9504    0.0414
   -5.9504    0.0654
   -4.9504    0.0994
   -3.9504    0.1594
   -2.9504    0.2328
   -1.9504    0.3227
   -0.9504    0.4755
    0.0496    0.6249
    1.0496    0.8121
    2.0496    1.0533
    3.0496    1.2594
    4.0496    1.4955
    5.0496    1.7336
    6.0496    1.9575
    7.0496    2.1871
    8.0496    2.4005
    9.0496    2.5536
   10.0496    2.6856
   11.0496    2.7795
   12.0496    2.8183
   13.0496    2.8522
   14.0496    2.8760
   15.0496    2.8910
   16.0496    2.8960
   17.0496    2.8997
   18.0496    2.9045
   19.0496    2.9044
   }{\PAMrcosspanthree}
 
    \addplot [QAM,TDRC,mycolor5dark] table[y expr=\thisrowno{1}*1/(1+\scaleTUKEYspeceff)] {\QAMrcosspanseven};
    
    \addplot [QAM,densely dashed,mark=none,mycolor5dark] table[y expr=\thisrowno{1}*1/(1+\scaleTUKEYspeceff)] {\QAMrcosspanthree};

    \addplot [QAM,TDRC_ICI,densely dotted,mycolor5bright] table[y expr=\thisrowno{1}*1/(1+\scaleTUKEYspeceff)] {\QAMrcosspansevenICI};

    \addlegendentry{sinc, \smash{$\widetilde{N} \!=\! 7$}}
    \addlegendentry{sinc, \smash{$\widetilde{N} \!=\! 3$}}
    \addlegendentry{TD-RC, \smash{$\widetilde{N} \!=\! 7$} }
    \addlegendentry{TD-RC, \smash{$\widetilde{N} \!=\! 3$}}
    \addlegendentry{TD-RC, ICI}
    
    \begin{pgfonlayer}{fg}  
        \draw[arr,shadowed] (axis cs:5.1,1.65)node[font=\footnotesize,right,xshift=+0.8cm,yshift=-0.0cm,fill=white,opacity=0.4,text opacity=1]{$\approx 3 \;\text{dB}$} --  (axis cs:8.0,1.65) ;
        \draw[arr,shadowed] (axis cs:3.6,1.4)node[font=\footnotesize,right,xshift=+0.2cm,yshift=-0.0cm,fill=white,opacity=0.4,text opacity=1]{$\approx 0.7 \;\text{dB}$} --  (axis cs:4.3,1.4) ;
    \end{pgfonlayer}
    
    \end{axis}
    \end{tikzpicture}%
    \captionsetup{width=.86\linewidth,margin={1cm,0cm}}
    \caption{$8$-SQAM.}

\end{subfigure}
\caption{SEs for $L=0$, $Q=8$, a sinc pulse and a TD-RC pulse. }
\label{fig:comparison_tdrc_to_fdrc}
\end{figure*}

Fig.~\ref{fig:comparison_tdrc_to_fdrc} compares the performance for $L=0$, $Q=8$, the sinc pulse~\eqref{eq:tx_ps_filter} and the TD-RC pulse with roll-off factor $\alpha=0.9$.
For the latter pulse, the filter $g_\text{rx}(t)$ passes 99\% of the energy of $Z^\prime(t)$.  The filters $g_\text{c}(t)$ and $g_\text{rx}(t)$ cause additional ISI so we set the maximal detector memory to $\smash{\widetilde{N}=7}$ for both pulses. Fig.~\ref{fig:comparison_tdrc_to_fdrc} shows that ICI reduces the SE of TD-RC. Sinc pulses gain \SI{2.4}{dB} and \SI{3.5}{dB} over TD-RC with ICI for $8$-PAM and $8$-ASK, respectively, when operating at a SE of approximately \SI{1.9}{bit/s/Hz}. The gain is \SI{3}{dB} for $8$-SQAM when operating at a SE of approximately \SI{1.6}{bit/s/Hz}.

To avoid or limit ICI, we choose  $B_\text{c}=B$ for the sinc pulse and $B_\text{c} = 1.15\,B$ for the TD-RC pulse and the transmit filter $g_\text{c}(t)$. Fig.~\ref{fig:comparison_tdrc_to_fdrc} shows that the sinc pulse has better SEs for PAM and ASK, and these two constellations have higher rates than SQAM. Sinc pulses gain \SI{1.2}{dB} and  \SI{0.4}{dB} over TD-RC for $8$-PAM and $8$-ASK, respectively, when operating at a SE of approximately \SI{1.9}{bit/s/Hz}.
For $8$-SQAM and SEs near $\SI{1.4}{bit/s/Hz}$, sinc pulses gain \SI{0.7}{dB} over TD-RC. 
At high-SNR, the TD-RC pulse outperforms the sinc pulse because of the auxiliary model mismatch. 

Fig.~\ref{fig:comparison_tdrc_to_fdrc} also shows the SEs if the receiver uses a lower-complexity forward-backward algorithm with $\smash{\widetilde{N} = 3}$. For $8$-PAM, the sinc pulse outperforms the TD-RC pulse for all SNRs. For $8$-ASK, the two schemes perform almost the same. For $8$-SQAM, one might prefer the TD-RC pulse because it has a smaller auxiliary model mismatch. However, we expect that the sinc pulse has more potential if one can find low-complexity detection algorithms that perform almost as well as the forward-backward algorithm.

\section{Conclusions}
\label{sec:conclusions}
We investigated spectrally efficient communication for short-reach fiber-optic links with DD. CD mitigates DD phase ambiguities by distributing the amplitude and phase information onto receiver samples at both symbol- and half-symbol times. This increases the achievable rates and reduces error propagation for symbol-wise MAP estimation. FD-RC pulses outperform TD-RC pulses in terms of SE.

There are many questions to explore. For example, one could design precoders to increase rates via higher-order complex-valued modulation formats. One can further remove sequences that cause phase ambiguities~\cite{tasbihi2021direct} and this approach can readily be integrated in forward-backward processing. Next, one might consider pulse shaping with PSWFs that are band-limited and have the maximum energy in a given time interval, a property that should decrease the auxiliary model mismatch for a fixed memory $\smash{\widetilde{N}}$. The design of complex-valued pulse shapes~\cite{mecozzi_capacity_amdd_2018} can increase the number of distinct receiver strings and thereby the achievable rates. One can also optimize the achievable rates via geometric and/or probabilistic shaping of the constellation points. Finally, one can design codes and simplified detection and decoding algorithms.
\section*{Acknowledgment}
\noindent The authors wish to thank F. J. Garc\'{i}a-G\'{o}mez and the reviewers for useful suggestions.

\section*{Appendix}
\noindent 
This appendix shows how to minimize the divergence $\Div{p_{\bm{X}\bm{Y}}}{p_{\bm{X}} \cdot q_{\bm{Y}\lvert\bm{X}}}$ specified in Sec.~\ref{sec:computing-rates}.
Consider the auxiliary channel $q(\bm{y}\lvert \bm{x}) = \mathcal{N}(\bm{y}-\lvert \bm{\Psi^\prime} \tilde{\bm{x}}' \rvert^{\circ 2} ;\bm{\upmu}_{\bm{Q}} , \bm{C}_{\bm{QQ}})$ and define the interference-plus-noise signal
\begin{align}
    \bm{w} = \bm{y}-\lvert \bm{\Psi^\prime} \tilde{\bm{x}}' \rvert^{\circ 2}
\end{align}
where $\tilde{\bm{x}}'$ is an oversampled vector that is extended by the initial channel state $\bm{s}_0$ as in \eqref{eq:Zvector}. Observe that $\tilde{\bm{x}}'$ is a function of $\bm{x}$ and $\bm{s}_0$. The optimization problem we wish to solve is 
\begin{equation}
\begin{aligned}
    & \argmin_{\bm{\upmu}_{\bm{Q}}, \bm{C}_{\bm{QQ}}} \quad \Div{p_{\bm{X}\bm{Y}}}{p_{\bm{X}} \cdot q_{\bm{Y}\lvert\bm{X}}} \nonumber \\
    & = \argmin_{\bm{\upmu}_{\bm{Q}}, \bm{C}_{\bm{QQ}}} \; \E[- \log_2 q(\bm{Y}\lvert \bm{X})] \nonumber \\
    & = \argmin_{\bm{\upmu}_{\bm{Q}}, \bm{C}_{\bm{QQ}}} \; \ln \det{\bm{C}_{\bm{QQ}}} +
    \trace\left(\bm{C}_{\bm{QQ}}^{-1} \E\left[(\bm{W}-\bm{\upmu}_{\bm{Q}})(\bm{W}-\bm{\upmu}_{\bm{Q}})^\mathrm{T} \right] \right).
    \label{eq:objective_proxy_simplified}
\end{aligned}
\end{equation}
Taking derivatives with respect to $\bm{\upmu}_{\bm{Q}}$ and $\bm{C}_{\bm{QQ}}$, and setting them to zero, we obtain
\begin{align}
    \bm{\upmu}_{\bm{Q}} = \E[\bm{W}], \quad
    \bm{C}_{\bm{QQ}} = \E\left[(\bm{W}-\bm{\upmu}_{\bm{Q}})(\bm{W}-\bm{\upmu}_{\bm{Q}})^\mathrm{T} \right] .
    \label{eq:argmin_auxiliary_mean_variance}
\end{align}

\bibliographystyle{IEEEtran}
\bibliography{IEEEabrv,imdd}

\end{document}